\documentclass[prd,aps,reprint,nofootinbib,amsmath,amssymb,eqsecnum]{revtex4-2}

\usepackage[T1]{fontenc}
\usepackage[utf8]{inputenc}
\usepackage{booktabs}
\usepackage{array}[=2016-10-06]%
\usepackage{xcolor}
\usepackage{tensor}
\usepackage{tikz}
\usetikzlibrary{decorations.pathmorphing,decorations.markings,arrows.meta}
\tikzset{
  wig/.style={decorate,decoration={snake,amplitude=1.4pt,segment length=5pt,pre length=1pt,post length=1pt}},
  grav/.style={draw=none,postaction={draw,wig,line width=0.5pt}},
  aux/.style={draw=none,postaction={draw,wig,line width=1.6pt,draw=black},
    postaction={draw,wig,line width=0.7pt,draw=white}},
  gh/.style={line width=0.5pt,dash pattern=on 1.7pt off 1.7pt},
  ghar/.style={postaction={decorate,decoration={markings,
    mark=at position 0.55 with {\arrow{Triangle[length=3.5pt,width=3.5pt]}}}}},
  bgtick/.style={postaction={decorate,decoration={markings,
    mark=at position 0 with {\draw[line width=0.7pt,solid] (0,-5pt)--(0,5pt);}}}},
}
\usepackage[colorlinks=true,allcolors=blue]{hyperref}
\hypersetup{pdftitle={Off-shell one-loop counterterms of Palatini gravity with an independent background connection},
  pdfauthor={F. T. Brandt and G. A. Menezes}}

\makeatletter
\def\frontmatter@thefootnote{\expandafter\@fnsymbol\expandafter{\the\numexpr\csname c@\@mpfn\endcsname+1\relax}}
\makeatother

\newcommand{\Riem}{\mathrm{Riem}}
\newcommand{\Ric}{\mathrm{Ric}}

\makeatletter%
\begin{document}

\title{Off-shell one-loop counterterms of Palatini gravity with an independent background connection}

\author{F. T. Brandt}
\email{fbrandt@usp.br}
\author{G. A. Menezes}
\email{g_alves@usp.br}
\affiliation{Instituto de F\'{\i}sica, Universidade de S\~ao Paulo, S\~ao Paulo, SP 05508-090, Brazil}

\date{\today}

\begin{abstract}
We compute the one-loop divergences of pure gravity in the first-order (Palatini) formulation, using
the background field method in Goldberg variables, with a background connection independent of the
background metric. Background Ward--Takahashi identities are verified
at one loop. The divergences of the two-point functions, obtained in arbitrary dimension and gauge,
require, besides the curvature-squared terms, invariants built from the nonmetricity of the background
connection; in particular, the Gauss--Bonnet combination, which is not topological for a nonmetric
connection, does not suffice. Three- and four-point functions in the Feynman--de Donder gauge complete
the counterterm Lagrangian up to fourth order in the nonmetricity and provide parameter-free tests of
it. All the counterterms vanish on shell, as in the metric formulation, and off shell they can be
removed by field redefinitions.
\end{abstract}

\maketitle

\section{Introduction}
\label{sec:intro}

General relativity, defined by the Einstein--Hilbert action~\cite{Einstein1915,Hilbert1915,Wald1984}, is
not renormalizable in the usual sense: its coupling has dimension of length, and each order of the loop
expansion requires counterterms of higher dimension~\cite{Goroff1986}. This does not prevent a perturbative
treatment as an effective field theory below some scale~\cite{Donoghue1994}, but it makes the structure of
the divergences, which operators appear, with which coefficients and with which physical meaning, a central
question. The tools for this study are the Feynman rules of gauge
theories~\cite{Faddeev1967,tHooft1972b}, dimensional regularization~\cite{tHooft1972a,Bollini1972} and,
decisively, the background field method~\cite{DeWitt1967,tHooft1973,Abbott1981,Abbott1982}, in which the
gauge fixing is covariant with respect to a background field and the effective action is an invariant
functional of it. After the graviton self-energy of Capper, Leibbrandt and Ram\'on Medrano~\cite{Capper1973},
't~Hooft and Veltman obtained all the one-loop divergences of pure gravity~\cite{tHooft1974}. Power counting
and invariance restrict them to $R^{2}$, $\Ric_{\mu\nu}\Ric^{\mu\nu}$ and $\Riem_{\mu\nu\rho\sigma}\Riem^{\mu\nu\rho\sigma}$; the
Gauss--Bonnet combination is a total derivative in four dimensions, and the remaining two terms vanish by
the vacuum Einstein equations and can be removed by a field redefinition. Pure gravity is free of ultraviolet divergences on
shell at one loop.

Off shell the coefficients are not universal. They depend on the gauge, as the calculation in a general
background gauge of Ref.~\cite{Brandt2022} shows, and on the parametrization of the quantum
field~\cite{Kalmykov1998,Shapiro2018}. The on-shell one-loop effective action is nevertheless independent of the gauge-fixing parameters
and of the parametrization of the quantum fields, by the theorem of DeWitt and
Kallosh~\cite{DeWitt1967,Kallosh1974}; for the divergences of gravity coupled to matter in a general
background gauge, the cancellation of the gauge parameter on shell was verified explicitly at one loop by
Frenkel and Martins-Filho~\cite{Frenkel2026}.
The off-shell coefficients are nonetheless the counterterms of the background effective action in a given
formulation and gauge, and it is with them that the off-shell background functions are renormalized in that
setting.

In the first-order, or Palatini, formulation~\cite{Palatini1919,Ferraris1982} the metric and the affine
connection are independent variables. The equation of motion of the connection makes it the Levi-Civita
connection of the metric, and the two formulations are classically equivalent. Written with the Goldberg
density $h^{\mu\nu}=\sqrt{-g}\,g^{\mu\nu}$~\cite{Goldberg1958}, the action has a single cubic interaction, and a
shift of the connection removes the mixing between graviton and connection in the quadratic part, which
gives a diagonal set of Feynman rules~\cite{Brandt2016}; a closely related form of the action was studied by
Cheung and Remmen~\cite{Cheung2017a,Cheung2017}. With background fields the equivalence with the metric
formulation holds when the background connection is the Levi-Civita connection of the background
metric~\cite{Brandt2026}. Away from that point the background nonmetricity $Q$ is independent of the
background metric, and the effective action can contain structures with no metric counterpart.

These structures have been studied before, but their coefficients have not been fully determined. Buchbinder and
Shapiro~\cite{Buchbinder1984,Buchbinder1985} computed the one-loop divergences of first-order gravity in the
background field method with an independent background connection, in $d=4$ and in one gauge. They found
the classes of invariants that appear, showed that the result vanishes on shell and that renormalization
amounts to a nonlinear reparametrization of the metric and the connection, and gave explicitly the
coefficients of part of the two-point terms; the coefficients of the terms of higher order in the
distortion $\bar\Gamma-\Gamma(\bar g)$ (equivalently, in the nonmetricity) were not written. \'Alvarez, Anero and Santos-Garcia~\cite{Alvarez2017} repeated the
calculation with the heat kernel, taking the background connection to be the Levi-Civita connection of the
background metric, so that the terms with the nonmetricity do not arise. In affine-metric gravity,
Kalmykov, Pronin, Stepanyantz and collaborators computed one-loop counterterms in arbitrary gauge and
parametrization with the metric off shell and the connection on
shell~\cite{Kalmykov1994,Kalmykov1998,Baurov2018}. In the tetrad formulation with torsion, Benedetti and
Speziale~\cite{Benedetti2011} kept the contorsion off shell but did not give the corresponding logarithmic
terms of dimension four. Without background fields, \'Alvarez, Anero and Martin~\cite{Alvarez2025} found
that the mixing between graviton and connection is regenerated at one loop, and Martins-Filho gave a
covariant quantization of the first-order theory and showed its equivalence with the metric one when the
auxiliary connection is on shell~\cite{MartinsFilho2026}. In all these works the background connection is
either the Levi-Civita one or on shell, or the terms with the independent connection are left implicit.

In this paper we determine the one-loop counterterm Lagrangian of pure first-order gravity in the
background field method, off shell and with an independent background connection. We extend the diagonal
Feynman rules of Ref.~\cite{Brandt2016} to background fields for the metric and for the connection, with a
harmonic gauge covariant with respect to the independent background connection. In the diagonal variables
the shifted background connection transforms as a tensor only up to a term $\Xi$, linear in the background
graviton, and we show that, for $d\neq1,2$, $\Xi$, as an operator on the background graviton, vanishes exactly when the transformation is generated by a
Killing vector of flat space (Sec.~\ref{sec:ward} and Appendix~\ref{app:xi}). The background
Ward--Takahashi identities then take a modified form, and we verify five of them, at two and three points,
with the full amplitudes, in arbitrary dimension and, at three points, in the Feynman--de Donder gauge,
separately for the bosonic and ghost loops. The three background two-point functions, of the graviton, of
the connection and mixed, are computed in arbitrary $d$ and gauge parameter $\xi$. The mixed function does
not vanish, and its decomposition requires invariants beyond the curvature: the counterterms at two points
belong to three classes, quadratic in the curvature of the connection, bilinear in the curvature and in the
derivative of the nonmetricity, and quadratic in the latter, and none of them can be dispensed with. In
particular the Gauss--Bonnet combination, which is not a total derivative for a nonmetric connection, does
not suffice. The curvature-squared coefficients agree for every $\xi$ with those of Ref.~\cite{Brandt2026}.

At three and four points, in the gauge $\xi=1$, two functions with background graviton legs are
parameter-free predictions of the coefficients fixed at two points, and both are confirmed; so is a
four-point function with one background graviton leg, which provides an independent check of the pole of the three-point function of the background connection. The remaining
three- and four-point functions fix the classes with two, three and four factors of the nonmetricity. The
Lagrangian has $166$ coefficients: $22=2+4+16$ in the three classes that start at two points, known in a
general gauge, and $21+58+65$ in the three classes with two, three and four factors of $Q$, determined at $\xi=1$.
The basis of invariants was checked, in each calculated sector, against all the contractions of each class,
and it is complete as a basis of covariant functionals (Appendix~\ref{app:basis}). To our knowledge,
this work provides the first explicit determination of the higher-order nonmetricity counterterms in this
formulation, together with the two-point sector in a general gauge. On shell all the counterterms vanish,
and the one-loop on-shell ultraviolet finiteness of 't~Hooft and Veltman is recovered; off shell the terms with the nonmetricity
are removed by a redefinition of the connection, as anticipated by Buchbinder and Shapiro.

The off-shell structure matters beyond pure gravity at one loop. At two loops the Goroff--Sagnotti
divergence survives on shell~\cite{Goroff1986}; with matter, on-shell ultraviolet finiteness already fails at one
loop~\cite{tHooft1974}; and there are metric-affine theories with a dynamical connection. In each case the
off-shell counterterms of the first-order formulation, with an independent background connection, are the
natural starting point for extensions of the present calculation, which does not address these
applications.

The paper is organized as follows. Section~\ref{sec:formalism} presents the formalism and the background
split, Sec.~\ref{sec:ward} the background transformations and the Ward--Takahashi identities, and
Sec.~\ref{sec:feynman} the Feynman rules and the method. The two-point functions are given in
Sec.~\ref{sec:twopoint}, and their decomposition in invariants in Sec.~\ref{sec:basis}. Section~\ref{sec:vertices}
treats the three- and four-point functions, Sec.~\ref{sec:lagrangian} collects the Lagrangian and discusses
its interpretation, and Sec.~\ref{sec:conclusions} concludes. The appendices contain the derivations, the
Feynman rules, the diagrams, the method of extraction of the poles and the tables of coefficients.
\section{First-order formalism with background fields}
\label{sec:formalism}

\subsection{Geometry}
\label{sec:geometry}

We work in $d$ dimensions with signature $(-+\cdots+)$. In the first-order formulation the
metric and the affine connection $\Gamma^{\lambda}_{\mu\nu}$ are independent fields. We take the
connection to be symmetric, $\Gamma^{\lambda}_{\mu\nu}=\Gamma^{\lambda}_{\nu\mu}$, and define
\begin{gather}
\nabla_{\lambda}V^{\mu}\equiv\partial_{\lambda}V^{\mu}+\Gamma^{\mu}_{\lambda\rho}V^{\rho},\nonumber\\
\Riem^{\rho}{}_{\sigma\mu\nu}\equiv\partial_{\mu}\Gamma^{\rho}_{\nu\sigma}-\partial_{\nu}\Gamma^{\rho}_{\mu\sigma}
+\Gamma^{\rho}_{\mu\lambda}\Gamma^{\lambda}_{\nu\sigma}-\Gamma^{\rho}_{\nu\lambda}\Gamma^{\lambda}_{\mu\sigma},
\label{eq:riem}
\end{gather}
with $\Ric_{\mu\nu}\equiv\Riem^{\rho}{}_{\mu\rho\nu}$ and $R\equiv g^{\mu\nu}\Ric_{\mu\nu}$. A tensor density of weight $w$ receives the
additional term $-w\,\Gamma^{\rho}_{\rho\lambda}$ in its covariant derivative. The curvature depends
on the connection alone; the metric enters only through contractions. Objects built from the
Levi-Civita connection $\mathring{\Gamma}$ of the metric carry a circle.

Instead of the metric we use the Goldberg density~\cite{Goldberg1958}
\begin{equation}
h^{\mu\nu}\equiv\sqrt{-g}\,g^{\mu\nu},
\label{eq:goldberg}
\end{equation}
of weight one, with inverse $h_{\mu\nu}$. Its nonmetricity,
\begin{equation}
Q_{\lambda}{}^{\mu\nu}\equiv\nabla_{\lambda}h^{\mu\nu},
\label{eq:Q}
\end{equation}
is a symmetric density of weight one, and for $d\neq2$ it vanishes if and only if
$\Gamma=\mathring{\Gamma}$. It has two independent traces,
\begin{equation}
W_{\lambda}\equiv h_{\mu\nu}Q_{\lambda}{}^{\mu\nu},
\qquad
V^{\nu}\equiv Q_{\mu}{}^{\mu\nu}=\nabla_{\mu}h^{\mu\nu},
\label{eq:traces}
\end{equation}
of weights zero and one. Any symmetric connection differs from $\mathring{\Gamma}$ by a distortion
tensor $L$, algebraic and linear in $Q$, so that $\Riem=\mathring{\Riem}+\mathring{\nabla}L+LL$:
every local invariant can be written in terms of $\Riem$, $Q$ and their derivatives. For $Q\neq0$
the curvature has fewer symmetries than in the metric case. Only the antisymmetry in the last pair
and the first Bianchi identity survive, and the Ricci tensor is no longer symmetric,
\begin{equation}
\Ric_{\mu\nu}-\Ric_{\nu\mu}=-\frac{1}{d-2}\left(\partial_{\mu}W_{\nu}-\partial_{\nu}W_{\mu}\right).
\label{eq:ricanti}
\end{equation}
There are therefore more independent invariants quadratic in the curvature (for instance,
$\Ric_{\mu\nu}\Ric^{\mu\nu}$ and $\Ric_{\mu\nu}\Ric^{\nu\mu}$ differ), and the Gauss--Bonnet
combination is no longer a total derivative in $d=4$~\cite{Janssen2019}.

\subsection{Action and field equations}
\label{sec:action}

The Einstein--Hilbert action in first-order form~\cite{Palatini1919,Ferraris1982} is
\begin{equation}
S=\frac{1}{\kappa^{2}}\int d^{d}x\,h^{\mu\nu}\Ric_{\mu\nu}(\Gamma),
\qquad \kappa^{2}\equiv16\pi G,
\label{eq:action}
\end{equation}
linear in $h$ and with no derivatives of the metric; by construction $S[h,\mathring{\Gamma}(h)]$ is
the metric action. Varying $h$ gives $\Ric_{(\mu\nu)}=0$. Varying $\Gamma$, with the Palatini
identity
$\delta\Ric_{\mu\nu}=\nabla_{\rho}\delta\Gamma^{\rho}_{\nu\mu}-\nabla_{\nu}\delta\Gamma^{\rho}_{\rho\mu}$,
gives
\begin{equation}
Q_{\rho}{}^{\mu\nu}=\tfrac{1}{2}\left(\delta^{\mu}_{\rho}V^{\nu}+\delta^{\nu}_{\rho}V^{\mu}\right).
\label{eq:eomGamma}
\end{equation}
Its trace gives $(d-1)V^{\nu}=0$, hence $Q=0$ and $\Gamma=\mathring{\Gamma}(h)$: the connection is
an auxiliary field, the vacuum Einstein equations follow, and the two formulations are classically
equivalent for $d\neq1,2$. The equivalence persists in the quantum theory when the sources couple
only to the metric, since the connection enters the action quadratically and integrating it out
leaves an ultralocal determinant, equal to one in dimensional regularization~\cite{Brandt2016}. With background fields the statement is more
restricted: the background effective actions of the two formulations coincide when the background
connection is the Levi-Civita connection of the background metric~\cite{Brandt2026}. Away from
that point the background nonmetricity $\bar Q$ produces terms with no metric counterpart, and
these are the subject of this paper.

\subsection{Auxiliary field and diagonal form}
\label{sec:diagonal}

Following Ref.~\cite{Brandt2016}, we expand around flat space,
$h^{\mu\nu}=\eta^{\mu\nu}+\kappa\phi^{\mu\nu}$, and replace the connection by the auxiliary field
\begin{equation}
\mathcal{G}^{\lambda}_{\mu\nu}\equiv\frac{1}{\kappa}\left[\Gamma^{\lambda}_{\mu\nu}
-\tfrac{1}{2}\left(\delta^{\lambda}_{\mu}\Gamma^{\rho}_{\rho\nu}+\delta^{\lambda}_{\nu}\Gamma^{\rho}_{\mu\rho}\right)\right],
\label{eq:Gcal}
\end{equation}
whose factor $1/\kappa$ removes $\kappa$ from its quadratic term. With $\kappa^{2}=16\pi G$ the field
$\phi$ is not canonically normalized: the transverse-traceless part of its quadratic term carries half
the canonical coefficient. Results in the canonical convention $\kappa^{2}=32\pi G$~\cite{Donoghue1994}
follow from ours by $\kappa^{2}\to\kappa^{2}/2$, with $\kappa\phi$ and $\kappa\mathcal{G}$ held fixed. After an
integration by parts the action
becomes
\begin{equation}
S=\int d^{d}x\left[-\mathcal{G}^{\lambda}_{\mu\nu}\,\partial_{\lambda}\phi^{\mu\nu}
+\tfrac{1}{2}\,\mathcal{G}\star M(h)\star\mathcal{G}\right],
\label{eq:actionG}
\end{equation}
where $\star$ contracts the index triples of $\mathcal{G}$ with those of
\begin{multline}
M^{\mu\nu}{}_{\lambda}{}^{\rho\sigma}{}_{\pi}(h)\equiv\frac{1}{2}\Big[\frac{1}{d-1}
\big(\delta^{\nu}_{\lambda}\delta^{\sigma}_{\pi}h^{\mu\rho}+\delta^{\mu}_{\lambda}\delta^{\sigma}_{\pi}h^{\nu\rho}\\
+\delta^{\mu}_{\lambda}\delta^{\rho}_{\pi}h^{\nu\sigma}+\delta^{\nu}_{\lambda}\delta^{\rho}_{\pi}h^{\mu\sigma}\big)
-\big(\delta^{\sigma}_{\lambda}\delta^{\nu}_{\pi}h^{\mu\rho}+\delta^{\sigma}_{\lambda}\delta^{\mu}_{\pi}h^{\nu\rho}\\
+\delta^{\rho}_{\lambda}\delta^{\mu}_{\pi}h^{\nu\sigma}+\delta^{\rho}_{\lambda}\delta^{\nu}_{\pi}h^{\mu\sigma}\big)\Big].
\label{eq:M}
\end{multline}
The bilinear form $M$ is linear in $h$, so $M(\eta+\kappa\phi)=M(\eta)+\kappa M(\phi)$. Its inverse at
$h=\eta$,
\begin{multline}
D^{\lambda}{}_{\mu\nu}{}^{\pi}{}_{\rho\sigma}\equiv\left(M^{-1}\right)^{\lambda}{}_{\mu\nu}{}^{\pi}{}_{\rho\sigma}(\eta)\\
=\frac{1}{2}\eta^{\lambda\pi}\Big(-\frac{\eta_{\mu\nu}\eta_{\rho\sigma}}{d-2}
+\frac{\eta_{\mu\rho}\eta_{\nu\sigma}+\eta_{\mu\sigma}\eta_{\nu\rho}}{2}\Big)
-\frac{1}{4}\big(\delta^{\lambda}_{\rho}\delta^{\pi}_{\mu}\eta_{\nu\sigma}\\
+\delta^{\lambda}_{\rho}\delta^{\pi}_{\nu}\eta_{\mu\sigma}
+\delta^{\lambda}_{\sigma}\delta^{\pi}_{\nu}\eta_{\mu\rho}+\delta^{\lambda}_{\sigma}\delta^{\pi}_{\mu}\eta_{\nu\rho}\big),
\label{eq:D}
\end{multline}
is the inverse of the quadratic kernel of the auxiliary field, whose propagator is the constant
$iD$ (Sec.~\ref{sec:feynman}). In the variables $(\phi,\mathcal{G})$ the
quadratic action mixes the two fields, and the free propagators include a mixed
$\langle\phi\,\mathcal{G}\rangle$ component. The shift~\cite{Brandt2016}
\begin{equation}
\mathcal{G}\to\mathcal{G}+D\star\partial\phi,
\qquad
(D\star\partial\phi)^{\lambda}_{\mu\nu}\equiv D^{\lambda}{}_{\mu\nu}{}^{\pi}{}_{\rho\sigma}\,\partial_{\pi}\phi^{\rho\sigma},
\label{eq:shiftflat}
\end{equation}
removes the mixing and gives the diagonal form
\begin{multline}
S=\int d^{d}x\Big[\tfrac{1}{2}\,\mathcal{G}\star D^{-1}\star\mathcal{G}-\tfrac{1}{2}\,\partial\phi\star D\star\partial\phi\\
+\tfrac{\kappa}{2}\left(\mathcal{G}+\partial\phi\star D\right)\star M(\phi)\star\left(\mathcal{G}+D\star\partial\phi\right)\Big].
\label{eq:actiondiag}
\end{multline}
Neither quadratic term contains $\kappa$, and $\kappa$ appears once, in front of the only interaction of
the classical action, which is cubic; gauge fixing adds further vertices (Sec.~\ref{sec:gauge}). The shift uses the linearized inverse $D=M^{-1}(\eta)$ rather than $M^{-1}(h)$; the
reason, and the price paid for it, are discussed in Appendix~\ref{app:shift}.

\subsection{Background--quantum split}
\label{sec:split}

In the background field method~\cite{DeWitt1967,Abbott1981,Abbott1982} we split the Goldberg
density into a background and a quantum part, and expand the background around flat space,
\begin{equation}
h^{\mu\nu}=\bar h^{\mu\nu}+\kappa\phi^{\mu\nu},
\qquad
\bar h^{\mu\nu}=\eta^{\mu\nu}+\kappa\bar\phi^{\mu\nu}.
\label{eq:splith}
\end{equation}
The background graviton $\bar\phi$ supplies the external legs of the diagrams. Since the action
depends only on $h=\eta+\kappa(\bar\phi+\phi)$, its background form follows from
Eq.~\eqref{eq:actiondiag} by $\phi\to\bar\phi+\phi$, and the diagonalizing shift becomes a shift by
$D\star\partial(\bar\phi+\phi)$. Splitting also the shifted auxiliary field, the original variable
reads
\begin{equation}
\mathcal{G}=\bar G+G+D\star\partial(\bar\phi+\phi),
\label{eq:splitG}
\end{equation}
where $\bar G$ and $G$ are the background and quantum parts of the shifted field; they are the
only connection variables in the propagators and vertices.

The counterterms are covariant functionals of the background metric and of the background
connection $\bar\Gamma$, which Eqs.~\eqref{eq:Gcal}, \eqref{eq:splith} and \eqref{eq:splitG} give
in terms of the diagram variables. Inverting Eq.~\eqref{eq:Gcal},
\begin{multline}
\bar\Gamma^{\lambda}_{\mu\nu}=\kappa\Big[(\bar G+D\star\partial\bar\phi)^{\lambda}_{\mu\nu}\\
+\frac{\delta^{\lambda}_{\mu}(\bar G+D\star\partial\bar\phi)^{\rho}_{\rho\nu}
+\delta^{\lambda}_{\nu}(\bar G+D\star\partial\bar\phi)^{\rho}_{\rho\mu}}{1-d}\Big].
\label{eq:ctparam}
\end{multline}
This relation is exactly linear, and all the $n$-point functions of this paper are compared with
the expansion of the invariants in this parametrization. Its $\bar\phi$-dependent part is the
Levi-Civita connection of $\bar h$ only at linear order. Consequently the background nonmetricity
$\bar Q$ is, at linear order, an invertible linear function of $\bar G$ alone, while at quadratic
order it acquires terms $\bar\phi\,\partial\bar\phi$ that matter at three points
(Sec.~\ref{sec:vertices}).

We denote by $S_{\rm eff}[\bar\phi,\bar G]$ the background effective action, the generating
functional of one-particle-irreducible diagrams with background external legs, taken at vanishing
quantum expectation values. Its functional derivatives at $\bar\phi=\bar G=0$ are the background
$n$-point functions, written $\delta^{2}_{\bar\phi}S_{\rm eff}$,
$\delta_{\bar\phi}\delta^{2}_{\bar G}S_{\rm eff}$, and so on. The same object in the unshifted
variable $\bar{\mathcal{G}}\equiv\bar G+D\star\partial\bar\phi$, the one used in
Ref.~\cite{Brandt2026}, is related to $S_{\rm eff}[\bar\phi,\bar G]$ by a linear change of
background coordinates (Appendix~\ref{app:shift}).

\subsection{Gauge fixing and ghosts}
\label{sec:gauge}

We fix the gauge of the quantum field with the background-covariant harmonic condition
\begin{gather}
\mathcal{L}_{\rm gf}\equiv-\frac{1}{2\xi}\,\bar h_{\nu\sigma}\left(\bar\nabla_{\mu}\phi^{\mu\nu}\right)
\left(\bar\nabla_{\rho}\phi^{\rho\sigma}\right),\nonumber\\
\bar\nabla_{\mu}\phi^{\mu\nu}=\partial_{\mu}\phi^{\mu\nu}+\bar\Gamma^{\nu}_{\mu\lambda}\,\phi^{\mu\lambda},
\label{eq:gf}
\end{gather}
where $\bar\nabla$ is built with the independent background connection~\eqref{eq:ctparam}, and
the second form uses that $\phi$ is a density of weight one and $\bar\Gamma$ is symmetric;
$\xi=1$ is the Feynman--de Donder gauge.

The ghost action is built from the response of the gauge condition
$\mathcal{F}^{\nu}\equiv\kappa\bar\nabla_{\mu}\phi^{\mu\nu}$ to the quantum diffeomorphisms, the
transformations of the quantum field with the background held fixed~\cite{Faddeev1967}. Under a
diffeomorphism generated by $\omega^{\lambda}$ the full density $h=\bar h+\kappa\phi$ changes by
its Lie derivative, and since $\bar h$ does not change, the whole variation falls on $\phi$:
\begin{equation}
\delta\phi^{\mu\nu}=\frac{1}{\kappa}\left[\partial_{\lambda}(h^{\mu\nu}\omega^{\lambda})
-h^{\mu\lambda}\partial_{\lambda}\omega^{\nu}-h^{\lambda\nu}\partial_{\lambda}\omega^{\mu}\right].
\label{eq:deltaphi}
\end{equation}
Using Eq.~\eqref{eq:deltaphi} to vary $\mathcal{F}^{\nu}$ gives the Faddeev--Popov operator
$\delta\mathcal{F}^{\alpha}/\delta\omega_{\beta}$, and the ghost action is
\begin{equation}
S_{\rm gh}\equiv-\int d^{d}x\,\bar c_{\alpha}\,\frac{\delta\mathcal{F}^{\alpha}}{\delta\omega_{\beta}}\,c_{\beta}.
\label{eq:ghost}
\end{equation}
With $h=\eta+\kappa(\bar\phi+\phi)$, the part of Eq.~\eqref{eq:deltaphi} with $\eta$ gives the
ghost kinetic term, and the parts linear in $\bar\phi$ and $\phi$, together with the background
connection in $\bar\nabla$, give the ghost vertices. Since $\mathcal{F}$ is linear in $\phi$ and in
$\bar\Gamma$, and $\delta\phi$ is linear in $h$, the ghost action is polynomial: each ghost vertex
contains at most two additional fields.
Both terms are invariant under background diffeomorphisms, under which $\bar h$ and $\phi$
transform as densities, $\bar\Gamma$ as a connection and the ghosts as vectors; this invariance
is the origin of the Ward--Takahashi identities of Sec.~\ref{sec:ward}. When the background fields
vanish, Eq.~\eqref{eq:gf} reduces to the flat harmonic gauge, so the graviton propagator is
unchanged and all the new terms are vertices (Sec.~\ref{sec:feynman}).

The covariant derivative in Eq.~\eqref{eq:gf} uses the independent background connection, as in
Ref.~\cite{Brandt2026}, rather than the Levi-Civita connection of $\bar h$, used in
Ref.~\cite{Buchbinder1985}; in Ref.~\cite{Alvarez2017} the background connection itself is taken
to be the Levi-Civita one. The two gauge conditions differ by terms proportional to $\bar Q$,
which vanish on shell; off shell the choice changes the counterterm coefficients, as any change
of gauge does (Sec.~\ref{sec:lagrangian}).
\section{Background transformations and Ward--Takahashi identities}
\label{sec:ward}

\subsection{Transformations of the background fields}
\label{sec:bgtransf}

Under a background diffeomorphism generated by $\omega^{\lambda}(x)$ the Goldberg density changes by
its Lie derivative,
$\delta\bar h^{\mu\nu}=\partial_{\lambda}(\bar h^{\mu\nu}\omega^{\lambda})-\bar h^{\mu\lambda}\partial_{\lambda}\omega^{\nu}
-\bar h^{\lambda\nu}\partial_{\lambda}\omega^{\mu}$, and with $\bar h=\eta+\kappa\bar\phi$ the background
graviton acquires an inhomogeneous term,
\begin{multline}
\delta\bar\phi^{\mu\nu}=\frac{1}{\kappa}\left(\eta^{\mu\nu}\partial_{\lambda}\omega^{\lambda}
-\eta^{\mu\lambda}\partial_{\lambda}\omega^{\nu}-\eta^{\lambda\nu}\partial_{\lambda}\omega^{\mu}\right)
+\mathcal{L}_{\omega}\bar\phi^{\mu\nu}\\
\equiv\delta\bar\phi^{\mu\nu}\big|_{\rm inh}+\mathcal{L}_{\omega}\bar\phi^{\mu\nu}.
\label{eq:deltaphibar}
\end{multline}
The inhomogeneous term is $\kappa^{-1}\mathcal{L}_{\omega}\eta^{\mu\nu}$, with $\eta$ treated as a density
of weight one, and it is the only one that survives when the background fields vanish.

The unshifted variable $\bar{\mathcal{G}}$ of Eq.~\eqref{eq:Gcal} is a connection variable: its
variation is its Lie derivative plus an inhomogeneous term
$\kappa^{-1}[\partial_{\mu}\partial_{\nu}\omega^{\lambda}-\frac{1}{2}(\delta^{\lambda}_{\mu}\partial_{\nu}\partial_{\rho}\omega^{\rho}
+\delta^{\lambda}_{\nu}\partial_{\mu}\partial_{\rho}\omega^{\rho})]$ that does not depend on the fields.
The shift $\bar G=\bar{\mathcal{G}}-D\star\partial\bar\phi$ removes this term exactly, because
$D\star\partial(\delta\bar\phi|_{\rm inh})$ is the linearized Levi-Civita variation of the connection
variable. It does not, however, make $\bar G$ a tensor. As shown in Appendix~\ref{app:xi},
\begin{equation}
\delta\bar G=\mathcal{L}_{\omega}\bar G+\Xi,
\qquad
\Xi\equiv\mathcal{L}_{\omega}(D\star\partial\bar\phi)-D\star\partial(\mathcal{L}_{\omega}\bar\phi),
\label{eq:deltaGbar}
\end{equation}
where the residue $\Xi$ is linear in $\bar\phi$ and independent of $\bar G$. The shift has traded a
field-independent inhomogeneous term for a homogeneous term that depends on the background
graviton. This is what matters for the identities below: through $\Xi$, the variation of $\bar G$
involves $\bar\phi$, and the identities acquire terms that tie the graviton and connection
sectors. (As an operator on $\bar\phi$, $\Xi$ vanishes exactly when $\omega$ is a Killing vector of
$\eta$, so $\bar G$ transforms covariantly only under the Poincar\'e group.) This does not prevent the use of $\bar G$ as a
background variable: $(\bar h,\bar\Gamma)\mapsto(\bar\phi,\bar G)$ is an invertible field
redefinition, so every invariant functional can be written in terms of $\bar\phi$ and $\bar G$,
with the invariance no longer manifest term by term.

\subsection{The identities}
\label{sec:identities}

Since the gauge fixing~\eqref{eq:gf} and the ghost action~\eqref{eq:ghost} are background
invariant and dimensional regularization preserves this invariance, the background effective action
satisfies
\begin{equation}
\delta S_{\rm eff}=\int d^{d}x\left[\frac{\delta S_{\rm eff}}{\delta\bar\phi^{\mu\nu}}\,\delta\bar\phi^{\mu\nu}
+\frac{\delta S_{\rm eff}}{\delta\bar G^{\lambda}{}_{\mu\nu}}\,\delta\bar G^{\lambda}{}_{\mu\nu}\right]=0,
\label{eq:deltaSeff}
\end{equation}
with the variations~\eqref{eq:deltaphibar} and~\eqref{eq:deltaGbar}. Functional derivatives of
Eq.~\eqref{eq:deltaSeff} at vanishing background fields give Ward--Takahashi identities. Because
the quantum fields transform homogeneously, these are genuine Ward identities, with a single
contraction; for the graviton self-energy, for instance, the background identity is more
restrictive than the corresponding identity of the conventional formulation~\cite{Brandt2016},
which involves a double contraction with the transverse projector and is implied by
Eq.~\eqref{eq:ward2} below, but not conversely.

We work with a plane-wave parameter, $\omega^{\lambda}(x)=\omega^{\lambda}e^{ik\cdot x}$, and keep the
coefficient of $\omega^{\varepsilon}$, so that $\varepsilon$ is a free index. With all momenta
incoming, the inhomogeneous variation of a $\bar\phi$ leg of momentum $k$ is
\begin{equation}
I^{\rho\sigma}_{\varepsilon}\equiv\frac{\partial\,\delta\bar\phi^{\rho\sigma}|_{\rm inh}}{\partial\omega^{\varepsilon}}
=\frac{i}{\kappa}\left(\eta^{\rho\sigma}k_{\varepsilon}-k^{\rho}\delta^{\sigma}_{\varepsilon}
-k^{\sigma}\delta^{\rho}_{\varepsilon}\right),
\label{eq:inhmom}
\end{equation}
and an $n$-point function is called transverse in a $\bar\phi$ leg when its contraction with
Eq.~\eqref{eq:inhmom} vanishes. We write $\mathcal{L}_{\varepsilon}|_{a}$ for the coefficient of
$\omega^{\varepsilon}$ in the Lie derivative inserted on leg $a$, and $\delta\Xi_{\varepsilon}/\delta\bar\phi$
for that of the residue; their momentum-space forms are given in Appendix~\ref{app:xi}.

At two points only the inhomogeneous term contributes, and both functions with a $\bar\phi$ leg are
transverse in it:
\begin{align}
\delta^{2}_{\bar\phi}S_{\rm eff}\star I_{\varepsilon}&=0,
\label{eq:ward2}\\
\delta_{\bar G}\delta_{\bar\phi}S_{\rm eff}\star I_{\varepsilon}&=0.
\label{eq:ward2mix}
\end{align}
At three points the contraction of a three-point function with Eq.~\eqref{eq:inhmom} is balanced by
two-point functions with the homogeneous variations inserted:
\begin{align}
&\delta_{\bar\phi}\delta_{\bar G_{a}}\delta_{\bar G_{b}}S_{\rm eff}\star I_{\varepsilon}
+\delta^{2}_{\bar G}S_{\rm eff}\star\mathcal{L}_{\varepsilon}|_{a}\nonumber\\
&\qquad+\delta^{2}_{\bar G}S_{\rm eff}\star\mathcal{L}_{\varepsilon}|_{b}=0,
\label{eq:ward3GG}\\[2pt]
&\delta^{2}_{\bar\phi}\delta_{\bar G}S_{\rm eff}\star I_{\varepsilon}
+\delta_{\bar\phi}\delta_{\bar G}S_{\rm eff}\star\mathcal{L}_{\varepsilon}|_{\bar\phi}
+\delta_{\bar\phi}\delta_{\bar G}S_{\rm eff}\star\mathcal{L}_{\varepsilon}|_{\bar G}\nonumber\\
&\qquad+\delta^{2}_{\bar G}S_{\rm eff}\star\frac{\delta\Xi_{\varepsilon}}{\delta\bar\phi}=0,
\label{eq:ward3phiG}\\[2pt]
&\delta^{3}_{\bar\phi}S_{\rm eff}\star I_{\varepsilon}
+\sum_{j=1,2}\Big[\delta^{2}_{\bar\phi}S_{\rm eff}\star\mathcal{L}_{\varepsilon}|_{j}\nonumber\\
&\qquad+\delta_{\bar\phi}\delta_{\bar G}S_{\rm eff}\star\frac{\delta\Xi_{\varepsilon}}{\delta\bar\phi}\Big|_{j}\Big]=0.
\label{eq:ward3phiphi}
\end{align}
In Eq.~\eqref{eq:ward3GG}, $a$ and $b$ label the two $\bar G$ legs; in Eq.~\eqref{eq:ward3phiphi},
$j$ labels the two $\bar\phi$ legs that do not carry the inhomogeneous variation. Table~\ref{tab:ward}
summarizes the five identities.

\begin{table}[t]
\caption{The background Ward--Takahashi identities, labeled by the legs of the function they
constrain ($\bar\phi\bar G$ stands for $\delta_{\bar\phi}\delta_{\bar G}S_{\rm eff}$, and so on). The third
column lists the functions of lower order that enter each identity, and the fourth whether the
residue $\Xi$ of Eq.~\eqref{eq:deltaGbar} contributes.}
\label{tab:ward}
\begin{ruledtabular}
\begin{tabular}{lccc}
Function & Eq. & Lower-order functions & $\Xi$\\
\hline
$\bar\phi\bar\phi$ & \eqref{eq:ward2} & none & no\\
$\bar\phi\bar G$ & \eqref{eq:ward2mix} & none & no\\
$\bar\phi\bar G\bar G$ & \eqref{eq:ward3GG} & $\bar G\bar G$ & no\\
$\bar\phi\bar\phi\bar G$ & \eqref{eq:ward3phiG} & $\bar\phi\bar G$, $\bar G\bar G$ & yes\\
$\bar\phi\bar\phi\bar\phi$ & \eqref{eq:ward3phiphi} & $\bar\phi\bar\phi$, $\bar\phi\bar G$ & yes\\
\end{tabular}
\end{ruledtabular}
\end{table}

The residue does not enter the first three identities. It vanishes at $\bar\phi=0$, which suffices
for Eq.~\eqref{eq:ward2mix}; it is annihilated by the two derivatives with respect to $\bar G$ in
Eq.~\eqref{eq:ward3GG}; and in Eq.~\eqref{eq:ward2} it would multiply the one-point function
$\delta_{\bar G}S_{\rm eff}$, which vanishes. Since $\delta\bar G$ has no inhomogeneous part, no
three-point function of the connection sector appears in Eq.~\eqref{eq:ward3phiG}.

\subsection{Structure of the identities}
\label{sec:wardstructure}

\paragraph{Bosonic and ghost loops.} Because the gauge condition is background covariant, the
bosonic operator $S^{(2)}$, which comprises the graviton and the auxiliary field, and the
Faddeev--Popov operator are each covariant under background transformations, so that
\begin{equation}
\delta\ln\det S^{(2)}=0,\qquad
\delta\ln\det\frac{\delta\mathcal{F}}{\delta\omega}=0 .
\label{eq:splitdet}
\end{equation}
The one-loop contributions with a bosonic loop, which may contain graviton and $G$ lines, and with a
ghost loop therefore satisfy every identity separately. This does not require $S^{(2)}$ to be block diagonal, and it is not: the shift
removes the mixed block only for vanishing backgrounds, and the vertices with background legs
restore mixed blocks proportional to $\bar\phi$ and $\bar G$.

\paragraph{Scalar integrals.} The lower-order terms of Eqs.~\eqref{eq:ward3GG}--\eqref{eq:ward3phiphi}
are two-point functions and cannot produce triangle integrals. The scalar function $C_{0}$ must
therefore cancel in each three-point identity after the contraction with
Eq.~\eqref{eq:inhmom}, although not in the amplitude itself.

\paragraph{Tree level.} At tree level the identities reduce to the Noether identities of the classical
action, which were used in Ref.~\cite{BrandtCarvalho2026} to reconstruct the action itself. At this order the
diagonalization makes the mixed function $\delta_{\bar\phi}\delta_{\bar G}S$ vanish, and the terms
proportional to it drop out of the identities. The residue still contributes at this order, through the term
$\delta^{2}_{\bar G}S\star\delta\Xi_{\varepsilon}/\delta\bar\phi$ of Eq.~\eqref{eq:ward3phiG}, with
$\delta^{2}_{\bar G}S=D^{-1}$. At one loop the mixing is regenerated and the terms with
$\delta_{\bar\phi}\delta_{\bar G}S_{\rm eff}$ become indispensable.

\paragraph{What the identities do not test.} Each identity contracts the amplitude, in the chosen
$\bar\phi$ leg, with the inhomogeneous variation of that leg. A contribution that is transverse in
that leg satisfies the identity by itself, whatever its coefficient. Some diagrams of the
three-point functions are of this kind (Sec.~\ref{sec:vertices}); when their ultraviolet poles do not
vanish they contribute to the counterterms, but an error in them would go undetected. Moreover, there is no
autonomous identity for the functions of the connection sector alone: $\delta^{2}_{\bar G}S_{\rm eff}$
is constrained only through Eqs.~\eqref{eq:ward3GG} and~\eqref{eq:ward3phiG}, and
$\delta^{3}_{\bar G}S_{\rm eff}$ and $\delta^{4}_{\bar G}S_{\rm eff}$ only through identities with
an additional $\bar\phi$ leg, at four and five points, which we have not checked.

The identities were verified explicitly. At two points, Eqs.~\eqref{eq:ward2} and~\eqref{eq:ward2mix}
hold identically in $d$ and $\xi$, separately for the bosonic and ghost loops (Sec.~\ref{sec:twopoint}).
At three points, Eqs.~\eqref{eq:ward3GG}--\eqref{eq:ward3phiphi} were verified with the full
amplitudes, in arbitrary dimension and in the Feynman--de Donder gauge, as exact identities in the
basis of scalar integrals $\{B_{0},C_{0}\}$ and not only for the poles. Since they relate vertices
with one and with two background insertions, they test the Feynman rules of Sec.~\ref{sec:feynman}
independently.
\section{Feynman rules and method}
\label{sec:feynman}

\subsection{Propagators}
\label{sec:propagators}

Vertex rules are functional derivatives of the action without the factor $i$ of $e^{iS}$, and
the sum of the amputated one-loop diagrams with $n$ external legs is identified with
$i\,\delta^{n}S_{\rm eff}$. All external momenta are incoming, and a field of incoming momentum
$p$ behaves as $e^{ip\cdot x}$. With the background fields set to zero, the gauge
condition~\eqref{eq:gf} reduces to the flat harmonic gauge, and the propagators are
\begin{gather}
\langle\phi^{\mu\nu}(p)\,\phi^{\rho\sigma}(-p)\rangle\nonumber\\
=-\frac{i}{p^{2}}\left(\eta^{\mu\sigma}\eta^{\nu\rho}+\eta^{\mu\rho}\eta^{\nu\sigma}
-(2-\xi)\eta^{\mu\nu}\eta^{\rho\sigma}\right)\nonumber\\
+\frac{i(1-\xi)}{p^{4}}\big(p^{\nu}p^{\sigma}\eta^{\mu\rho}+p^{\nu}p^{\rho}\eta^{\mu\sigma}
+p^{\mu}p^{\sigma}\eta^{\nu\rho}+p^{\mu}p^{\rho}\eta^{\nu\sigma}\nonumber\\
-2p^{\rho}p^{\sigma}\eta^{\mu\nu}-2p^{\mu}p^{\nu}\eta^{\rho\sigma}\big),
\label{eq:propphi}\\
\langle c^{\mu}(p)\,\bar c^{\nu}(-p)\rangle=-\frac{i}{p^{2}}\,\eta^{\mu\nu},
\nonumber\\
\langle G^{\lambda}_{\mu\nu}\,G^{\pi}_{\rho\sigma}\rangle=i\,D^{\lambda}{}_{\mu\nu}{}^{\pi}{}_{\rho\sigma}.
\label{eq:propcG}
\end{gather}
The graviton propagator depends on $\xi$ both through the coefficient $2-\xi$ of
$\eta^{\mu\nu}\eta^{\rho\sigma}/p^{2}$ and through the longitudinal part, proportional to $(1-\xi)/p^{4}$; in
the Feynman--de Donder gauge, $\xi=1$, it reduces to
$-\frac{i}{p^{2}}(\eta^{\mu\sigma}\eta^{\nu\rho}+\eta^{\mu\rho}\eta^{\nu\sigma}-\eta^{\mu\nu}\eta^{\rho\sigma})$. The
ghost propagator is independent of $\xi$, and the propagator of the auxiliary field is
independent of the momentum.

\subsection{Vertices}
\label{sec:vertices4}

Substituting Eq.~\eqref{eq:splitG} in the background form of Eq.~\eqref{eq:actiondiag}, the
classical action yields only trivertices, all of order $\kappa$ and all of the form
$\frac{\kappa}{2}[\,\cdot\,]\star M(\cdot)\star[\,\cdot\,]$, in which the argument of $M$ is always a
graviton. The gauge-fixing term and the ghost action add trivertices and vertices of higher
order. Through $\bar h_{\nu\sigma}$, the gauge-fixing term contains vertices
$\bar\phi^{n}\phi\phi$ for every $n$, with one or two $\bar G$ legs at most, since
$\bar\nabla\phi$ is linear in the connection; the ghost action is polynomial
(Sec.~\ref{sec:gauge}). At one loop a vertex whose $n$ background legs are all those of an
$n$-point function can only close a tadpole, which vanishes, so the functions of Secs.~\ref{sec:twopoint}--\ref{sec:vertices} need
no vertex beyond quartic order, except the four-point function $\delta_{\bar\phi}\delta^{3}_{\bar G}S_{\rm eff}$
of Sec.~\ref{sec:fourpointtest}, which also uses the quintic gauge-fixing vertex $\bar\phi\bar G\bar G\phi\phi$. Table~\ref{tab:vertices} lists the vertices that enter them; their
explicit forms are given in Appendix~\ref{app:feynman}. The trivertex $\bar\phi GG$, whose two quantum
connection legs are always internal, appears up to three points only in diagrams without scale
(Sec.~\ref{sec:counting}); it first contributes to $\delta_{\bar\phi}\delta^{3}_{\bar G}S_{\rm eff}$, where a box
with two auxiliary lines becomes a bubble.

\begin{table}[t]
\caption{Vertices that enter the background functions computed in this paper: origin (classical
action, gauge fixing or ghost action) and the first function to which each contributes. Overbars
denote background fields.}
\label{tab:vertices}
\begin{ruledtabular}
\begin{tabular}{lll}
Vertex & Origin & First function\\
\hline
$\bar\phi\phi\phi$ & action, gauge fixing & $\bar\phi\bar\phi$\\
$\bar\phi\bar cc$ & ghosts & $\bar\phi\bar\phi$\\
$\bar G\phi\phi$ & action, gauge fixing & $\bar\phi\bar G$\\
$\bar G\bar cc$ & ghosts & $\bar\phi\bar G$\\
$G\bar\phi\phi$, $\bar GG\phi$ & action & three points\\
$\bar\phi GG$ & action & $\bar\phi\bar G\bar G\bar G$\\
$\phi\phi\bar G\bar G$, $\bar\phi\bar\phi\phi\phi$, $\bar\phi\bar G\phi\phi$ & gauge fixing & three points\\
$\bar\phi\bar\phi\bar cc$, $\bar\phi\bar G\bar cc$ & ghosts & three points\\
$\bar\phi\bar G\bar G\phi\phi$ & gauge fixing & $\bar\phi\bar G\bar G\bar G$\\
\end{tabular}
\end{ruledtabular}
\end{table}

\subsection{Counting rules}
\label{sec:counting}

Four rules organize the enumeration of one-loop diagrams. (i) The propagator of $G$ is
independent of the momentum: an internal $G$ line transports momentum but adds no denominator.
(ii) In a two-point function every quartic vertex closes a tadpole with a single propagator,
which vanishes in dimensional regularization; at three points the same vertex closes a bubble.
(iii) Every trivertex of the classical action has at least one graviton leg, the argument of $M$,
so there is no $\bar GGG$ vertex; this holds in the bosonic sector only, since
$\bar G\bar cc$ has no graviton leg. (iv) By (i), what counts in a one-loop diagram is the number
of internal graviton and ghost lines: each internal $G$ line merges the two vertices it joins and
moves the diagram one step down the sequence box, triangle, bubble, tadpole, and a diagram with at
most one internal graviton or ghost line vanishes. Up to three vertices, every diagram with two
or more internal auxiliary lines therefore vanishes; at four points a box with two auxiliary
lines becomes a bubble and survives. Rule (iv) concerns internal lines only: an external $\bar G$
leg attaches to a single vertex and merges nothing, so that, for instance,
$\delta_{\bar\phi}\delta_{\bar G}S_{\rm eff}$ is a genuine bubble.

\subsection{Regularization and extraction of the poles}
\label{sec:poles}

We use dimensional regularization~\cite{tHooft1972a,Bollini1972}, which preserves the background
invariance. With the measure
$\mu^{4-d}d^{d}l/(2\pi)^{d}$, the massless bubble
\begin{multline}
B_{0}(k^{2})\equiv\mu^{4-d}\int\frac{d^{d}l}{(2\pi)^{d}}\,\frac{1}{l^{2}(l+k)^{2}}\\
=\frac{i}{16\pi^{2}}\Big[\frac{2}{4-d}-\ln\frac{k^{2}}{\mu^{2}}+c\Big]+O(4-d)
\label{eq:B0}
\end{multline}
has the pole $i/[8\pi^{2}(4-d)]$, where $c$ is a scheme-dependent constant. Coefficients that
multiply $B_{0}$ are kept in arbitrary $d$ until the pole is extracted, since a factor $d-4$
turns the pole into a finite term, which is local when the structure it multiplies is polynomial in the
momenta (Sec.~\ref{sec:twopointstructure}).

The functions were computed at three levels of completeness.
\begin{itemize}
\item At two points, the amplitudes were computed in full, in arbitrary $d$ and $\xi$. The
Passarino--Veltman reduction~\cite{Passarino1979} and the reduction of
$\int d^{d}p\,(p^{2})^{-a}[(p-k)^{2}]^{-b}$ to $B_{0}(k^{2})$ write them as $B_{0}$ times
coefficients rational in $d$ and polynomial in $\xi$. For $\xi\neq1$ the reduction involves integrals with higher powers of the
propagators, which can contain infrared poles; these must cancel, and their cancellation was checked in the bosonic loops of the three
two-point functions.
\item At three points, the amplitudes were computed in full in the Feynman--de Donder gauge, in the
basis $\{B_{0},C_{0}\}$; this is the form in which the Ward--Takahashi identities of
Sec.~\ref{sec:ward} were verified. For the counterterms only their poles were decomposed.
\item At four points, only the poles of $\delta^{4}_{\bar G}S_{\rm eff}$ and
$\delta_{\bar\phi}\delta^{3}_{\bar G}S_{\rm eff}$ were computed, in the same gauge.
\end{itemize}
The poles are obtained without tensor reduction. Being local, the pole of a function with $n_{G}$
external $\bar G$ legs is a homogeneous polynomial of degree $4-n_{G}$ in the external momenta,
whatever the number of $\bar\phi$ legs, since each $\bar G$ counts as one derivative of $\bar\phi$.
It is extracted by expanding the integrand in the external momenta to that order and isolating the
ultraviolet pole of the resulting logarithmic integrals (Appendix~\ref{app:poles}). In particular,
$\delta^{4}_{\bar G}S_{\rm eff}$ has a pole of degree zero and can be computed at vanishing external
momenta. At two points the procedure reproduces the poles of the full amplitudes of the three functions, for
all $\xi$.

The calculations were performed in Mathematica~\cite{Mathematica} with the packages
\textsc{FeynCalc}~\cite{Mertig1991,Shtabovenko2016,Shtabovenko2020,Shtabovenko2025} and
\textsc{xAct}~\cite{xAct,MartinGarcia2008}, including \textsc{xTras}~\cite{Nutma2014}.
In this work, the authors used Claude Opus 5.1 and Claude Opus 5.5 (Anthropic), accessed through Claude Code, for literature
search, for drafting and editing the text and its LaTeX source, for writing and debugging part of the
Mathematica code, and in formulating the completeness argument of Appendix~\ref{app:basis}. The authors
directed this use, reviewed all AI-assisted text, code and arguments, verified the results through the
independent checks described in Secs.~\ref{sec:twopointchecks} and \ref{sec:fourpointtest} and
Appendix~\ref{app:basis}, and take full responsibility for the content.
\section{Two-point functions}
\label{sec:twopoint}

\begin{figure}[t]
\centering
\setlength{\tabcolsep}{2pt}
\begin{tabular}{ccc}
\begin{tikzpicture}[scale=0.42]
\coordinate (e1) at (-2.5,0);
\coordinate (L) at (-1,0);
\coordinate (R) at (1,0);
\coordinate (e2) at (2.5,0);
\draw[grav,bgtick] (e1) -- (L);
\draw[grav] (L) to[bend left=90,looseness=1.5] (R);
\draw[grav] (R) to[bend left=90,looseness=1.5] (L);
\draw[grav,bgtick] (e2) -- (R);
\node at (0,-2.55) {\scriptsize (a)};
\end{tikzpicture} &
\begin{tikzpicture}[scale=0.42]
\coordinate (e1) at (-2.5,0);
\coordinate (L) at (-1,0);
\coordinate (R) at (1,0);
\coordinate (e2) at (2.5,0);
\draw[grav,bgtick] (e1) -- (L);
\draw[grav] (L) to[bend left=90,looseness=1.5] (R);
\draw[grav] (R) to[bend left=90,looseness=1.5] (L);
\draw[aux,bgtick] (e2) -- (R);
\node at (0,-2.55) {\scriptsize (b)};
\end{tikzpicture} &
\begin{tikzpicture}[scale=0.42]
\coordinate (e1) at (-2.5,0);
\coordinate (L) at (-1,0);
\coordinate (R) at (1,0);
\coordinate (e2) at (2.5,0);
\draw[aux,bgtick] (e1) -- (L);
\draw[grav] (L) to[bend left=90,looseness=1.5] (R);
\draw[grav] (R) to[bend left=90,looseness=1.5] (L);
\draw[aux,bgtick] (e2) -- (R);
\node at (0,-2.55) {\scriptsize (c)};
\end{tikzpicture} \\[2pt]
\begin{tikzpicture}[scale=0.42]
\coordinate (e1) at (-2.5,0);
\coordinate (L) at (-1,0);
\coordinate (R) at (1,0);
\coordinate (e2) at (2.5,0);
\draw[grav,bgtick] (e1) -- (L);
\draw[gh,ghar] (L) to[bend left=90,looseness=1.5] (R);
\draw[gh,ghar] (R) to[bend left=90,looseness=1.5] (L);
\draw[grav,bgtick] (e2) -- (R);
\node at (0,-2.55) {\scriptsize (d)};
\end{tikzpicture} &
\begin{tikzpicture}[scale=0.42]
\coordinate (e1) at (-2.5,0);
\coordinate (L) at (-1,0);
\coordinate (R) at (1,0);
\coordinate (e2) at (2.5,0);
\draw[grav,bgtick] (e1) -- (L);
\draw[gh,ghar] (L) to[bend left=90,looseness=1.5] (R);
\draw[gh,ghar] (R) to[bend left=90,looseness=1.5] (L);
\draw[aux,bgtick] (e2) -- (R);
\node at (0,-2.55) {\scriptsize (e)};
\end{tikzpicture} &
\begin{tikzpicture}[scale=0.42]
\coordinate (e1) at (-2.5,0);
\coordinate (L) at (-1,0);
\coordinate (R) at (1,0);
\coordinate (e2) at (2.5,0);
\draw[aux,bgtick] (e1) -- (L);
\draw[gh,ghar] (L) to[bend left=90,looseness=1.5] (R);
\draw[gh,ghar] (R) to[bend left=90,looseness=1.5] (L);
\draw[aux,bgtick] (e2) -- (R);
\node at (0,-2.55) {\scriptsize (f)};
\end{tikzpicture}
\end{tabular}
\caption{One-loop diagrams of the background two-point functions: (a), (d) $\delta^{2}_{\bar\phi}S_{\rm eff}$;
(b), (e) $\delta_{\bar\phi}\delta_{\bar G}S_{\rm eff}$; (c), (f) $\delta^{2}_{\bar G}S_{\rm eff}$. The upper row has a bosonic
loop and the lower row a ghost loop. Wavy lines are gravitons, double wavy lines the connection field $G$,
dashed lines with arrows ghosts; a bar across a line marks a background leg.}
\label{fig:selfenergies}
\end{figure}
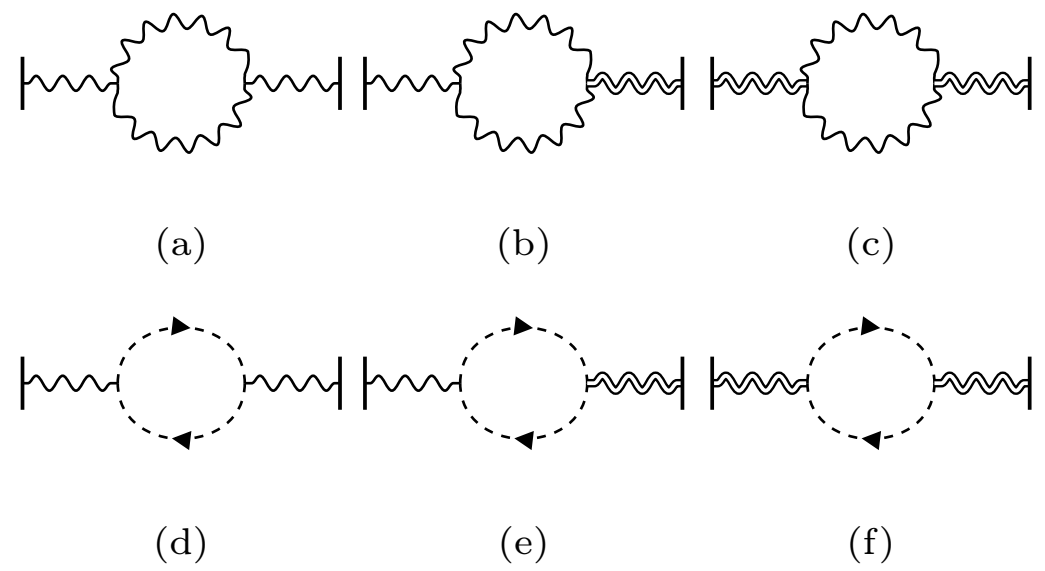
 
\subsection{Structure}
\label{sec:twopointstructure}

Away from the Feynman--de Donder gauge two ingredients enter: the gauge-fixing vertices carry a
factor $1/\xi$, and the graviton propagator acquires its $\xi$-dependent terms. After the reduction
to $B_{0}$ (Sec.~\ref{sec:poles}), the three background two-point functions (Fig.~\ref{fig:selfenergies}) are proportional to
$B_{0}(k^{2})$, with coefficients rational in $d$ and polynomial in $\xi$: the powers of $1/\xi$ from
the vertices cancel against the powers of $\xi$ from the propagator.

For each local invariant $I$ we define its kernel in the parametrization~\eqref{eq:ctparam},
\begin{equation}
K_{I}^{ab}(k)\equiv\frac{\delta^{2}}{\delta\Phi_{a}(k)\,\delta\Phi_{b}(-k)}\int d^{d}x\,I,
\qquad \Phi\in\{\bar\phi,\bar G\}.
\label{eq:kernel}
\end{equation}
Each two-point function is then written as
\begin{multline}
i\,\delta_{a}\delta_{b}S_{\rm eff}(k)=B_{0}(k^{2})\sum_{I}c_{I}\,K_{I}^{ab}(k)\\
+(\text{nonlocal structures}),
\label{eq:gammaK}
\end{multline}
where the left-hand side is the sum of the amputated one-loop diagrams. Both sides are of order
$\kappa^{2}$; in the explicit expressions of the two-point functions and kernels below, in
Sec.~\ref{sec:basis} and in Appendix~\ref{app:tensor}, this common factor is omitted, that is, they stand
for $\delta_{a}\delta_{b}S_{\rm eff}/\kappa^{2}$ and $K_{I}/\kappa^{2}$. The coefficients $c_{I}$ are not
affected. Since the pole of $B_{0}$ is
$i/[8\pi^{2}(4-d)]$, the divergent part of the effective action is
\begin{equation}
S^{\rm div}_{\rm eff}=\frac{1}{8\pi^{2}(4-d)}\sum_{I}c_{I}\Big|_{d=4}\int d^{d}x\,I,
\qquad
S_{\rm ct}\equiv-S^{\rm div}_{\rm eff}.
\label{eq:Sdiv}
\end{equation}
The values of the coefficients at $d=4$ fix the residue of the pole and also the coefficient of
the logarithm in $B_{0}$. Their terms of order $d-4$, multiplied by the pole, give finite
contributions. Those multiplying polynomial structures are local and can be changed by finite local
counterterms.

Structures with $1/k^{2}$ or $1/k^{4}$ appear in $\delta_{\bar\phi}\delta_{\bar G}S_{\rm eff}$ and
$\delta^{2}_{\bar G}S_{\rm eff}$, but all their coefficients are proportional to $(d-4)(\xi-1)$: they
vanish in the Feynman--de Donder gauge in any dimension, and in $d=4$ in any gauge. Multiplied by
the pole of $B_{0}$ they give finite terms, and the pole is local, as it must be to be absorbed by
local counterterms. In $\delta^{2}_{\bar\phi}S_{\rm eff}$ all tensor structures are polynomial in the
momentum. The amplitudes are nevertheless not local, since they contain $B_{0}(k^{2})$ with its
logarithmic dependence on $k^{2}$; only their ultraviolet poles are. The tensor bases and the
coefficients in arbitrary $d$ and $\xi$, separately for the bosonic and ghost loops, are given in
Appendix~\ref{app:tensor}.

\subsection{Graviton self-energy}
\label{sec:phiphi}

With $k$ the external momentum, the self-energy is expanded on five structures $T^{\phi\phi}_{i}$,
Eq.~\eqref{eq:Tphiphi}, as
\begin{multline}
i\,\delta^{2}_{\bar\phi}S_{\rm eff}^{\alpha\beta,\mu\nu}(k)=B_{0}(k^{2})\big[a_{1}k^{4}T^{\phi\phi}_{1}+a_{2}k^{4}T^{\phi\phi}_{2}
+a_{3}k^{2}T^{\phi\phi}_{3}\\
-a_{1}k^{2}T^{\phi\phi}_{4}+(d-2)\,a_{3}T^{\phi\phi}_{5}\big],
\label{eq:Piphi}
\end{multline}
where the relations $a_{4}=-a_{1}$, $a_{5}=(d-2)a_{3}$ and $a_{3}=2a_{1}+(d-2)a_{2}$ are the three
conditions imposed by the Ward--Takahashi identity~\eqref{eq:ward2}. Both loops satisfy them
identically in $d$ and $\xi$. The transverse space has dimension two and is spanned by the kernels
of $R^{2}$ and $\Ric_{\mu\nu}\Ric^{\mu\nu}$,
\begin{align}
K^{\phi\phi}_{R^{2}}&=2\Big[\frac{k^{4}T^{\phi\phi}_{2}}{(d-2)^{2}}+\frac{k^{2}T^{\phi\phi}_{3}}{d-2}+T^{\phi\phi}_{5}\Big],
\nonumber\\
K^{\phi\phi}_{\Ric^{2}}&=\frac{k^{4}T^{\phi\phi}_{1}}{4}-\frac{(d-4)k^{4}T^{\phi\phi}_{2}}{2(d-2)^{2}}
+\frac{k^{2}T^{\phi\phi}_{3}}{d-2}\nonumber\\
&\quad-\frac{k^{2}T^{\phi\phi}_{4}}{4}+T^{\phi\phi}_{5};
\label{eq:Kphiphi}
\end{align}
the kernel of $\Riem_{\mu\nu\rho\sigma}\Riem^{\mu\nu\rho\sigma}$ is not independent, since the
quadratic part of the Gauss--Bonnet combination is a total derivative in any $d$, so that
$K^{\phi\phi}_{\Riem^{2}}=4K^{\phi\phi}_{\Ric^{2}}-K^{\phi\phi}_{R^{2}}$.\footnote{In the metric case,
the quadratic part of $\sqrt{-g}\,\mathcal{E}$ around flat space is a total derivative in any
dimension, by the linearized Bianchi identities, so two-point functions cannot detect the
dimension. The cubic part is a total derivative only for $d\le4$: its variation is the quadratic
part of the Lanczos tensor, proportional to
$\delta^{\mu\mu_{1}\cdots\mu_{4}}_{\nu\nu_{1}\cdots\nu_{4}}R^{\nu_{1}\nu_{2}}{}_{\mu_{1}\mu_{2}}R^{\nu_{3}\nu_{4}}{}_{\mu_{3}\mu_{4}}$,
which vanishes identically for $d\le4$~\cite{Zumino1986}. We checked this explicitly:
in $d=4$ the cubic vertex of $\sqrt{-g}\,\mathcal{E}$ is proportional to the Gram determinant of the
five vectors of the vertex (two independent momenta and three polarizations), which vanishes
identically in four dimensions and not for $d\ge5$. This is the origin of the four-dimensional
identity of the $\phi\phi\phi$ sector used in Sec.~\ref{sec:vertices}.}
The coefficients of the two invariants are
\begin{equation}
c^{(\phi\phi)}_{\Ric}=4a_{1},\qquad
c^{(\phi\phi)}_{R}=\tfrac{1}{2}(d-2)^{2}a_{2}+(d-4)\,a_{1}.
\label{eq:cphiphi}
\end{equation}
In $d=4$ the bosonic loop gives $c_{R}=-(4\xi+13)/24$ and $c_{\Ric}=\xi^{2}-\xi+55/12$, and the
ghost loop $c_{R}=-17/60$ and $c_{\Ric}=-7/30$, so that
\begin{equation}
c^{(\phi\phi)}_{R}=-\frac{20\xi+99}{120},\qquad
c^{(\phi\phi)}_{\Ric}=\xi^{2}-\xi+\frac{87}{20}.
\label{eq:cphiphi4}
\end{equation}

These coefficients agree, for every $\xi$, with those of Eq.~(3.7) of Ref.~\cite{Brandt2026}, obtained
in the nondiagonal first-order formulation with the background connection on shell. The comparison
is meaningful because of Eq.~\eqref{eq:twopointmap}: at linear order $\bar G=0$ corresponds to the
on-shell background connection, so $\delta^{2}_{\bar\phi}S_{\rm eff}$ is that self-energy. Since the
Feynman rules of the two formulations are different, the agreement is a nontrivial test of the
calculation. In the gauge $\xi=1$ the result also agrees, in the Goldberg parametrization, with the
independent calculation of Gon\c{c}alves, Netto and Shapiro~\cite{Shapiro2018}, up to an overall
factor of two from the normalization: they write the divergent part as
$-\epsilon^{-1}\int d^{4}x\sqrt{-g}\,(\cdots)$ with $\epsilon\equiv(4\pi)^{2}(d-4)$, while
Eq.~\eqref{eq:Sdiv} uses $1/[8\pi^{2}(4-d)]=-2/\epsilon$.

\subsection{Mixed function}
\label{sec:phiG}

Let $\alpha\beta$ be the indices of the $\bar\phi$ leg, $(\lambda;\mu\nu)$ those of the $\bar G$ leg and
$k$ the momentum entering through the $\bar\phi$ leg. On the twelve structures $T^{\phi G}_{j}$ of
Eq.~\eqref{eq:TphiG} we write
$i\,\delta_{\bar\phi}\delta_{\bar G}S_{\rm eff}=-iB_{0}(k^{2})\sum_{j}k_{j}T^{\phi G}_{j}$. The
identity~\eqref{eq:ward2mix} imposes
\begin{gather}
k_{6}=k_{10}=0,\qquad k_{5}=2k_{1}+(d-2)k_{2},\nonumber\\
k_{8}=-k_{1}-k_{7},\qquad k_{4}=(d-2)k_{3}+2k_{7},\nonumber\\
k_{9}=-2k_{7},\qquad k_{12}=(d-2)k_{11},
\label{eq:wardS}
\end{gather}
leaving five free coefficients, $k_{1}$, $k_{2}$, $k_{3}$, $k_{7}$ and $k_{11}$. The two loops satisfy
Eqs.~\eqref{eq:wardS} separately, identically in $d$ and $\xi$. The solution of Eqs.~\eqref{eq:wardS}
spans five directions. Four of them are local, and their decomposition into curvature and
nonmetricity invariants is the subject of Sec.~\ref{sec:basis}.

The fifth direction deserves a comment. The combination
\begin{equation}
T^{\phi G}_{11}+(d-2)T^{\phi G}_{12}=\frac{k^{\lambda}k^{\mu}k^{\nu}}{k^{2}}
\big[(d-2)k^{\alpha}k^{\beta}+k^{2}\eta^{\alpha\beta}\big]
\label{eq:nonlocaldir}
\end{equation}
is transverse by itself, owing to the ratio $k_{12}=(d-2)k_{11}$, but it is not local: $T^{\phi G}_{11}$
is polynomial in $k$, while $T^{\phi G}_{12}$ carries a factor $1/k^{2}$. The kernel of any local
invariant is polynomial in $k$, so no counterterm reaches this direction, and since the one-loop
divergence is local, the pole of $k_{11}$ must vanish, although $T^{\phi G}_{11}$ alone looks local.
The bracket in Eq.~\eqref{eq:nonlocaldir} is, up to normalization, the $\bar\phi$ kernel of the
linearized Levi-Civita scalar curvature, $\mathring{R}\propto k_{\alpha}k_{\beta}\bar\phi^{\alpha\beta}+k^{2}\bar\phi/(d-2)$,
so the term has the form of a nonlocal form factor,
$\mathring{R}\,\partial^{-2}\partial_{\lambda}\partial^{\mu}\partial^{\nu}\bar G^{\lambda}{}_{\mu\nu}$. Its
coefficient,
\begin{equation}
k_{11}=-\frac{(d-4)(\xi-1)(d^{2}\xi-d^{2}-4d\xi+d+2)}{8(d-2)(d-1)},
\label{eq:k11}
\end{equation}
comes only from the bosonic loop and carries the factor $(d-4)(\xi-1)$: the pole cancels, but the
coefficient does not. As $d\to4$ the term is finite,
\begin{equation}
i\,\delta_{\bar\phi}\delta_{\bar G}S_{\rm eff}\supset-\frac{5(\xi-1)}{192\pi^{2}}
\big[T^{\phi G}_{11}+2T^{\phi G}_{12}\big],
\label{eq:k11finite}
\end{equation}
a finite, nonlocal and gauge-dependent part of the off-shell effective action, which requires no
counterterm and vanishes in the Feynman--de Donder gauge.

\subsection{Connection self-energy}
\label{sec:GG}

With $(\lambda;\mu\nu)$ and $(\gamma;\alpha\beta)$ the indices of the two legs, the self-energy of the
background connection is expanded on sixteen local structures $T^{GG}_{i}$ and six nonlocal ones
$N_{j}$, Eqs.~\eqref{eq:TGG} and~\eqref{eq:NGG},
\begin{equation}
i\,\delta^{2}_{\bar G}S_{\rm eff}=B_{0}(k^{2})\Big[\sum_{i=1}^{16}g_{i}\langle T^{GG}_{i}\rangle
+\sum_{j=1}^{6}n_{j}\langle N_{j}\rangle\Big],
\label{eq:GammaGG}
\end{equation}
where $\langle\,\cdot\,\rangle$ averages over $\mu\leftrightarrow\nu$, $\alpha\leftrightarrow\beta$ and
the exchange of the two legs. There is no autonomous two-point identity for this function. In the other
two sectors the identities reduce the structures from five to two and from twelve to five because
$\bar\phi$ has an inhomogeneous variation, whose contraction with a two-point function survives at
vanishing fields. The variation of $\bar G$, Eq.~\eqref{eq:deltaGbar}, has no inhomogeneous part;
it is linear in the fields, and in the variation of a functional quadratic in $\bar G$ it gives
quadratic terms, which enter the three-point identities~\eqref{eq:ward3GG}
and~\eqref{eq:ward3phiG} through the insertions $\mathcal{L}_{\varepsilon}$ and
$\delta\Xi_{\varepsilon}/\delta\bar\phi$. The function is thus constrained only through its relation
to three-point functions with a $\bar\phi$ leg, and its twenty-two coefficients are independent. The ghost loop is independent of $\xi$, and its tensor coefficients are polynomial in
the momentum, without the structures $N_{j}$; the coefficients of $N_{j}$ in the bosonic loop are
proportional to $(d-4)(\xi-1)$.

\subsection{Checks}
\label{sec:twopointchecks}

Besides the Ward--Takahashi identities, satisfied identically in $d$ and $\xi$ and separately by the
two loops, the three self-energies were recomputed independently with
\textsc{xAct}~\cite{MartinGarcia2008,xAct}, in arbitrary $d$ and $\xi$, in an independent
implementation based on projections onto tensor bases. The results agree exactly with those of
this section, separately for the bosonic and ghost loops. In the graviton sector the comparisons
with Refs.~\cite{Brandt2026,Shapiro2018} above provide external checks. For the mixed and connection
sectors, Eq.~\eqref{eq:twopointmap} relates our functions to the off-shell self-energies of the
nondiagonal formulation tabulated in Ref.~\cite{Brandt2026}. Mapping our results with it, we find exact
agreement in $d=4$, for every $\xi$, in all three sectors (graviton, mixed and connection), separately
for the bosonic and ghost loops.
\section{Counterterm basis at two points}
\label{sec:basis}

\subsection{Building blocks and the triangular structure}
\label{sec:blocks}

The counterterms are local functionals of the background fields $\bar h^{\mu\nu}$ and
$\bar\Gamma^{\lambda}_{\mu\nu}$, invariant under diffeomorphisms, which is what guarantees the
background invariance of the effective action, and densities of weight one, like the
Einstein--Hilbert Lagrangian. Since $\bar h$ and $\bar\Gamma$ are independent, there are two
covariant building blocks: the curvature $\Riem^{\rho}{}_{\sigma\mu\nu}$ of the connection, with its
contractions, and the nonmetricity $Q_{\lambda}{}^{\mu\nu}$, Eq.~\eqref{eq:Q}. We use the curvature of
the independent connection rather than $\mathring{\Riem}$, because it is polynomial in the
connection, hence in $\bar G$ and $\partial\bar\phi$; the complete invariants still involve inverse
metrics and density factors. In what follows all fields are background fields, and we drop the
overbars.

Dimensional counting restricts the list. The one-loop two-point functions are of order $\kappa^{2}$
with four derivatives, each $G$ leg counting as one derivative of $\phi$: in $d=4$, $G$ has mass
dimension two, like $\partial\phi$. The invariants with dimensionless coefficients and four derivatives
are, schematically and after integrations by parts and curvature identities,
\begin{equation}
\mathcal{R}^{2},\quad \mathcal{R}\nabla Q,\quad (\nabla Q)^{2},\quad \mathcal{R}QQ,\quad QQ\nabla Q,\quad Q^{4},
\label{eq:classes}
\end{equation}
where $\mathcal{R}$ stands for the curvature tensor and any of its contractions, and $Q$ includes
its traces $V$ and $W$. Since $\mathcal{R}$ and $Q$ vanish on the flat background, the first three
classes start at quadratic order in the fields, while the other three start at cubic or quartic
order and are fixed only by three- and four-point functions. Weights are assigned as follows: the
curvature with indices moved by $g$ has weight zero, and $Q$ and the quantities built from it and
from $g$ have weight one, except $W_{\lambda}$, of weight zero; an invariant of weight $w$ carries
the factor $(\sqrt{-g})^{1-w}$. These factors start at order $\kappa\phi$ and do not affect the
two-point kernels.

Since the connection is not metric, moving indices does not commute with $\nabla$. We fix that
$\nabla$ acts on the natural objects $Q_{\lambda}{}^{\mu\nu}$, $V^{\nu}$ and $W_{\lambda}$, Eqs.~\eqref{eq:Q}
and~\eqref{eq:traces}, and only afterwards are indices moved with $g$; thus $\nabla_{\alpha}W_{\lambda}$ is
the derivative of $W_{\lambda}$ itself, not $h_{\mu\nu}\nabla_{\alpha}Q_{\lambda}{}^{\mu\nu}$. Likewise, the
Riemann tensor with all indices down is $\Riem_{\rho\sigma\mu\nu}\equiv g_{\rho\lambda}\Riem^{\lambda}{}_{\sigma\mu\nu}$,
and $\Ric_{\mu\nu}\equiv\Riem^{\rho}{}_{\mu\rho\nu}$. Expressions written below with lowered indices are to be
read in this sense. Other orderings differ by terms of order $Q^{2}$, which leave the two-point
kernels unchanged but shift the cubic parts of the invariants (Sec.~\ref{sec:vertices}).

At linear order the $\phi$-dependent part of the connection~\eqref{eq:ctparam} is the Levi-Civita
connection, so $Q$ depends only on $G$, while $R=\mathring{R}[\phi]+\kappa\,\partial_{\lambda}G^{\lambda\mu}{}_{\mu}$.
This gives a triangular structure, which makes the problem solvable sector by sector:
\begin{center}
\begin{tabular}{lccc}
\hline
& $\phi\phi$ & $\phi G$ & $GG$\\
\hline
$\mathcal{R}^{2}$ & yes & yes & yes\\
$\mathcal{R}\nabla Q$ & --- & yes & yes\\
$(\nabla Q)^{2}$ & --- & --- & yes\\
\hline
\end{tabular}
\end{center}
The graviton self-energy fixes the curvature-squared terms, $a_{R}=c^{(\phi\phi)}_{R}$ and
$a_{\Ric}=c^{(\phi\phi)}_{\Ric}$ of Eq.~\eqref{eq:cphiphi4}, the kernel of $\Riem^{2}$ being dependent
in this sector (Sec.~\ref{sec:phiphi}).

\subsection{The mixed sector and the direction \texorpdfstring{$X$}{X}}
\label{sec:X}

The local part of the mixed function spans the four directions allowed by Eqs.~\eqref{eq:wardS}.
Three of them are the curvature-squared kernels,
\begin{align}
K^{\phi G}_{R^{2}}&=i\Big[-\frac{2T^{\phi G}_{2}}{d-2}-2T^{\phi G}_{5}\Big],\nonumber\\
K^{\phi G}_{\Ric^{2}}&=i\Big[\frac{T^{\phi G}_{1}}{2}-\frac{T^{\phi G}_{2}}{d-2}-\frac{T^{\phi G}_{8}}{2}\Big],
\nonumber\\
K^{\phi G}_{\Riem^{2}}&=i\Big[T^{\phi G}_{1}-\frac{2T^{\phi G}_{2}}{d-2}+\frac{T^{\phi G}_{3}-T^{\phi G}_{4}}{d-1}
\nonumber\\
&\qquad-\frac{T^{\phi G}_{7}+T^{\phi G}_{8}}{2}+T^{\phi G}_{9}\Big],
\label{eq:KphiG}
\end{align}
and the fourth is
\begin{multline}
X\equiv i\big[T^{\phi G}_{3}+(d-2)T^{\phi G}_{4}\big]\\
=i\,(\eta^{\mu\lambda}k^{\nu}+\eta^{\nu\lambda}k^{\mu})\big[(d-2)k^{\alpha}k^{\beta}+k^{2}\eta^{\alpha\beta}\big].
\label{eq:Xdef}
\end{multline}
Thus
\begin{multline}
i\,\delta_{\bar\phi}\delta_{\bar G}S_{\rm eff}=B_{0}\big[C_{R}K_{R^{2}}+C_{\Ric}K_{\Ric^{2}}\\
+C_{\Riem}K_{\Riem^{2}}+C_{X}X\big]-iB_{0}\,k_{11}\big[T^{\phi G}_{11}+(d-2)T^{\phi G}_{12}\big],
\label{eq:decmix}
\end{multline}
with $C_{R}=k_{1}+\frac{1}{2}(d-2)k_{2}$, $C_{\Ric}=-2k_{1}-4k_{7}$, $C_{\Riem}=2k_{7}$ and
$C_{X}=-k_{3}-2k_{7}/(d-1)$ (Appendix~\ref{app:tensor}). In $d=4$,
\begin{gather}
C_{R}=\frac{(\xi-2)(3\xi-2)}{24},\qquad C_{\Ric}=\frac{3\xi^{2}-\xi+1}{6},\nonumber\\
C_{\Riem}=\frac{2\xi^{2}-2\xi+17}{12},\qquad C_{X}=-\frac{\xi^{2}-7\xi+4}{72}.
\label{eq:Cmix4}
\end{gather}

Two facts follow. First, no contraction quadratic in the curvature of $\Gamma$ produces $X$. We
examined $R^{2}$; the two contractions of the Ricci tensor with itself, $\Ric_{\mu\nu}\Ric^{\mu\nu}$ and
$\Ric_{\mu\nu}\Ric^{\nu\mu}$, distinct because $\Ric_{\mu\nu}$ is not symmetric; the analogous contractions
with the co-Ricci tensor, obtained by contracting the second and third indices of the Riemann
tensor with the metric, which for a metric connection coincides with the Ricci tensor up to a sign
but is independent here; $\Riem_{\mu\nu\rho\sigma}\Riem^{\mu\nu\rho\sigma}$,
$\Riem_{\mu\nu\rho\sigma}\Riem^{\rho\sigma\mu\nu}$ and $\Riem_{\mu\nu\rho\sigma}\Riem^{\mu\rho\nu\sigma}$; and the
Hodge duals, such as the Pontryagin density and the $\varepsilon\varepsilon$ form of the Euler density.
All have a vanishing component along $X$.\footnote{With $\widehat{\Ric}_{\mu\nu}$ the co-Ricci tensor,
the mixed kernels of the contractions not in the basis are
$K_{\widehat{\Ric}\cdot\widehat{\Ric}}=\frac{1}{2}K_{R^{2}}-K_{\Ric^{2}}+\frac{1}{2}K_{\Riem^{2}}$,
$K_{\Ric\cdot\widehat{\Ric}}=-\frac{1}{4}K_{R^{2}}-\frac{1}{4}K_{\Riem^{2}}$, and, for the other two orders
of contraction of the Riemann tensor, $K_{\Riem_{\mu\nu\rho\sigma}\Riem^{\rho\sigma\mu\nu}}=K_{\Riem^{2}}$ and
$K_{\Riem_{\mu\nu\rho\sigma}\Riem^{\mu\rho\nu\sigma}}=\frac{1}{2}K_{\Riem^{2}}$, identically in $d$.}
Since $C_{X}\neq0$ for generic $\xi$ (it vanishes only at $\xi=(7\pm\sqrt{33})/2$), curvature squared
does not suffice in a generic gauge. Second, the curvature coefficients themselves do not fit. With
$a_{R}$ and $a_{\Ric}$ fixed by the graviton sector, the only freedom left in the curvature-squared
class is a multiple $\lambda$ of the Gauss--Bonnet combination $\mathcal{E}\equiv R^{2}-4\Ric_{\mu\nu}\Ric^{\mu\nu}
+\Riem_{\mu\nu\rho\sigma}\Riem^{\mu\nu\rho\sigma}$, which does not change the graviton sector. Matching
$C_{\Riem}$ requires $\lambda=C_{\Riem}$, and then both $C_{R}=a_{R}+\lambda$ and $C_{\Ric}=a_{\Ric}-4\lambda$ fail:
in $d=4$, $C_{R}-a_{R}-C_{\Riem}=-(5\xi^{2}+51)/120$ and $C_{\Ric}-a_{\Ric}+4C_{\Riem}=(10\xi^{2}+10\xi+89)/60$,
neither of which vanishes for real $\xi$. Both facts require terms
with the nonmetricity.

\subsection{The invariants \texorpdfstring{$\mathcal{R}\nabla Q$}{R∇Q}}
\label{sec:RQ}

At linear order the two traces of $Q$ are
\begin{equation}
W_{\lambda}=\frac{2(d-2)}{d-1}\,\kappa\,G^{\rho}{}_{\rho\lambda},\qquad
\partial_{\mu}V^{\mu}=\kappa\,\partial_{\lambda}G^{\lambda\mu}{}_{\mu}-\frac{\partial^{\lambda}W_{\lambda}}{d-2}.
\label{eq:VWlinear}
\end{equation}
Since $\kappa\,\partial_{\lambda}G^{\lambda\mu}{}_{\mu}=R-\mathring{R}$ at this order, $R\,\nabla_{\mu}V^{\mu}$ is,
up to cubic terms, $R(R-\mathring{R})-R\,\nabla^{\lambda}W_{\lambda}/(d-2)$. On the Ricci side we use
\begin{equation}
Y\equiv\Ric^{\mu}{}_{\nu}\,\nabla_{\rho}Q_{\mu}{}^{\rho\nu},\qquad
Z\equiv\Ric_{\mu\nu}\,\nabla^{\lambda}Q_{\lambda}{}^{\mu\nu},
\label{eq:YZ}
\end{equation}
with the first index of $\Ric^{\mu\nu}$ on the derivative index of $Q$ in $Y$. The contractions
$\Ric^{\mu\nu}\nabla_{\mu}W_{\nu}$ and $\Ric^{\mu\nu}\nabla_{\mu}V_{\nu}$ are not independent in the mixed
sector: there the Ricci tensor is the linearized Levi-Civita one, and the linearized Bianchi
identity $k_{\mu}\mathring{\Ric}^{\mu\nu}=\frac{1}{2}k^{\nu}\mathring{R}$ reduces both to half the invariants with
$R$; their differences contribute only to $GG$, where they are absorbed by $(\nabla Q)^{2}$. The mixed
kernels are
\begin{align}
R\,\nabla_{\mu}V^{\mu}&\to\tfrac{1}{2}K_{R^{2}}+\frac{X}{(d-1)(d-2)},\nonumber\\
\sqrt{-g}\,R\,\nabla^{\lambda}W_{\lambda}&\to-\frac{X}{d-1},\nonumber\\
Y&\to K_{\Ric^{2}}-\tfrac{1}{4}K_{\Riem^{2}}+\frac{X}{2(d-1)(d-2)},\nonumber\\
Z&\to K_{\Ric^{2}}-\tfrac{1}{2}K_{\Riem^{2}},
\label{eq:kernelsRQ}
\end{align}
all transverse, and the four span the whole local space of the mixed sector. That an invariant built
from the Ricci tensor reaches the direction $K_{\Riem^{2}}$ has a simple reason: in the mixed sector
the curvature of $\phi$ appears contracted with $k$ and with $G$, and the divergence of the
linearized Weyl tensor is proportional to the Cotton tensor, that is, to derivatives of the Ricci
tensor.

This makes $\Riem_{\mu\nu\rho\sigma}\Riem^{\mu\nu\rho\sigma}$ redundant. Since $\mathring{\mathcal{E}}$ is a
total derivative in $d=4$ and $\mathcal{E}-\mathring{\mathcal{E}}$ vanishes with $Q$, $\mathcal{E}$ is, up to a
boundary term, a combination of invariants of the classes $\mathcal{R}\nabla Q$, $(\nabla Q)^{2}$,
$\mathcal{R}QQ$, $QQ\nabla Q$ and $Q^{4}$. In the sectors computed here its vertices lie in the span of the
basis; that it is a combination of the specific representatives chosen to all orders depends on the
completeness of the basis (Appendix~\ref{app:basis}). We may therefore drop
$\Riem_{\mu\nu\rho\sigma}\Riem^{\mu\nu\rho\sigma}$, as in the metric case, but for a different reason: not
because $\mathcal{E}$ is topological, which it is not (Sec.~\ref{sec:necessity}), but because its
difference from $\mathring{\mathcal{E}}$ lies in the classes with $Q$. Within the calculated sectors the total
counterterm Lagrangian does not depend on this choice; only its distribution among the classes does.

\subsection{Solution in the graviton and mixed sectors}
\label{sec:solution}

The part of the counterterm basis that contributes at two points is
\begin{multline}
\sum_{I}c_{I}I=a_{R}\sqrt{-g}R^{2}+a_{\Ric}\sqrt{-g}\Ric_{\mu\nu}\Ric^{\mu\nu}\\
+b_{V}R\,\nabla_{\mu}V^{\mu}+b_{W}\sqrt{-g}R\,\nabla^{\lambda}W_{\lambda}\\
+b_{Y}Y+b_{Z}Z+\mathcal{L}_{(\nabla Q)^{2}},
\label{eq:Lct2}
\end{multline}
where $\mathcal{L}_{(\nabla Q)^{2}}$ collects the terms quadratic in $\nabla Q$ (Sec.~\ref{sec:dQ2}). The
graviton sector fixes $a_{R}$ and $a_{\Ric}$; by Eqs.~\eqref{eq:kernelsRQ} the mixed sector gives
\begin{gather}
a_{R}+\tfrac{1}{2}b_{V}=C_{R},\qquad a_{\Ric}+b_{Y}+b_{Z}=C_{\Ric},\nonumber\\
-\tfrac{1}{4}b_{Y}-\tfrac{1}{2}b_{Z}=C_{\Riem},\nonumber\\
\frac{b_{V}}{(d-1)(d-2)}-\frac{b_{W}}{d-1}+\frac{b_{Y}}{2(d-1)(d-2)}=C_{X},
\label{eq:mixsystem}
\end{gather}
with a unique solution, given in Table~\ref{tab:coef2pt} for $d=4$ and in Appendix~\ref{app:tensor}
for arbitrary $d$. No internal consistency test remains in these two sectors: six numbers fix six
coefficients. The tests come from outside this counting: the Ward--Takahashi identities, satisfied
separately by the two loops, the locality of the pole, and the comparisons of Sec.~\ref{sec:phiphi}.

\begin{table}[t]
\caption{Coefficients of the counterterms~\eqref{eq:Lct2} fixed by the graviton and mixed sectors,
in $d=4$.}
\label{tab:coef2pt}
\renewcommand{\arraystretch}{1.5}
\begin{ruledtabular}
\begin{tabular}{lcc}
Coefficient & arbitrary $\xi$ & $\xi=1$\\
\hline
$a_{R}$ & $-\frac{1}{120}(20\xi+99)$ & $-\frac{119}{120}$\\
$a_{\Ric}$ & $\frac{1}{20}(20\xi^{2}-20\xi+87)$ & $\frac{87}{20}$\\
$b_{V}$ & $\frac{1}{60}(15\xi^{2}-20\xi+119)$ & $\frac{19}{10}$\\
$b_{W}$ & $\frac{1}{120}(10\xi^{2}-25\xi+58)$ & $\frac{43}{120}$\\
$b_{Y}$ & $\frac{1}{30}(-10\xi^{2}+30\xi-81)$ & $-\frac{61}{30}$\\
$b_{Z}$ & $\frac{1}{60}(-10\xi^{2}-10\xi-89)$ & $-\frac{109}{60}$\\
\end{tabular}
\end{ruledtabular}
\end{table}

\subsection{The connection sector and \texorpdfstring{$(\nabla Q)^{2}$}{(∇Q)²}}
\label{sec:dQ2}

Subtracting from $\delta^{2}_{\bar G}S_{\rm eff}$ the contributions of $R^{2}$, $\Ric_{\mu\nu}\Ric^{\mu\nu}$ and of
the four invariants $\mathcal{R}\nabla Q$, with the coefficients of Table~\ref{tab:coef2pt}, the remainder
is the $(\nabla Q)^{2}$ content. The $105$ contractions of $\partial Q\,\partial Q$ have kernels of rank $16$,
in $d=4$ and in generic $d$, all local: they span the sixteen local structures $T^{GG}_{i}$, and the
remainder fixes their coefficients without further condition. To write a Lagrangian we choose
sixteen invariants whose quadratic kernels form a basis of that space, favoring the traces of $Q$.
Besides $W_{\lambda}$ and $V_{\nu}$ we use the two divergences
\begin{equation}
A^{\mu\nu}\equiv\nabla^{\lambda}Q_{\lambda}{}^{\mu\nu},\qquad B_{\lambda}{}^{\nu}\equiv\nabla_{\mu}Q_{\lambda}{}^{\mu\nu},
\label{eq:AB}
\end{equation}
with $A$ symmetric and $B$ not; the first index of $B$ is the derivative index of $Q$, and
$\nabla^{\lambda}=g^{\lambda\alpha}\nabla_{\alpha}$. The table lists the monomials $\mathcal{O}_{j}$, in the shorthand explained in its caption, without
the weight factor; in the Lagrangian each carries $(\sqrt{-g})^{1-w_{j}}$:
$\sqrt{-g}$ for $\mathcal{O}_{1}$ and $\mathcal{O}_{6}$, none for $\mathcal{O}_{2}$, $\mathcal{O}_{4}$,
$\mathcal{O}_{7}$ and $\mathcal{O}_{9}$, and $1/\sqrt{-g}$ for the others. In $d=4$,
\begin{equation}
\mathcal{L}_{(\nabla Q)^{2}}\equiv\sum_{j=1}^{16}q_{j}\,(\sqrt{-g})^{1-w_{j}}\,\mathcal{O}_{j},
\label{eq:LdQ}
\end{equation}
with the operators and coefficients of Table~\ref{tab:dQ2}. The choice of basis is not unique.
Another choice reproducing the same two-point kernels differs by combinations whose quadratic kernels
vanish; such combinations are distinguished only at three points, and replacing one basis by another
requires readjusting the coefficients of the higher classes (Sec.~\ref{sec:vertices}).

\begin{table}[t]
\caption{Basis of the $(\nabla Q)^{2}$ invariants and their coefficients in $d=4$. Shorthand, not the
Einstein convention: all indices are written down and each repeated pair is contracted with the
inverse metric, e.g.\ $\nabla_{\alpha}V_{\gamma}\nabla_{\alpha}W_{\gamma}=g^{\alpha\beta}g^{\gamma\delta}\nabla_{\alpha}V_{\gamma}\nabla_{\beta}W_{\delta}$.
The derivative acts first on the natural objects $Q_{\lambda}{}^{\mu\nu}$, $V^{\nu}$, $W_{\lambda}$, and indices are
moved afterwards (Sec.~\ref{sec:blocks}): $\nabla_{\alpha}V_{\gamma}=g_{\gamma\delta}\nabla_{\alpha}V^{\delta}$, and $A$, $B$
are lowered from Eq.~\eqref{eq:AB}.}
\label{tab:dQ2}
\begin{ruledtabular}
\begin{tabular}{llcc}
$j$ & $\mathcal{O}_{j}$ & $q_{j}$ (arbitrary $\xi$) & $\xi=1$\\
\hline
1 & $(\nabla\!\cdot\!W)^{2}$ & $\frac{1}{240}(31\xi^{2}-67\xi+167)$ & $\frac{131}{240}$\\
2 & $(\nabla\!\cdot\!W)(\nabla\!\cdot\!V)$ & $\frac{1}{120}(-3\xi^{2}+26\xi-61)$ & $-\frac{19}{60}$\\
3 & $(\nabla\!\cdot\!V)^{2}$ & $\frac{1}{120}(2\xi^{2}+6\xi-117)$ & $-\frac{109}{120}$\\
4 & $\nabla_{\alpha}V_{\gamma}\nabla_{\alpha}W_{\gamma}$ & $\frac{1}{240}(\xi-1)(2\xi-77)$ & $0$\\
5 & $\nabla_{\alpha}V_{\gamma}\nabla_{\alpha}V_{\gamma}$ & $\frac{1}{240}(2\xi^{2}+41\xi-23)$ & $\frac{1}{12}$\\
6 & $\nabla_{\alpha}W_{\beta}\nabla_{\alpha}W_{\beta}$ & $\frac{1}{240}(-32\xi^{2}+64\xi-165)$ & $-\frac{133}{240}$\\
7 & $A_{\beta\gamma}\nabla_{\beta}W_{\gamma}$ & $\frac{1}{120}(-13\xi^{2}+71\xi-119)$ & $-\frac{61}{120}$\\
8 & $A_{\beta\gamma}\nabla_{\beta}V_{\gamma}$ & $\frac{1}{60}(-3\xi^{2}-9\xi-8)$ & $-\frac{1}{3}$\\
9 & $B_{\beta\gamma}\nabla_{\gamma}W_{\beta}$ & $\frac{1}{240}(17\xi^{2}-109\xi+214)$ & $\frac{61}{120}$\\
10 & $B_{\beta\gamma}\nabla_{\gamma}V_{\beta}$ & $\frac{1}{120}(7\xi^{2}-14\xi+107)$ & $\frac{5}{6}$\\
11 & $A_{\beta\gamma}B_{\beta\gamma}$ & $\frac{1}{60}(2\xi^{2}-59\xi+96)$ & $\frac{13}{20}$\\
12 & $A_{\beta\gamma}A_{\beta\gamma}$ & $\frac{1}{240}(2\xi^{2}+96\xi-77)$ & $\frac{7}{80}$\\
13 & $B_{\beta\gamma}B_{\beta\gamma}$ & $\frac{1}{120}(12\xi^{2}+41\xi-72)$ & $-\frac{19}{120}$\\
14 & $B_{\beta\gamma}B_{\gamma\beta}$ & $\frac{1}{120}(2\xi^{2}+6\xi-17)$ & $-\frac{3}{40}$\\
15 & $\nabla_{\alpha}Q_{\beta\gamma\delta}\nabla_{\alpha}Q_{\beta\gamma\delta}$ & $\frac{1}{120}(-2\xi^{2}-41\xi+48)$ & $\frac{1}{24}$\\
16 & $\nabla_{\alpha}Q_{\beta\gamma\delta}\nabla_{\alpha}Q_{\gamma\beta\delta}$ & $\frac{1}{240}(2\xi^{2}+61\xi-113)$ & $-\frac{5}{24}$\\
\end{tabular}
\end{ruledtabular}
\end{table}

\subsection{Necessity of each class and the Gauss--Bonnet combination}
\label{sec:necessity}

In the organization of invariants adopted here (identities and integrations by parts can move terms
between classes), the argument is cumulative. The graviton self-energy requires both $R^{2}$ and
$\Ric_{\mu\nu}\Ric^{\mu\nu}$ in a generic gauge: in $d=4$, $c^{(\phi\phi)}_{\Ric}$ vanishes for no real
$\xi$ and $c^{(\phi\phi)}_{R}$ only at $\xi=-99/20$. The mixed sector requires the four invariants
$\mathcal{R}\nabla Q$ (Sec.~\ref{sec:X}), which exactly cover its four local directions. The connection
sector, after the contributions of the previous classes are subtracted, retains nonvanishing
components along all sixteen local structures, and only terms that reach neither $\phi\phi$ nor
$\phi G$, the $(\nabla Q)^{2}$, can absorb them. The classes $\mathcal{R}QQ$, $QQ\nabla Q$ and $Q^{4}$ do not
enter at two points; their role is decided in Sec.~\ref{sec:vertices}.

In the metric formulation the Gauss--Bonnet combination is a total derivative in $d=4$, which is
what allows the one-loop counterterms to be written with $R^{2}$ and $\Ric_{\mu\nu}\Ric^{\mu\nu}$ alone and
fixed by the graviton self-energy~\cite{tHooft1974,Brandt2022}. With an independent connection this
fails. Neither $\mathcal{E}$ nor its $\varepsilon\varepsilon$ form,
\begin{equation}
\sqrt{-g}\,\mathcal{E}_{\varepsilon\varepsilon}\equiv\frac{\sqrt{-g}}{4}\,\delta^{\mu\nu\rho\sigma}_{\alpha\beta\gamma\delta}\,
\Riem^{\alpha\beta}{}_{\mu\nu}\Riem^{\gamma\delta}{}_{\rho\sigma},
\label{eq:Eepseps}
\end{equation}
with $\Riem^{\alpha\beta}{}_{\mu\nu}\equiv\Riem^{\alpha}{}_{\lambda\mu\nu}\,g^{\lambda\beta}$ and the generalized delta not normalized (the signed sum over the $24$ permutations), is a total derivative when the connection is
not metric~\cite{Zumino1986,Janssen2019}; for a metric connection the two coincide, but their
reduction to each other uses symmetries of the Riemann tensor that only metric compatibility
guarantees. Equation~\eqref{eq:Eepseps} involves no Levi-Civita symbols; it equals
$-\frac{1}{4}\sqrt{-g}\,\varepsilon^{\mu\nu\rho\sigma}\varepsilon^{\alpha\beta\gamma\delta}\Riem_{\mu\nu\alpha\beta}\Riem_{\rho\sigma\gamma\delta}$
with the Levi-Civita tensor of $g$ in Lorentzian signature, and it is the form used in the vertex
checks below.
We checked this with the vertices of the two invariants in
plane waves, in the parametrization~\eqref{eq:ctparam} and in $d=4$: with the Levi-Civita connection
of $h$, used as a control, the vertices of $\mathcal{E}$ vanish in the $\phi\phi$ and $\phi\phi\phi$
sectors, the latter modulo the cubic Gauss--Bonnet identity of that sector
(Appendix~\ref{app:basis}), but with the independent
connection the $\phi G$ and $GG$ sectors do not vanish; for $\mathcal{E}_{\varepsilon\varepsilon}$ the quadratic
vertices and the cubic $\phi\phi\phi$ and $\phi\phi G$ ones vanish, but not the $GGG$ and $\phi GG$ ones. The
Pontryagin density, which does not involve the metric and is locally the divergence of a
Chern--Simons form for any linear connection~\cite{ChernSimons1974,Eguchi1980}, vanishes identically in
all these sectors and serves as a second control. Gauss--Bonnet has, like any invariant quadratic in
the curvature, the number of derivatives the one-loop divergences require, but it does not suffice:
it has no component along $X$, it cannot reconcile the curvature coefficients of the graviton and
mixed sectors, and the connection sector requires sixteen independent $(\nabla Q)^{2}$ coefficients. It
is one particular combination within the family $\mathcal{R}^{2}+\mathcal{R}\nabla Q+(\nabla Q)^{2}$, completed
at cubic and quartic order in Sec.~\ref{sec:vertices}.
\section{Vertex functions and higher orders}
\label{sec:vertices}

The classes $\mathcal{R}QQ$, $QQ\nabla Q$ and $Q^{4}$ of Eq.~\eqref{eq:classes} first appear at three and
four points. Five background functions are needed, and they play different roles.
$\delta^{3}_{\bar\phi}S_{\rm eff}$ and $\delta^{2}_{\bar\phi}\delta_{\bar G}S_{\rm eff}$ receive no
contribution from the new classes and are parameter-free predictions of the coefficients fixed at two
points. $\delta_{\bar\phi}\delta^{2}_{\bar G}S_{\rm eff}$ fixes the $\mathcal{R}QQ$ coefficients, but only after
passing a nontrivial condition: its remainder must lie in a proper subspace of the local vertices.
$\delta^{3}_{\bar G}S_{\rm eff}$ and $\delta^{4}_{\bar G}S_{\rm eff}$ fix the $QQ\nabla Q$ and $Q^{4}$
coefficients; the four-point function $\delta_{\bar\phi}\delta^{3}_{\bar G}S_{\rm eff}$, computed as a further
test, checks the pole of the first of these functions (Sec.~\ref{sec:fourpointtest}). The first three functions also enter the Ward--Takahashi identities of
Sec.~\ref{sec:ward}, which were verified with the full amplitudes; the comparison with the
counterterms uses only their poles. All the statements of this section hold in $d=4$ and in the
Feynman--de Donder gauge $\xi=1$.

\subsection{Three-point functions}
\label{sec:threepoint}

We label the momenta so that the leg of the unique species carries $k$ and the two legs of the same
species carry $p_{1}$, incoming, and $p_{2}$, outgoing, with $k+p_{1}=p_{2}$; in
$\delta^{3}_{\bar\phi}S_{\rm eff}$ the leg $k$ is the one that carries the inhomogeneous variation in
Eq.~\eqref{eq:ward3phiphi}. By the counting rules of Sec.~\ref{sec:counting}, the functions
$\delta_{\bar\phi}\delta^{2}_{\bar G}S_{\rm eff}$, $\delta^{2}_{\bar\phi}\delta_{\bar G}S_{\rm eff}$ and
$\delta^{3}_{\bar\phi}S_{\rm eff}$ have $12$, $14$ and $16$ diagrams, of which $11$, $12$ and $12$ do not
vanish; they are listed in Appendix~\ref{app:diagrams}. Only three diagrams in each function have three
momentum-dependent denominators and produce the triangle function $C_{0}$: the graviton triangle,
which is a single diagram because graviton lines are not oriented, and the ghost triangle in its two
orientations. After the contraction with Eq.~\eqref{eq:inhmom}, $C_{0}$ therefore cancels within the
graviton triangle by itself and between the two ghost orientations. Every triangle with one internal
$G$ line is a bubble, by rule (iv), and the remaining nonvanishing diagrams are bubbles with a quartic
vertex.

Grouping the diagrams by the momentum that flows through their most singular denominator, their
\emph{channel}, each identity closes channel by channel. The lower-order terms of
Eqs.~\eqref{eq:ward3GG}--\eqref{eq:ward3phiphi} are attached to the legs that do not carry the
variation, so they never reach the channel of the leg that does: in that channel the contracted
three-point diagrams must cancel among themselves. In
$\delta_{\bar\phi}\delta^{2}_{\bar G}S_{\rm eff}$ and $\delta^{2}_{\bar\phi}\delta_{\bar G}S_{\rm eff}$ the
calculation shows more: the diagrams of that channel are transverse one by one. In
$\delta_{\bar\phi}\delta^{2}_{\bar G}S_{\rm eff}$, where the variation is on the $\bar\phi$ leg $k$, they are
the bosonic triangle whose internal $G$ line joins the vertices of the two $\bar G$ legs, and the
bubble formed by the quartic gauge-fixing vertex $\bar G\bar G\phi\phi$, which carries both $\bar G$
legs, and the trivertex $\bar\phi\phi\phi$. In $\delta^{2}_{\bar\phi}\delta_{\bar G}S_{\rm eff}$, where the
variation is on the $\bar\phi$ leg $p_{2}$, they are the three diagrams in which the $\bar G$ leg and
the $\bar\phi$ leg $p_{1}$ meet, directly or through an internal $G$ line: the bosonic triangle whose $G$
line joins their vertices, the ghost bubble with the quartic vertex $\bar\phi\bar G\bar cc$ and the
bosonic bubble with $\bar\phi\bar G\phi\phi$, each closed by a trivertex on the leg $p_{2}$. These five
diagrams satisfy the identities whatever their coefficients, including their symmetry factors, and
contribute to the counterterms when their poles do not vanish; a nonvanishing pole was confirmed for the
two diagrams of $\delta_{\bar\phi}\delta^{2}_{\bar G}S_{\rm eff}$, while for the three of
$\delta^{2}_{\bar\phi}\delta_{\bar G}S_{\rm eff}$ only the transversality was checked. In $\delta^{3}_{\bar\phi}S_{\rm eff}$ the
channel $k$ also closes by itself, which gives a test of the inhomogeneous variation and of the
diagrams of that channel independent of the two-point terms.

The remaining symmetries of the enumeration are exchange symmetries. In
$\delta^{2}_{\bar\phi}\delta_{\bar G}S_{\rm eff}$ the exchange of the two $\bar\phi$ legs maps the set of
diagrams onto itself, which tests its completeness, but it is not a symmetry of
Eq.~\eqref{eq:ward3phiG}, where the leg $p_{2}$ is distinguished. In $\delta^{3}_{\bar\phi}S_{\rm eff}$
neither of the two exchanged legs carries the variation, and the exchange is a symmetry of the
identity itself.

\subsection{Cubic order: the unknowns}
\label{sec:cubicunknowns}

The comparison with the counterterms uses the exact parametrization~\eqref{eq:ctparam}, the same as at
two points. Its $\bar\phi$-dependent part is the Levi-Civita connection only at linear order, so the
nonmetricity, an invertible linear function $Q^{(1)}[G]$ of $G$ at first order, acquires at second order
the part $Q^{(2)}[\phi,\phi]\propto\phi\,\partial\phi$, besides terms in $\phi G$. This restricted part is
what brings the lower classes into the parameter-free predictions: $\mathcal{R}\nabla Q$ contributes to
$\delta^{3}_{\bar\phi}S_{\rm eff}$ through $\mathcal{R}^{(1)}[\phi]\,\nabla Q^{(2)}[\phi,\phi]$, and
$(\nabla Q)^{2}$ to $\delta^{2}_{\bar\phi}\delta_{\bar G}S_{\rm eff}$ through
$\nabla Q^{(1)}[G]\,\nabla Q^{(2)}[\phi,\phi]$. A covariant parametrization, with the full Levi-Civita
connection of $\bar h$, would not have these terms.

With the two-point coefficients fixed, the only cubic unknowns are the coefficients of $\mathcal{R}QQ$ and
$QQ\nabla Q$, since $Q^{4}$ is quartic. Because each factor $Q$ starts with $Q^{(1)}[G]$, none of them
reaches $\phi\phi\phi$ or $\phi\phi G$. In $\phi GG$ only the part $\mathcal{R}^{(1)}[\phi]\,QQ$ of
$\mathcal{R}QQ$ enters, and in $GGG$ both $\mathcal{R}^{(1)}[G]\,QQ$ and $QQ\nabla Q$ enter. The cubic vertices of
all the complete contractions of the two classes (Appendix~\ref{app:basis}) have the following exact
ranks in $d=4$:
\begin{center}
\begin{tabular}{lc}
\hline\hline
Independent directions & Rank\\
\hline
$\mathcal{R}QQ$ & $51$\\
$QQ\nabla Q$ & $58$\\
$\mathcal{R}QQ$ and $QQ\nabla Q$ together & $79$\\
\quad visible in $\delta_{\bar\phi}\delta^{2}_{\bar G}S_{\rm eff}$ & $21$\\
\quad visible only in $\delta^{3}_{\bar G}S_{\rm eff}$ & $58$\\
\hline\hline
\end{tabular}
\end{center}
The two families overlap: together they have rank $79$, not $51+58=109$. In $GGG$ the part
$\mathcal{R}^{(1)}[G]\,QQ$ adds nothing to the span of $QQ\nabla Q$, and the $21$ directions visible in $\phi GG$,
those of $\mathcal{R}^{(1)}[\phi]\,QQ$, do not overlap with the other $58$. We therefore choose $21$
invariants $\mathcal{R}QQ$ that complement the $58$ directions of $QQ\nabla Q$. The function
$\delta_{\bar\phi}\delta^{2}_{\bar G}S_{\rm eff}$ fixes their coefficients, and $\delta^{3}_{\bar G}S_{\rm eff}$,
by itself, fixes the other $58$.

\subsection{Parameter-free predictions}
\label{sec:predictions}

We expanded $\sum_{I}c_{I}I$ to cubic order, with the two-point coefficients of Sec.~\ref{sec:basis} at
$d=4$ and $\xi=1$ and the conventions of Sec.~\ref{sec:blocks}. The expansion reproduces all the
two-point kernels, including the $(\nabla Q)^{2}$ content of Table~\ref{tab:dQ2}, and the trivertex
$G\bar\phi\phi$, with a quantum connection leg, obtained in the same way from the action~\eqref{eq:actiondiag} coincides with the
Feynman rule, which fixes the momentum conventions. The poles of $\delta^{3}_{\bar\phi}S_{\rm eff}$ and
$\delta^{2}_{\bar\phi}\delta_{\bar G}S_{\rm eff}$, bosonic loop plus ghost loop, coincide exactly with
this expansion, with no free parameter; the first does so modulo the four-dimensional identity of that
sector, which is the cubic Gauss--Bonnet identity (Appendix~\ref{app:basis}). With the sign of the expansion reversed, both equalities fail. The
two functions thus test the coefficients of the three classes of Sec.~\ref{sec:basis} beyond the
kernels from which they were obtained, the $\mathcal{R}\nabla Q$ and $(\nabla Q)^{2}$ coefficients in particular
through the cubic terms described above.

\subsection{The function \texorpdfstring{$\delta_{\bar\phi}\delta^{2}_{\bar G}S_{\rm eff}$}{δφ δ²G Seff} and the invariants \texorpdfstring{$\mathcal{R}QQ$}{RQQ}}
\label{sec:RQQ}

We subtract from the pole of $\delta_{\bar\phi}\delta^{2}_{\bar G}S_{\rm eff}$ the cubic expansion of the
classes fixed at two points. The remainder lies entirely in the $21$-dimensional space of
$\mathcal{R}^{(1)}[\phi]\,QQ$: adding it to that space does not increase the rank. The condition is not
trivial: with the sign of the expansion reversed the remainder does increase the rank. Unlike
$\delta^{3}_{\bar G}S_{\rm eff}$ and $\delta^{4}_{\bar G}S_{\rm eff}$ below, this function is therefore a test
as well as a determination. Choosing
the $21$ directions from the simplest to the most complex, with the scalar curvature first, then the
Ricci tensor, then the Riemann tensor, we obtain the coefficients $r_{j}$ of Table~\ref{tab:RQQ}, in the
sense
\begin{equation}
\sum_{I}c_{I}I\supset\sum_{j=1}^{21}r_{j}\,(\sqrt{-g})^{1-w_{j}}\,\mathcal{O}^{\mathcal{R}QQ}_{j},
\label{eq:RQQsum}
\end{equation}
with the normalization of Eq.~\eqref{eq:Sdiv}.

\begin{table}[t]
\caption{Basis of the $\mathcal{R}QQ$ invariants and their coefficients, in $d=4$ and $\xi=1$. Shorthand, not the
Einstein convention: all indices are written down and each repeated pair is contracted with the
inverse metric, e.g.\ $R\,W_{\lambda}W_{\lambda}=R\,g^{\lambda\rho}W_{\lambda}W_{\rho}=R\,W_{\lambda}W^{\lambda}$;
$\Riem_{\alpha\beta\gamma\delta}=g_{\alpha\epsilon}\Riem^{\epsilon}{}_{\beta\gamma\delta}$,
$\Ric_{\mu\nu}=\Riem^{\rho}{}_{\mu\rho\nu}$, and $Q_{\lambda\mu\nu}=g_{\mu\alpha}g_{\nu\beta}Q_{\lambda}{}^{\alpha\beta}$,
with the derivative index first. The weight $w_{j}$ is the number of factors $Q$ and $V$.}
\label{tab:RQQ}
\renewcommand{\arraystretch}{1.5}
\begin{ruledtabular}
\begin{tabular}{rlc}
$j$ & $\mathcal{O}^{\mathcal{R}QQ}_{j}$ & $r_{j}$\\
\hline
1 & $R\,W_{\lambda}W_{\lambda}$ & $-\frac{77}{240}$\\
2 & $R\,W_{\lambda}V_{\lambda}$ & $\frac{23}{120}$\\
3 & $R\,V_{\lambda}V_{\lambda}$ & $\frac{1}{8}$\\
4 & $R\,Q_{\lambda\mu\nu}Q_{\lambda\mu\nu}$ & $\frac{33}{80}$\\
5 & $R\,Q_{\lambda\mu\nu}Q_{\mu\lambda\nu}$ & $-\frac{33}{40}$\\
6 & $\Ric_{\mu\nu}W_{\mu}W_{\nu}$ & $-\frac{2}{15}$\\
7 & $\Ric_{\mu\nu}W_{\mu}V_{\nu}$ & $\frac{1}{2}$\\
8 & $\Ric_{\mu\nu}V_{\mu}V_{\nu}$ & $\frac{3}{2}$\\
9 & $\Ric_{\mu\nu}W_{\alpha}Q_{\alpha\mu\nu}$ & $\frac{11}{12}$\\
10 & $\Ric_{\mu\nu}V_{\alpha}Q_{\alpha\mu\nu}$ & $-\frac{89}{60}$\\
11 & $\Ric_{\mu\nu}W_{\alpha}Q_{\mu\alpha\nu}$ & $-\frac{11}{12}$\\
12 & $\Ric_{\mu\nu}V_{\alpha}Q_{\mu\alpha\nu}$ & $-\frac{5}{6}$\\
13 & $\Ric_{\mu\nu}Q_{\mu\alpha\beta}Q_{\nu\alpha\beta}$ & $-\frac{11}{20}$\\
14 & $\Ric_{\mu\nu}Q_{\mu\alpha\beta}Q_{\alpha\nu\beta}$ & $\frac{29}{20}$\\
15 & $\Ric_{\mu\nu}Q_{\alpha\mu\beta}Q_{\alpha\nu\beta}$ & $\frac{37}{30}$\\
16 & $\Ric_{\mu\nu}Q_{\alpha\mu\beta}Q_{\beta\nu\alpha}$ & $-\frac{97}{120}$\\
17 & $\Riem_{\alpha\beta\gamma\delta}Q_{\alpha\beta\gamma}W_{\delta}$ & $\frac{29}{120}$\\
18 & $\Riem_{\alpha\beta\gamma\delta}Q_{\alpha\beta\gamma}V_{\delta}$ & $\frac{5}{6}$\\
19 & $\Riem_{\alpha\beta\gamma\delta}Q_{\alpha\beta\epsilon}Q_{\gamma\delta\epsilon}$ & $-\frac{1}{2}$\\
20 & $\Riem_{\alpha\beta\gamma\delta}Q_{\alpha\gamma\epsilon}Q_{\beta\delta\epsilon}$ & $\frac{1}{12}$\\
21 & $\Riem_{\alpha\beta\gamma\delta}Q_{\alpha\gamma\epsilon}Q_{\epsilon\beta\delta}$ & $-\frac{1}{6}$\\
\end{tabular}
\end{ruledtabular}
\end{table}

The $r_{j}$ depend on the conventions of Sec.~\ref{sec:blocks} and on the basis~\eqref{eq:LdQ}; what does
not depend on them is that the remainder fits in the $21$ directions. They include the two transverse
diagrams of $\delta_{\bar\phi}\delta^{2}_{\bar G}S_{\rm eff}$ (Sec.~\ref{sec:threepoint}), which belong to this
space: omitting them would leave the fit consistent but change every $r_{j}$. The table fixes the $\phi GG$
projection of each invariant; the $GGG$ parts follow and enter Sec.~\ref{sec:QQdQ} as known terms.

\subsection{The function \texorpdfstring{$\delta^{3}_{\bar G}S_{\rm eff}$}{δ³G Seff} and the invariants \texorpdfstring{$QQ\nabla Q$}{QQ∇Q}}
\label{sec:QQdQ}

The pole of $\delta^{3}_{\bar G}S_{\rm eff}$ is linear in the external momenta. Its nonvanishing diagrams
are the ghost triangle in its two orientations, the three bubbles formed by the quartic gauge-fixing
vertex $\bar G\bar G\phi\phi$ and a trivertex $\bar G\phi\phi$, with symmetry factor $1/2$, the three
triangles with one internal $G$ line, and the graviton triangle. Triangles with two or more internal
$G$ lines vanish because they are scaleless, and there is no $\bar GGG$ vertex.

This function has no autonomous identity: the identity that constrains it relates it to
$\delta_{\bar\phi}\delta^{3}_{\bar G}S_{\rm eff}$, a four-point function whose pole we computed
(Sec.~\ref{sec:fourpointtest}) but whose identity we have not checked. Nor is the comparison with the
counterterms a test: since $Q^{(1)}$ is an invertible linear function of $G$, the contractions $QQ\nabla Q$
span the whole $58$-dimensional space of local $GGG$ vertices with one derivative. The available tests at
this stage are internal: locality, and Bose symmetry under the permutations of the three $\bar G$ legs,
which must hold separately in each topological family (bubbles, triangles with one $G$ line, graviton
triangle, ghost triangles), since the permutations preserve the topology of a diagram. The pole is
symmetric in each family.

We subtract from the pole the cubic parts, already fixed, of $R^{2}$, $\Ric_{\mu\nu}\Ric^{\mu\nu}$, the four
$\mathcal{R}\nabla Q$, the sixteen $(\nabla Q)^{2}$ and the $21$ $\mathcal{R}QQ$ invariants. The remainder fits
exactly in the $58$ directions of $QQ\nabla Q$, modulo the four $d=4$ identities of the $GGG$
sector, and determines their coefficients $s_{k}$ uniquely; they are listed in
Appendix~\ref{app:tables}. With them the counterterm Lagrangian is determined to cubic order at
$\xi=1$.

\subsection{The function \texorpdfstring{$\delta^{4}_{\bar G}S_{\rm eff}$}{δ⁴G Seff} and the invariants \texorpdfstring{$Q^{4}$}{Q⁴}}
\label{sec:Q4}

Four $\bar G$ legs already exhaust the mass dimension of a counterterm, so the pole of
$\delta^{4}_{\bar G}S_{\rm eff}$ has degree zero in the external momenta and is computed with them set to
zero (Appendix~\ref{app:poles}); all its diagrams are logarithmically divergent. They are built from the
vertices $\bar G\phi\phi$, $\bar GG\phi$, $\bar G\bar cc$ and the quartic gauge-fixing vertex
$\bar G\bar G\phi\phi$. Since $\bar GG\phi$ has one internal auxiliary leg and one graviton leg, and
there is no $\bar GGG$ vertex, the auxiliary lines of a box are never adjacent, and each joins two
vertices $\bar GG\phi$. At $\xi=1$ there remain $42$ diagrams in seven families:
\begin{center}
\begin{tabular}{lccc}
\hline\hline
Family & Vertices & Diagrams & Factor\\
\hline
ghost box & $4\,\bar G\bar cc$ & $6$ & $-1$\\
graviton box & $4\,\bar G\phi\phi$ & $3$ & $1$\\
box, one $G$ line & $2\,\bar G\phi\phi+2\,\bar GG\phi$ & $12$ & $1$\\
box, two $G$ lines & $4\,\bar GG\phi$ & $6$ & $1$\\
graviton triangle & $\bar G\bar G\phi\phi+2\,\bar G\phi\phi$ & $6$ & $1$\\
triangle, one $G$ line & $\bar G\bar G\phi\phi+2\,\bar GG\phi$ & $6$ & $1$\\
bubble & $2\,\bar G\bar G\phi\phi$ & $3$ & $1/2$\\
\hline\hline
\end{tabular}
\end{center}
The factor $-1$ of the ghost box is the Grassmann sign of the closed loop. The ghosts are complex, and
the two orientations of their loop count separately, while in the bosonic loops a reflection identifies
the two orientations. The box with two $G$ lines survives by rule (iv) of Sec.~\ref{sec:counting}.

Of the internal tests of $\delta^{3}_{\bar G}S_{\rm eff}$ only Bose symmetry in each family remains, since
locality is automatic at vanishing external momenta. To it we add an independent recomputation, with
the exact angular average of Appendix~\ref{app:poles}: the four legs are contracted with random integer
tensors and the average over the $24$ vertices of the $24$-cell is computed in exact arithmetic. For each
of the three sets of tensors tested, the recomputation agrees exactly with the symbolic result, family
by family and for the sum. The recomputation tests the sum of the families, which Bose symmetry,
holding in each family separately, cannot do.

At $\phi=0$ and vanishing momentum every invariant of the basis is a homogeneous quartic polynomial in
$G$ without derivatives: the curvature reduces to $\Gamma\Gamma$ and $\nabla Q$ to $\Gamma Q$. All the
classes therefore contribute to $\delta^{4}_{\bar G}S_{\rm eff}$ through their quartic parts, already
fixed, and $Q^{4}$ is the only unknown. As for $QQ\nabla Q$, the
contractions $Q^{4}$ span all the local $GGGG$ vertices without derivatives, so the comparison only
determines the coefficients. Of the $10395$ complete contractions of four tensors $Q_{\lambda}{}^{\mu\nu}$,
symmetric in the upper indices, $69$ are distinct up to these symmetries. In generic $d$ they are
independent; in $d=4$ their rank is $65$, because four combinations vanish identically in that dimension
(Appendix~\ref{app:basis}). After the quartic parts of all the lower invariants are subtracted, the
remainder fits exactly in the $65$ directions, modulo these identities, and determines the coefficients
$u_{k}$ uniquely (Appendix~\ref{app:tables}). Two coefficients, $s_{19}$ and $u_{10}$, vanish; these
zeros depend on the choice of basis.

A dependence on the conventions appears here that the cubic parts of $QQ\nabla Q$ did not see. Since the
connection is not metric, the definition~\eqref{eq:traces} of $W$ gives
\begin{equation}
\nabla_{\alpha}W_{\lambda}-h_{\mu\nu}\nabla_{\alpha}Q_{\lambda}{}^{\mu\nu}=(\nabla_{\alpha}h_{\mu\nu})\,Q_{\lambda}{}^{\mu\nu},
\label{eq:nablaW}
\end{equation}
which is of order $Q^{2}$, since $\nabla_{\alpha}h_{\mu\nu}=-h_{\mu\rho}h_{\nu\sigma}Q_{\alpha}{}^{\rho\sigma}$; in the
invariants $QQ\nabla Q$ this difference is a $Q^{4}$ term. With the convention of Sec.~\ref{sec:blocks}, $\nabla$ acts on $W_{\lambda}$
itself in the eight invariants of the $QQ\nabla Q$ basis that contain $\nabla\cdot W$ or
$\nabla_{\alpha}W_{\beta}$. The $s_{k}$ do not depend on this choice; the $u_{k}$ do, and they hold only together
with it.

\subsection{A four-point prediction}
\label{sec:fourpointtest}

With all the coefficients fixed, the poles of the remaining functions are predictions. We computed one
of them, $\delta_{\bar\phi}\delta^{3}_{\bar G}S_{\rm eff}$, at $\xi=1$. Its $54$ diagrams are built from the
vertices of Table~\ref{tab:vertices} and from the quintic gauge-fixing vertex $\bar\phi\bar G\bar G\phi\phi$,
which follows from the same expansion of Eq.~\eqref{eq:gf}; Bose symmetry under the exchange of the
$\bar G$ legs holds exactly. All the classes contribute to the prediction except $Q^{4}$, whose expansion
starts with four $G$ legs; in particular, the $\phi GGG$ parts of the invariants $\mathcal{R}QQ$ and
$QQ\nabla Q$ enter with the coefficients $r_{j}$ and $s_{k}$ of Tables~\ref{tab:RQQ} and~\ref{tab:QQdQ}. The
pole coincides with the expansion of Eq.~\eqref{eq:Lctfull}, with no free parameter, modulo the
four-dimensional identities of the $\phi GGG$ sector: with rank-one polarizations the difference is a
combination of the $d=4$ Gram identities of that sector, of which $53$ are independent. The agreement
provides an independent check of the $\delta^{3}_{\bar G}S_{\rm eff}$ pole through its parameter-free
prediction for the $\phi GGG$ pole. In the chosen basis, this prediction involves both the $s_{k}$,
previously checked only internally in Sec.~\ref{sec:QQdQ}, and the $r_{j}$. The agreement also shows
that the pole of a quartic function with a $\bar\phi$ leg fits in the basis.

\subsection{Completeness of the basis}
\label{sec:completeness}

The basis of invariants of dimension four and even parity was checked sector by sector. Each class of Eq.~\eqref{eq:classes} first appears in a sector of its own, so it suffices to compare,
sector by sector, all the complete contractions of each class with the basis used here. There are $16$,
$27$, $25$, $75$, $95$ and $69$ distinct contractions in the classes $\mathcal{R}^{2}$, $\mathcal{R}\nabla Q$,
$(\nabla Q)^{2}$, $\mathcal{R}QQ$, $QQ\nabla Q$ and $Q^{4}$; $3$ of those of $\mathcal{R}\nabla Q$ and $14$ of those
of $\mathcal{R}QQ$ vanish identically, because the antisymmetric pair of the curvature is contracted with a
symmetric pair of $Q$, which leaves $24$ and $61$. In the quadratic sectors the $65$ nonvanishing
contractions of the first three classes have rank $22$, the rank of the basis. Up to cubic order the $221$
nonvanishing contractions of the first five classes, expanded with all their metric and density factors, have rank $101=22+79$
in the joint matrix of the seven sectors $\phi\phi$, $\phi G$, $GG$, $\phi\phi\phi$, $\phi\phi G$, $\phi GG$ and $GGG$,
again that of the basis, with the rank of each sector also equal to that of the basis. In the $GGGG$
sector all the contractions, including those of $Q^{4}$, have rank $65$, which is that of the basis. The
three ranks are exact; the method is described in Appendix~\ref{app:basis}. In each of the sectors
computed, the basis therefore spans the same space as all the contractions of the classes that reach it.

This extends to all orders and to every invariant functional, whatever its form (Appendix~\ref{app:basis}).
The lowest term of a background-invariant functional is invariant under the inhomogeneous part of the
background transformations, Eq.~\eqref{eq:deltaphibar}, which acts on $\bar\phi$ alone. Completeness therefore
reduces to showing that, at each order in the fields, the vertices invariant under that transformation are
spanned by the lowest terms of combinations of the basis. The Euler--Lagrange derivative of such a vertex with
respect to $\bar\phi$ is a polynomial in the linearized curvature, in $\bar G$ and in their derivatives, and it
satisfies the Noether identity of the transformation. Dimensional counting and this identity, together with
exact counts in the sectors $\phi G$, $\phi\phi\phi$, $\phi\phi G$ and $\phi GG$, leave $2$, $4$ and $16$ invariant
vertices in the quadratic sectors; $0$, $0$, $21$ and $58$ in $\phi\phi\phi$, $\phi\phi G$, $\phi GG$ and $GGG$; $65$ in
$GGGG$, none in the other quartic sectors, and none beyond. These are exactly the $22$, $79$ and $65$
directions that the basis supplies at each order. The basis is therefore complete: in $d=4$, every background-invariant local
functional of dimension four and even parity is, up to a boundary term, a combination of the $166$
invariants.

The $166$ coefficients of the six classes are thus all fixed; the Lagrangian they define is collected in
Sec.~\ref{sec:lagrangian}.
\section{The one-loop counterterm Lagrangian and its interpretation}
\label{sec:lagrangian}

\subsection{The counterterm Lagrangian}
\label{sec:Lct}

Collecting the results of Secs.~\ref{sec:basis} and~\ref{sec:vertices}, with $S_{\rm ct}=-S^{\rm div}_{\rm eff}$
and the normalization of Eq.~\eqref{eq:Sdiv}, the one-loop counterterm Lagrangian of first-order gravity
in the background field method, with an independent background connection, is
\begin{multline}
\mathcal{L}_{\rm ct}=-\frac{1}{8\pi^{2}(4-d)}\Big[a_{R}\sqrt{-g}R^{2}+a_{\Ric}\sqrt{-g}\Ric_{\mu\nu}\Ric^{\mu\nu}\\
+b_{V}R\,\nabla_{\mu}V^{\mu}+b_{W}\sqrt{-g}R\,\nabla^{\lambda}W_{\lambda}+b_{Y}Y+b_{Z}Z
+\mathcal{L}_{(\nabla Q)^{2}}\\
+\sum_{j=1}^{21}r_{j}(\sqrt{-g})^{1-w_{j}}\mathcal{O}^{\mathcal{R}QQ}_{j}
+\sum_{k=1}^{58}s_{k}(\sqrt{-g})^{1-w_{k}}\mathcal{O}^{QQ\nabla Q}_{k}\\
+\sum_{k=1}^{65}u_{k}(\sqrt{-g})^{1-w_{k}}\mathcal{O}^{Q^{4}}_{k}\Big],
\label{eq:Lctfull}
\end{multline}
with $166$ coefficients distributed as in Table~\ref{tab:classes}. By the triangular structure of
Sec.~\ref{sec:blocks}, the coefficients fixed in one sector enter the next one as known terms. The
coefficients of the first three classes are known in arbitrary gauge; those of the last three were
determined at $\xi=1$. The divergent part of the one-loop background effective action is thus
determined at $\xi=1$ in a complete basis (Sec.~\ref{sec:completeness}). Since the reduction of the basis uses identities
that hold only in $d=4$, Eq.~\eqref{eq:delta5}, the result fixes the four-dimensional residue modulo these
identities; it does not fix the coefficients of evanescent operators, which vanish in $d=4$ but can
contribute through subdivergences at higher loops~\cite{Bern2015}.

\begin{table}[t]
\caption{The six classes of invariants in Eq.~\eqref{eq:Lctfull}: number of representatives retained in the combined basis, the
function that fixes their coefficients, the gauge in which these are known, and where they are given.}
\label{tab:classes}
\begin{ruledtabular}
\begin{tabular}{lcccc}
Class & Invariants & Fixed by & Gauge & Values\\
\hline
$\mathcal{R}^{2}$ & $2$ & $\delta^{2}_{\bar\phi}S_{\rm eff}$ & any $\xi$ & Table~\ref{tab:coef2pt}\\
$\mathcal{R}\nabla Q$ & $4$ & $\delta_{\bar\phi}\delta_{\bar G}S_{\rm eff}$ & any $\xi$ & Table~\ref{tab:coef2pt}\\
$(\nabla Q)^{2}$ & $16$ & $\delta^{2}_{\bar G}S_{\rm eff}$ & any $\xi$ & Table~\ref{tab:dQ2}\\
$\mathcal{R}QQ$ & $21$ & $\delta_{\bar\phi}\delta^{2}_{\bar G}S_{\rm eff}$ & $\xi=1$ & Table~\ref{tab:RQQ}\\
$QQ\nabla Q$ & $58$ & $\delta^{3}_{\bar G}S_{\rm eff}$ & $\xi=1$ & Table~\ref{tab:QQdQ}\\
$Q^{4}$ & $65$ & $\delta^{4}_{\bar G}S_{\rm eff}$ & $\xi=1$ & Table~\ref{tab:Q4}\\
\end{tabular}
\end{ruledtabular}
\end{table}

An invariant of dimension four has at most four factors $Q$, so no unknowns remain. Completeness then
fixes, without free parameters, the poles of all four-point functions with $\bar\phi$ legs and all
functions with five or more points through Eq.~\eqref{eq:Lctfull}. Of these, only
$\delta_{\bar\phi}\delta^{3}_{\bar G}S_{\rm eff}$ was computed, confirming the prediction
(Sec.~\ref{sec:fourpointtest}).

\subsection{On-shell content and field redefinitions}
\label{sec:onshell}

On shell the counterterms vanish. The equation of motion~\eqref{eq:eomGamma} of the connection gives
$Q=0$, so every class with $Q$ vanishes, and the curvature of $\Gamma$ becomes that of the Levi-Civita
connection; $R^{2}$ and $\Ric_{\mu\nu}\Ric^{\mu\nu}$ then vanish by the vacuum Einstein equations. This recovers,
in the first-order formulation, the one-loop on-shell ultraviolet finiteness of pure gravity found by 't~Hooft and
Veltman~\cite{tHooft1974}.

Off shell the counterterms can be removed, perturbatively and at one-loop order, by redefinitions of the
fields, in two steps. The variation of the action~\eqref{eq:action} with respect to $\Gamma$ is a linear
function of $Q$, and Eq.~\eqref{eq:eomGamma} shows that it vanishes only for $Q=0$ when $d\neq1$; the map is
therefore invertible, and $Q$ is a local linear function of $\delta S/\delta\Gamma$. Terms with derivatives of
$Q$ are first integrated by parts, which for a symmetric connection and densities of weight one costs only
boundary terms, so that every term of Eq.~\eqref{eq:Lctfull} with at least one factor $Q$ takes the form
$(\delta S/\delta\Gamma)\cdot\mathcal{X}$ with $\mathcal{X}$ local. This includes the parts with $Q$ of the curvature invariants,
since $\Gamma=\mathring{\Gamma}(h)+L$ with the distortion $L$ linear in $Q$, and
$\Riem(\Gamma)=\mathring{\Riem}+\mathring{\nabla}L+LL$. A redefinition $\Gamma\to\Gamma+\delta\Gamma$, with $\delta\Gamma$ local, of one-loop order and
proportional to $\mathcal{X}$, removes all these terms. What remains are the metric invariants $\mathring{R}^{2}$ and
$\mathring{\Ric}_{\mu\nu}\mathring{\Ric}^{\mu\nu}$, proportional to the metric equations of motion. They are removed by a
redefinition of $h$, as in the metric formulation~\cite{tHooft1974}; this step refers to the reduced metric
action $S[h,\mathring{\Gamma}(h)]$. In the first-order action the same redefinition of $h$ produces, besides
the terms it is designed to cancel, terms proportional to the equation of motion of the connection, which
are absorbed by a further one-loop adjustment of $\Gamma$. That the counterterms with the
independent connection can be absorbed by a reparametrization of the connection was observed by Buchbinder
and Shapiro~\cite{Buchbinder1984,Buchbinder1985}.

The values of the off-shell coefficients depend on the gauge and on the parametrization. At two points,
where they are known for arbitrary $\xi$, they depend on $\xi$, with isolated zeros such as those of
$C_{X}$ at $\xi=(7\pm\sqrt{33})/2$; the covariant form of the counterterms does not fix them by itself. They are
the counterterms of the background effective action in this formulation and gauge, and it is in this
setting that they renormalize the off-shell functions and predict the poles of the functions not computed
here.

This is the situation described by the theorem of DeWitt and Kallosh~\cite{DeWitt1967,Kallosh1974}: the
one-loop effective action, and with it the counterterms, is independent of the gauge-fixing parameters and of
the parametrization of the quantum fields on shell, that is, once the classical field equations are imposed.
For the divergent part, the vanishing of all the counterterms on shell in pure gravity satisfies the
theorem in the simplest way; it says nothing by itself about the finite part of the effective action;
at two points, where the dependence on $\xi$ is known, this holds for every $\xi$, while at three and four
points our coefficients, determined at $\xi=1$, do not test it. When matter is present, the equations to be
imposed are those of the coupled system, including the gravitational equation with the matter source.
For the one-loop divergences in this case, Frenkel and Martins-Filho~\cite{Frenkel2026} verified explicitly,
in the metric formulation and in a general background gauge, that the dependence on the gauge parameter
cancels on shell. Gauge independence does not imply finiteness: with
matter, on-shell divergences independent of the gauge parameters survive~\cite{tHooft1974}.

If the background connection is taken to be the Levi-Civita connection of the background metric,
as in Ref.~\cite{Alvarez2017}, all the classes with $Q$ disappear from the start and only the curvature
part survives. It is the independent background connection that makes the classes $\mathcal{R}\nabla Q$,
$(\nabla Q)^{2}$, $\mathcal{R}QQ$, $QQ\nabla Q$ and $Q^{4}$ visible.

\subsection{Comparison with Buchbinder and Shapiro}
\label{sec:BS}

Buchbinder and Shapiro~\cite{Buchbinder1984,Buchbinder1985} computed the one-loop divergences of
first-order gravity in the background field method with an independent background connection, off shell,
in $d=4$ and in a single gauge, with a cutoff $L$, the quantum field defined from the inverse metric,
$g^{\mu\nu}\to g^{\mu\nu}+\kappa\tilde h^{\mu\nu}$, the distortion $\mathcal{A}=\bar\Gamma-\Gamma(\bar g)$, and the
Levi-Civita covariant derivative of the background metric in the gauge fixing and in the ghost sector. They gave the structure
of the result, with the classes $R^{2}$, $R\nabla\mathcal{A}$, $(\nabla\mathcal{A})^{2}$, $R\mathcal{A}\mathcal{A}$,
$\mathcal{A}\mathcal{A}\nabla\mathcal{A}$ and $\mathcal{A}^{4}$, showed that
it vanishes on shell and that renormalization amounts to a nonlinear reparametrization of the metric and of
the connection, and gave explicitly the coefficients of part of the two-point terms. The coefficients of
the higher orders in $\mathcal{A}$, denoted $F_{1},\ldots,F_{5}$ there, were not written, and the completeness
of the basis was not addressed. Equation~\eqref{eq:Lctfull} gives the corresponding coefficients in the
present gauge and parametrization, with the two-point part in arbitrary $d$ and $\xi$, in a formulation in
which each class is checked by the tests of Secs.~\ref{sec:ward}--\ref{sec:vertices}. Being off-shell
quantities, they are not the coefficients $F_{1},\ldots,F_{5}$ of Ref.~\cite{Buchbinder1985}, which refer to a
different gauge condition and quantum variable.

In the sector $\mathcal{A}=0$, Eq.~(11) of Ref.~\cite{Buchbinder1985} gives the logarithmic part of the
counterterms as
\begin{equation}
-\frac{\ln(L^{2}/\mu^{2})}{32\pi^{2}}\int d^{4}x\sqrt{-g}\,\Big(\frac{53}{3}\Ric_{\mu\nu}\Ric^{\mu\nu}+\frac{107}{60}R^{2}\Big).
\label{eq:BS11}
\end{equation}
This differs from the corresponding terms of 't~Hooft and Veltman~\cite{tHooft1974} and from ours,
Table~\ref{tab:coef2pt}. These terms vanish on shell, where the results are therefore compatible; attributing the
differences to the gauge and the parametrization would require translating the variables, gauges and
normalizations of Ref.~\cite{Buchbinder1985} into ours, which we do not attempt here.

\subsection{Relation to the metric formulation}
\label{sec:metriccomparison}

Section~\ref{sec:phiphi} compared the graviton self-energy with the first-order calculation of
Ref.~\cite{Brandt2026} and with the metric calculation of Ref.~\cite{Shapiro2018}. A further comparison is
with Ref.~\cite{Brandt2022}, which computes the background self-energy of metric gravity with the
covariant fluctuation $g_{\mu\nu}=\bar g_{\mu\nu}+\kappa h_{\mu\nu}$, $\bar g_{\mu\nu}=\eta_{\mu\nu}+\kappa\bar h_{\mu\nu}$, in a
background harmonic gauge with parameter $\xi$. The comparison is made in the sector $\bar G=0$, where the
background connection reduces, at the order needed, to the Levi-Civita connection of $\bar h$. The background
variables are related by $\bar\phi^{\mu\nu}=\mathcal{J}^{\mu\nu,\alpha\beta}\bar h_{\alpha\beta}+O(\kappa)$, with
\begin{equation}
\mathcal{J}^{\mu\nu,\alpha\beta}\equiv-\tfrac{1}{2}(\eta^{\mu\alpha}\eta^{\nu\beta}+\eta^{\mu\beta}\eta^{\nu\alpha})+\tfrac{1}{2}\eta^{\mu\nu}\eta^{\alpha\beta},
\label{eq:Jmap}
\end{equation}
and, since the one-point function vanishes in dimensional regularization around flat space, the two-point
functions of our effective action in the two background variables are related by
$\Pi^{\rm ours}_{\bar h}=\mathcal{J}^{\rm T}\Pi^{\rm ours}_{\bar\phi}\mathcal{J}$. Under this map our ghost loop
coincides exactly with that of Ref.~\cite{Brandt2022}, for every $d$. This is expected: the two gauge
conditions express the same harmonic condition, and for these two conditions the one-loop
Faddeev--Popov operator, obtained at vanishing quantum fields, depends only on the background and not on
the choice of quantum variable. The difference is
entirely in the bosonic loop, where the integration variables, the Goldberg fluctuation here and the
covariant one there, are related nonlinearly and in a background-dependent way, and where the
gauge-fixing terms have different metric weights. The resulting local difference of the poles is a
combination of $R^{2}$ and $\Ric_{\mu\nu}\Ric^{\mu\nu}$, which vanish on shell, and can be absorbed by a
background redefinition $\bar g_{\mu\nu}\to\bar g_{\mu\nu}+\Delta_{\mu\nu}$. Such a redefinition always exists for this
difference, so the comparison would be an independent test only if $\Delta_{\mu\nu}$ were obtained beforehand by
another procedure.
\section{Conclusions}
\label{sec:conclusions}

We have determined the one-loop divergences of pure first-order gravity in the background field method,
off shell and with an independent background connection. The three background two-point functions,
computed in arbitrary dimension and gauge, require three classes of invariants, quadratic in the curvature,
bilinear in the curvature and the derivative of the nonmetricity, and quadratic in the latter; none can be
dispensed with, and the Gauss--Bonnet combination does not suffice. At three and four points, in the
Feynman--de Donder gauge, two functions with background graviton legs are parameter-free predictions of
the two-point coefficients and are confirmed, as is a four-point function with one background graviton
leg, and the remaining functions fix the classes $\mathcal{R}QQ$,
$QQ\nabla Q$ and $Q^{4}$. The counterterm Lagrangian has $166$ coefficients fixed in the basis adopted here, which is complete as a
basis of covariant functionals (Appendix~\ref{app:basis}).
The counterterms vanish on shell and can be removed off shell by redefinitions of the connection and the
metric. The calculation rests on background Ward--Takahashi identities that take into account the
noncovariant term $\Xi$ of the shifted background connection, verified with the full amplitudes.

The results have limitations. Above two points the coefficients are known only at $\xi=1$. The full
three-point amplitudes were used to verify the identities, while the decomposition in invariants used
their poles; at four points only the poles of $\delta^{4}_{\bar G}S_{\rm eff}$ and $\delta_{\bar\phi}\delta^{3}_{\bar G}S_{\rm eff}$
were computed. The classes $QQ\nabla Q$ and $Q^{4}$ are fixed by functions with no autonomous identity and
span all the local vertices of their sectors; $QQ\nabla Q$ is also tested by the four-point function
$\delta_{\bar\phi}\delta^{3}_{\bar G}S_{\rm eff}$, while $Q^{4}$ is checked only internally, by Bose symmetry
and an independent recomputation. The values of the coefficients beyond two points depend on the conventions and
on the choice of basis described in Secs.~\ref{sec:basis} and~\ref{sec:vertices}.

Several extensions follow. With the pole of $\delta_{\bar\phi}\delta^{3}_{\bar G}S_{\rm eff}$ known, the identity that relates it to
$\delta^{3}_{\bar G}S_{\rm eff}$ would provide a further consistency test, and the
other four-point functions with $\bar\phi$ legs are parameter-free predictions of Eq.~\eqref{eq:Lctfull}. Extending the calculation to gravity coupled to matter would provide a nontrivial
test of the DeWitt--Kallosh theorem in the first-order formulation. The gauge-parameter dependence should
cancel upon imposing the coupled field equations, as verified at one loop in the metric formulation by
Frenkel and Martins-Filho~\cite{Frenkel2026}; gauge independence would not imply finiteness, since with
matter on-shell divergences survive~\cite{tHooft1974}. Further steps are the gauge dependence above two
points, the explicit field redefinitions that remove the counterterms, which would allow a comparison with
Refs.~\cite{Buchbinder1984,Buchbinder1985}, and the extension to two loops, which would also require the evanescent operators left undetermined
here, and to metric-affine theories with
a dynamical connection.
 
\begin{acknowledgments}
F.T.B. thanks J. Frenkel, S. Martins-Filho, D. G. C. McKeon and A. C. B. Carvalho for discussions.
F.T.B. acknowledges partial support from CNPq (grant No.~309411/2022-0). The work of G.A.M. was
financed in part by the Coordena\c{c}\~ao de Aperfei\c{c}oamento de Pessoal de N\'{\i}vel Superior,
Brasil (CAPES), Finance Code 001.
\end{acknowledgments}

\section*{Data availability}
The data that support the findings of this article are contained within the article and its
appendices. The computer-algebra code used to obtain them can be made available upon request.

\appendix
\section{The shift by \texorpdfstring{$D$}{D} and the background effective action}
\label{app:shift}

\paragraph{Why the linearized inverse.} The quadratic part of Eq.~\eqref{eq:actionG} in
$\mathcal{G}$ involves $M(h)$, and the square can be completed exactly by shifting
$\mathcal{G}\to\mathcal{G}+M^{-1}(h)\star\partial h/\kappa$. The diagonalization would then hold to
all orders, and the shifted field would be the difference between $\mathcal{G}$ and its Levi-Civita value, a
tensor of weight zero related to the nonmetricity by an algebraic, $h$-dependent map (Appendix~\ref{app:xi}). The remainder, however, is $-\frac{1}{2\kappa^{2}}\,\partial h\star M^{-1}(h)\star\partial h$,
and $M^{-1}(h)$ is rational in $h$: substituting the equation of motion of $\mathcal{G}$ into the
action is the same as eliminating the connection, and the exact shift returns the second-order
action with its infinite tower of graviton vertices, the decoupled auxiliary sector contributing
only $[\det M(h)]^{-1/2}$, ultralocal and equal to one in dimensional regularization. The shift by
$D=M^{-1}(\eta)$, Eq.~\eqref{eq:shiftflat}, keeps the action cubic. The price is a variable that is
not covariant: the variation of the shifted background field depends on the background graviton
(Sec.~\ref{sec:ward}), the mixing removed at tree level reappears at one loop, and the counterterm
invariants must be built with the connection $\bar\Gamma$ of Eq.~\eqref{eq:ctparam} and only then
expressed in terms of $\bar\phi$ and $\bar G$.

\paragraph{Equivalence of the effective actions.} Consider the background field method in the
unshifted variable, with $h=\bar h+\kappa\phi$ and
$\mathcal{G}=\bar{\mathcal{G}}+\mathcal{G}_{q}$, and denote its background effective action by
$S^{\rm orig}_{\rm eff}[\bar\phi,\bar{\mathcal{G}}]$. The variables of Sec.~\ref{sec:split} are
\begin{equation}
\bar{\mathcal{G}}\equiv\bar G+D\star\partial\bar\phi,
\qquad
\mathcal{G}_{q}\equiv G+D\star\partial\phi .
\label{eq:shiftsplit}
\end{equation}
Four facts make the two effective actions agree. (i) In the quantum sector the change
$\mathcal{G}_{q}\to G$ is linear in the quantum fields, with constant coefficients, and its
Jacobian is one. (ii) The effective action is taken at vanishing quantum expectation values, and a
homogeneous linear redefinition of the quantum fields preserves this condition, so the Legendre
transform follows it. (iii) The gauge-fixing term~\eqref{eq:gf} and the ghost
action~\eqref{eq:ghost} depend only on $\phi$ and on the background fields $\bar h$ and
$\bar\Gamma$, and $\bar\Gamma$ is the same function of the background in both variables; neither is
touched by the quantum shift. (iv) In the background sector the first of
Eqs.~\eqref{eq:shiftsplit} is a change of coordinates. Hence
\begin{equation}
S_{\rm eff}[\bar\phi,\bar G]=S^{\rm orig}_{\rm eff}\big[\bar\phi,\bar G+D\star\partial\bar\phi\big].
\label{eq:Seffrelation}
\end{equation}
The argument does not depend on the loop order. It relies on the linearity of the shift in the
quantum fields: a shift nonlinear in them, such as the one by $M^{-1}(h)$ with the full $h$, would
guarantee the equality of the effective actions only on shell.

\paragraph{Relation between the $n$-point functions.} Write $\Lambda=D\star\partial$ for the linear map
$\bar\phi\mapsto D\star\partial\bar\phi$ of Eq.~\eqref{eq:shiftsplit}, $\Lambda^{\rm T}$ for its functional
transpose, defined by $\int d^{d}x\,f\star(\Lambda\bar\phi)=\int d^{d}x\,(\Lambda^{\rm T}f)\,\bar\phi$ and hence
$\Lambda^{\rm T}=-\partial\star D$ after an integration by parts, and $\Pi_{ab}$ for the
two-point kernels of $S^{\rm orig}_{\rm eff}$, with $a,b\in\{\bar\phi,\bar{\mathcal{G}}\}$. Expanding
Eq.~\eqref{eq:Seffrelation} to second order in the background fields,
\begin{align}
\delta^{2}_{\bar G}S_{\rm eff}&=\Pi_{\bar{\mathcal{G}}\bar{\mathcal{G}}},\qquad
\delta_{\bar\phi}\delta_{\bar G}S_{\rm eff}=\Pi_{\bar\phi\bar{\mathcal{G}}}+\Lambda^{\rm T}\Pi_{\bar{\mathcal{G}}\bar{\mathcal{G}}},
\nonumber\\
\delta^{2}_{\bar\phi}S_{\rm eff}&=\Pi_{\bar\phi\bar\phi}+\Pi_{\bar\phi\bar{\mathcal{G}}}\Lambda
+\Lambda^{\rm T}\Pi_{\bar{\mathcal{G}}\bar\phi}+\Lambda^{\rm T}\Pi_{\bar{\mathcal{G}}\bar{\mathcal{G}}}\Lambda,
\label{eq:twopointmap}
\end{align}
where in momentum space $\Lambda$ becomes $D\star ip$, with $p$ the momentum entering the leg on which it
acts. At linear order the on-shell background connection,
$\bar{\mathcal{G}}=M^{-1}(\bar h)\star\partial\bar h/\kappa$, is $D\star\partial\bar\phi$, that is,
$\bar G=0$. The function $\delta^{2}_{\bar\phi}S_{\rm eff}$ is therefore the background graviton
self-energy with the background connection on shell, the quantity computed in
Ref.~\cite{Brandt2026} in the unshifted variable, and Eqs.~\eqref{eq:twopointmap} relate the
remaining two sectors to the off-shell self-energies of that reference.
\section{Transformation law of the shifted background connection}
\label{app:xi}

\paragraph{Derivation.} The unshifted variable transforms as
$\delta\bar{\mathcal{G}}=\mathcal{L}_{\omega}\bar{\mathcal{G}}+\delta\bar{\mathcal{G}}|_{\rm inh}$, with
$\mathcal{L}_{\omega}$ the Lie derivative of a tensor of type $(1,2)$ and the field-independent
inhomogeneous term of Sec.~\ref{sec:bgtransf}. The tensor $D$ is built from $\eta$ and Kronecker
deltas and is not varied, so $\delta(D\star\partial\bar\phi)=D\star\partial(\delta\bar\phi)$. Writing
$\bar G=\bar{\mathcal{G}}-D\star\partial\bar\phi$ and using Eq.~\eqref{eq:deltaphibar},
\begin{multline}
\delta\bar G=\mathcal{L}_{\omega}\bar G
+\big[\mathcal{L}_{\omega}(D\star\partial\bar\phi)-D\star\partial(\mathcal{L}_{\omega}\bar\phi)\big]\\
+\big[\delta\bar{\mathcal{G}}|_{\rm inh}-D\star\partial(\delta\bar\phi|_{\rm inh})\big].
\label{eq:deltaGderiv}
\end{multline}
The last bracket vanishes: $\delta\bar\phi|_{\rm inh}=\kappa^{-1}\mathcal{L}_{\omega}\eta$ is a pure-gauge
perturbation of flat space, and $D\star\partial$ of it is the change of the linearized Levi-Civita
connection variable, which is exactly the inhomogeneous term. The middle bracket is the residue
$\Xi$ of Eq.~\eqref{eq:deltaGbar}. Explicitly, with $\Theta\equiv D\star\partial\bar\phi$,
\begin{multline}
\Xi^{\lambda}{}_{\mu\nu}=-\Theta^{\alpha}{}_{\mu\nu}\partial_{\alpha}\omega^{\lambda}
+\Theta^{\lambda}{}_{\alpha\nu}\partial_{\mu}\omega^{\alpha}+\Theta^{\lambda}{}_{\mu\alpha}\partial_{\nu}\omega^{\alpha}\\
-D^{\lambda}{}_{\mu\nu}{}^{\pi}{}_{\rho\sigma}\Big[(\partial_{\pi}\omega^{\alpha})(\partial_{\alpha}\bar\phi^{\rho\sigma})
+(\partial_{\pi}\bar\phi^{\rho\sigma})(\partial_{\alpha}\omega^{\alpha})\\
-(\partial_{\pi}\bar\phi^{\alpha\sigma})(\partial_{\alpha}\omega^{\rho})
-(\partial_{\pi}\bar\phi^{\rho\alpha})(\partial_{\alpha}\omega^{\sigma})\\
+\bar\phi^{\rho\sigma}\partial_{\pi}\partial_{\alpha}\omega^{\alpha}
-\bar\phi^{\alpha\sigma}\partial_{\pi}\partial_{\alpha}\omega^{\rho}
-\bar\phi^{\rho\alpha}\partial_{\pi}\partial_{\alpha}\omega^{\sigma}\Big].
\label{eq:Xi}
\end{multline}
The residue is exactly linear in $\bar\phi$, because $\delta\bar\phi$ is, and independent of $\bar G$.

\paragraph{When the residue vanishes.} For the identities only the form of $\Xi$ matters; the
following characterization of the parameters for which it vanishes is a side remark. The terms of Eq.~\eqref{eq:Xi} with first derivatives of
$\omega$ combine into the variation of $D$ itself. Since $D=M^{-1}(\eta)$ and $M$ is linear in its
argument, a density of weight one,
\begin{gather}
\Xi=-D\star M(\ell_{\omega})\star D\star\partial\bar\phi+D\star C_{\omega}[\bar\phi],
\nonumber\\
\ell_{\omega}^{\mu\nu}\equiv\mathcal{L}_{\omega}\eta^{\mu\nu}=\eta^{\mu\nu}\partial_{\lambda}\omega^{\lambda}
-\partial^{\mu}\omega^{\nu}-\partial^{\nu}\omega^{\mu},
\label{eq:Xiclosed}
\end{gather}
where
\begin{equation}
\big(C_{\omega}[\bar\phi]\big)_{\pi}{}^{\rho\sigma}\equiv-\bar\phi^{\rho\sigma}\partial_{\pi}\partial_{\alpha}\omega^{\alpha}
+\bar\phi^{\alpha\sigma}\partial_{\pi}\partial_{\alpha}\omega^{\rho}
+\bar\phi^{\rho\alpha}\partial_{\pi}\partial_{\alpha}\omega^{\sigma}
\label{eq:Comega}
\end{equation}
is minus the bracket of the last line of Eq.~\eqref{eq:Xi}, linear in $\bar\phi$ and in
$\partial\partial\omega$, and $\star$ contracts it with the last index triple of $D$. We checked
Eq.~\eqref{eq:Xiclosed} against Eq.~\eqref{eq:Xi} component by component, in exact arithmetic. If $\omega$ is a Killing vector of
$\eta$, then $\ell_{\omega}=0$ and $\partial\partial\omega=0$, so $\Xi=0$ for every $\bar\phi$; this
holds for every $d\neq1,2$, where $D$ is defined. Conversely, suppose $\Xi=0$ for every $\bar\phi$. At a point, $\bar\phi$ and
$\partial\bar\phi$ can be chosen independently, so both terms of Eq.~\eqref{eq:Xiclosed} vanish
separately. Since $D$ is invertible, $M(\ell_{\omega})=0$, which forces $\ell_{\omega}=0$; and
$C_{\omega}[\bar\phi]=0$ for every $\bar\phi$, which forces $\partial\partial\omega=0$. Both linear maps are injective for every
$d\neq1,2$ (the second argument needs two distinct index values, and $D$ must exist). For the first, contracting Eq.~\eqref{eq:M} gives
$M^{\mu\nu}{}_{\nu}{}^{\rho\sigma}{}_{\sigma}(\ell)=2\ell^{\mu\rho}/(d-1)$. For the second, take
$\bar\phi^{\rho\sigma}=u^{\rho}u^{\sigma}$ and $v_{\pi}{}^{\rho}=u^{\alpha}\partial_{\pi}\partial_{\alpha}\omega^{\rho}$;
contracting $C_{\omega}=0$ with any $w_{\rho}$ orthogonal to $u$ gives $v_{\pi}{}^{\rho}=c_{\pi}u^{\rho}$.
Since this holds for every $u$, the linear map $u\mapsto v_{\pi}$ is a multiple of the identity,
$\partial_{\pi}\partial_{\alpha}\omega^{\rho}=c_{\pi}\delta^{\rho}_{\alpha}$, and the symmetry in $\pi\alpha$, with
$\pi\neq\alpha=\rho$, gives $c_{\pi}=0$. (We also confirmed both statements by computing the ranks of
the two maps in $d=4$.) Hence $\omega$ is a Killing vector: $\Xi$ vanishes as an operator on
$\bar\phi$ exactly for translations and Lorentz transformations. For a particular configuration,
$\bar\phi=0$ for instance, it may of course vanish for other $\omega$ as well.

\paragraph{Origin of the residue.} Write
\begin{equation}
\bar G=\big[\bar{\mathcal{G}}-M^{-1}(\bar h)\star\partial\bar\phi\big]
+\big[M^{-1}(\bar h)-D\big]\star\partial\bar\phi .
\label{eq:Gbardecomp}
\end{equation}
The first bracket is the difference between $\bar{\mathcal{G}}$ and its on-shell value: an
algebraic, invertible function of the background nonmetricity, with coefficients that depend on
$\bar h$, and hence a tensor. The second bracket is not covariant. It starts at order
$\bar\phi\,\partial\bar\phi$, but its variation starts at linear order, because $\delta\bar\phi$ has an
inhomogeneous part: $[M^{-1}(\bar h)-D]=O(\kappa\bar\phi)$ multiplies
$\partial(\delta\bar\phi|_{\rm inh})=O(\partial\partial\omega/\kappa)$, and the variation of
$M^{-1}(\bar h)$, of order $\partial\omega$, multiplies $\partial\bar\phi$. Since the first bracket
varies covariantly, $\Xi$ is exactly the noncovariant part of the variation of the second. At linear
order $\bar G$ coincides, up to normalization, with the linearized nonmetricity, in agreement with
Sec.~\ref{sec:split}.

\paragraph{Momentum space.} For the plane-wave parameter $\omega^{\lambda}e^{ik\cdot x}$, the
coefficients of $\omega^{\varepsilon}$ in the homogeneous variations of a leg of incoming momentum
$p$ are
\begin{align}
\mathcal{L}_{\varepsilon}\bar\phi^{\mu\nu}&=i\left[(p_{\varepsilon}+k_{\varepsilon})\bar\phi^{\mu\nu}
-k_{\alpha}\bar\phi^{\alpha\nu}\delta^{\mu}_{\varepsilon}-k_{\alpha}\bar\phi^{\mu\alpha}\delta^{\nu}_{\varepsilon}\right],
\nonumber\\
\mathcal{L}_{\varepsilon}\bar G^{\lambda}{}_{\mu\nu}&=i\left[p_{\varepsilon}\bar G^{\lambda}{}_{\mu\nu}
-k_{\rho}\bar G^{\rho}{}_{\mu\nu}\delta^{\lambda}_{\varepsilon}
+k_{\mu}\bar G^{\lambda}{}_{\varepsilon\nu}+k_{\nu}\bar G^{\lambda}{}_{\mu\varepsilon}\right],
\label{eq:Lmom}
\end{align}
where the transport term carries the momentum of the leg and the index terms that of the parameter.
The kernel $\delta\Xi_{\varepsilon}/\delta\bar\phi$ follows from Eq.~\eqref{eq:Xi} with
$\partial\to ik$ on $\omega$ and $\partial\to ip$ on $\bar\phi$; since $\Xi$ is linear in $\bar\phi$,
it is a field-independent kernel.
\onecolumngrid
\section{Feynman rules with background insertions}
\label{app:feynman}

\paragraph{Conventions.} The background metric is expanded around flat space,
Eq.~\eqref{eq:splith}, so that its whole dependence on $\bar\phi$ appears as background legs; the
remaining contractions use $\eta$, with which indices are raised and lowered. In each vertex the
free indices stand in the position shown; the position of the indices of the auxiliary tensors
below is chosen for convenience of contraction. Vertex rules are functional derivatives of the
action without the factor $i$ of $e^{iS}$; the explicit factors $i$ come from the Fourier transforms
of the derivatives. Symmetrizations act on index pairs together with the momenta attached to
them: exchanging a pair without its momentum changes the vertex.

\paragraph{Auxiliary tensors.} The background part of the covariant derivative in the gauge
condition, $\bar\nabla_{\mu}\phi^{\mu\nu}\supset\kappa\,\mathcal{D}^{\nu}{}_{\rho\sigma}{}^{\xi}{}_{\alpha\beta}
(\partial_{\xi}\bar\phi^{\alpha\beta})\phi^{\rho\sigma}$, defines
\begin{equation}
\tensor*{\mathcal{D}}{^{\kappa\,\,\,}_{\rho\sigma}^{\xi}_{\alpha\beta}}
\equiv\frac{2}{1-d}\,\delta^{\kappa}_{\sigma}\,\tensor*{D}{^{\lambda\,\,\,}_{\lambda\rho}^{\xi}_{\alpha\beta}}
+\tensor*{D}{^{\kappa\,\,\,}_{\rho\sigma}^{\xi}_{\alpha\beta}} .
\label{eq:Dcal}
\end{equation}
The vertices with auxiliary legs are built from $M(\phi)$ with the graviton leg opened. With
\begin{equation}
\tensor*{\mathcal{A}}{_{\zeta\,\,\,}^{\eta\theta}_{\rho\sigma\,\tau\,\,\,}^{\chi\psi}}
\equiv\delta^{\chi}_{\rho}\delta^{\theta}_{\sigma}
\Big(\frac{1}{d-1}\delta^{\psi}_{\tau}\delta^{\eta}_{\zeta}-\delta^{\psi}_{\zeta}\delta^{\eta}_{\tau}\Big),
\label{eq:Acal}
\end{equation}
we define its symmetrization
\begin{equation}\label{eq:Mcal}
\begin{split}
\tensor*{\mathcal{M}}{_{\zeta\,\,\,}^{\eta\theta}_{\rho\sigma\,\tau\,\,\,}^{\chi\psi}}
&\equiv\frac{1}{4}\Big\{\Big\{\Big[\tensor*{\mathcal{A}}{_{\zeta\,\,\,}^{\eta\theta}_{\rho\sigma\,\tau\,\,\,}^{\chi\psi}}
+(\chi\leftrightarrow\psi)\Big] +(\eta\leftrightarrow\theta)\Big\}+(\rho\leftrightarrow\sigma)\Big\},
\end{split}
\end{equation}
and the symmetrizer
$\tensor{I}{^{\lambda\xi}_{\rho\sigma}}=\frac{1}{2}(\delta^{\lambda}_{\rho}\delta^{\xi}_{\sigma}+\delta^{\xi}_{\rho}\delta^{\lambda}_{\sigma})$.

\subsection{Trivertices}

\paragraph{$\bar\phi\phi\phi$.} The contribution of the classical action is the same as without
background fields~\cite{Brandt2016}. With $(k;\mu\nu)$, $(p;\rho\sigma)$ and $(q;\lambda\pi)$ the
three legs,
\begin{equation}\label{eq:Vphiphiphi1}
\begin{split}
\bar\phi\phi\phi\big|_{\rm action}&\sim\frac{\kappa}{8}\Big\{\Big\{\Big[\Big(
\tensor*{D}{^{\alpha\,\,\,}_{\mu\nu}^{\gamma}_{\lambda\delta}}\tensor*{D}{^{\beta\,\,\,}_{\rho\sigma}^{\delta}_{\gamma\pi}}
-\frac{1}{d-1}\tensor*{D}{^{\alpha\,\,\,}_{\mu\nu}^{\gamma}_{\gamma\lambda}}\tensor*{D}{^{\beta\,\,\,}_{\rho\sigma}^{\delta}_{\delta\pi}}\Big) +\mu\leftrightarrow\nu\Big]+\rho\leftrightarrow\sigma\Big\}\\
&\quad+\lambda\leftrightarrow\pi\Big\}k_{\alpha}p_{\beta}
+\text{five permutations},
\end{split}
\end{equation}
the permutations acting on $\{(k;\mu,\nu),(p;\rho,\sigma),(q;\lambda,\pi)\}$. The background gauge
fixing adds
\begin{equation}\label{eq:Vphiphiphi2}
\begin{split}
\bar\phi\phi\phi\big|_{\rm gf}&\sim\frac{\kappa}{8\xi}\Big\{\Big\{\Big[q_{\lambda}k_{\beta}
\tensor*{\mathcal{D}}{_{\pi\rho\sigma}^{\beta}_{\mu\nu}}-\frac{1}{2}p_{\rho}q_{\lambda}\eta_{\mu\sigma}\eta_{\nu\pi}
+(\mu\leftrightarrow\nu)\Big] +(\rho\leftrightarrow\sigma)\Big\}+(\lambda\leftrightarrow\pi)\Big\}+(p;\rho,\sigma)\leftrightarrow(q;\lambda,\pi),
\end{split}
\end{equation}
and the vertex is the sum of the two.

\paragraph{$\bar\phi\bar cc$.} From the ghost action~\eqref{eq:ghost}; $(k;\mu\nu)$ is the background
graviton, $q$ the momentum of the ghost $c_{\beta}$ and $p$ that of the antighost $\bar c_{\alpha}$, all
incoming,
\begin{equation}\label{eq:Vphicc}
\begin{split}
\tensor*{(V_{\bar\phi\bar cc})}{_{\mu\nu}^{\alpha\beta}}&=\frac{\kappa}{2}\Big\{k_{\rho}q_{\sigma}\Big[
\frac{2}{d-1}\tensor*{D}{^{\pi\,\,\,}_{\lambda\pi}^{\rho}_{\mu\nu}}
\left(\eta^{\lambda\beta}\eta^{\alpha\sigma}-\eta^{\lambda\alpha}\eta^{\beta\sigma}+\eta^{\alpha\beta}\eta^{\lambda\sigma}\right)\\
&\quad+\tensor*{D}{^{\alpha\,\,\,}_{\lambda\pi}^{\rho}_{\mu\nu}}\left(\eta^{\lambda\pi}\eta^{\beta\sigma}-2\eta^{\lambda\beta}\eta^{\pi\sigma}\right)\Big]
-p_{\rho}k_{\sigma}\,\delta^{\rho}_{\mu}\delta^{\alpha}_{\nu}\eta^{\sigma\beta}\\
&\quad+p_{\rho}q_{\sigma}\left(\delta^{\rho}_{\mu}\delta^{\sigma}_{\nu}\eta^{\alpha\beta}
-\delta^{\rho}_{\mu}\delta^{\alpha}_{\nu}\eta^{\sigma\beta}+\delta^{\alpha}_{\mu}\delta^{\sigma}_{\nu}\eta^{\rho\beta}\right)\Big\}
+(\mu\leftrightarrow\nu).
\end{split}
\end{equation}

\paragraph{$\bar G\phi\phi$.} The contribution of the classical action is
\begin{equation}
\tensor*{(\mathcal{V}^{\bar G\phi\phi}_{1})}{_{\gamma\,\,\,}^{\alpha\beta}_{\mu\nu\rho\sigma}}(p)
=i\kappa\,p_{\upsilon}\,\tensor*{D}{^{\tau\,\,\,}_{\chi\psi}^{\upsilon}_{\mu\nu}}\,
\tensor*{\mathcal{M}}{_{\gamma\,\,\,}^{\alpha\beta}_{\rho\sigma\,\tau\,\,\,}^{\chi\psi}},
\label{eq:VGphiphi1}
\end{equation}
and that of the gauge fixing is built from
\begin{equation}
\tensor*{(\mathcal{V}^{\bar G\phi\phi}_{{\rm gf},1})}{_{\gamma\,\,\,}^{\alpha\beta}_{\mu\nu\rho\sigma}}(p)
=-\frac{2i\kappa}{\xi}\,p_{\mu}\Big(\frac{1}{1-d}\delta^{\beta}_{\rho}\delta^{\alpha}_{\gamma}\eta_{\nu\sigma}
+\frac{1}{2}\delta^{\beta}_{\sigma}\delta^{\alpha}_{\rho}\eta_{\nu\gamma}\Big),
\label{eq:VGphiphigf1}
\end{equation}
symmetrized in the pairs of the graviton and of the connection,
$\mathcal{V}^{\bar G\phi\phi}_{{\rm gf},2}=\frac{1}{8}\{\{[\mathcal{V}^{\bar G\phi\phi}_{{\rm gf},1}
+(\alpha\leftrightarrow\beta)]+(\mu\leftrightarrow\nu)\}+(\rho\leftrightarrow\sigma)\}$. With the quantum
gravitons of momenta $p$ and $q$,
\begin{equation}\label{eq:VGphiphi}
\begin{split}
\tensor*{(V_{\bar G\phi\phi})}{_{\gamma\,\,\,}^{\alpha\beta}_{\mu\nu\rho\sigma}}(p,q)
&=\tensor*{(\mathcal{V}^{\bar G\phi\phi}_{1})}{_{\gamma\,\,\,}^{\alpha\beta}_{\mu\nu\rho\sigma}}(p)
+\tensor*{(\mathcal{V}^{\bar G\phi\phi}_{1})}{_{\gamma\,\,\,}^{\alpha\beta}_{\rho\sigma\mu\nu}}(q)\\
&\quad+\tensor*{(\mathcal{V}^{\bar G\phi\phi}_{{\rm gf},2})}{_{\gamma\,\,\,}^{\alpha\beta}_{\mu\nu\rho\sigma}}(p)
+\tensor*{(\mathcal{V}^{\bar G\phi\phi}_{{\rm gf},2})}{_{\gamma\,\,\,}^{\alpha\beta}_{\rho\sigma\mu\nu}}(q).
\end{split}
\end{equation}

\paragraph{$\bar G\bar cc$.} From the ghost action; $(k;\lambda\mu\nu)$ is the background connection,
$q$ the momentum of the ghost $c_{\beta}$ and $p$ that of the antighost $\bar c_{\alpha}$,
\begin{equation}\label{eq:VGcc}
\begin{split}
(V_{\bar G\bar cc})\indices*{_\lambda^{\mu\nu\alpha\beta}}&=-\frac{i\kappa}{2}q_{\gamma}\Big[\frac{2}{1-d}
\delta^{\nu}_{\lambda}\left(\eta^{\mu\alpha}\eta^{\beta\gamma}-\eta^{\mu\gamma}\eta^{\alpha\beta}-\eta^{\alpha\gamma}\eta^{\mu\beta}\right) +\delta^{\alpha}_{\lambda}\left(\eta^{\mu\nu}\eta^{\beta\gamma}-2\eta^{\mu\gamma}\eta^{\nu\beta}\right)+(\mu\leftrightarrow\nu)\Big].
\end{split}
\end{equation}

\paragraph{Trivertices with a quantum connection.} With $k$ the momentum of the background
graviton and $p$ that of the quantum graviton,
\begin{equation}\label{eq:VGphibarphi}
\begin{split}
\tensor*{(V_{G\bar\phi\phi})}{_{\lambda\,\,\,}^{\mu\nu}_{\alpha\beta\rho\sigma}}(k,p)
&=i\kappa\Big(\tensor*{\mathcal{M}}{_{\lambda\,\,\,}^{\mu\nu}_{\alpha\beta\,\tau\,\,\,}^{\chi\psi}}\,
\tensor*{D}{^{\tau\,\,\,}_{\chi\psi}^{\pi}_{\rho\sigma}}\,p_{\pi} +\tensor*{\mathcal{M}}{_{\lambda\,\,\,}^{\mu\nu}_{\rho\sigma\,\tau\,\,\,}^{\chi\psi}}\,
\tensor*{D}{^{\tau\,\,\,}_{\chi\psi}^{\pi}_{\alpha\beta}}\,k_{\pi}\Big).
\end{split}
\end{equation}
The vertex $\bar GG\phi$ comes from $\kappa\,\bar G\star M(\phi)\star G$ and is momentum independent,
\begin{equation}
\tensor*{(V_{\bar GG\phi})}{_{\gamma\,\,\,}^{\alpha\beta}_{\rho\sigma\,\tau\,\,\,}^{\chi\psi}}
=\kappa\,\tensor*{\mathcal{M}}{_{\gamma\,\,\,}^{\alpha\beta}_{\rho\sigma\,\tau\,\,\,}^{\chi\psi}},
\label{eq:VGGphi}
\end{equation}
with $\gamma\alpha\beta$ the indices of the background connection, $\tau\chi\psi$ those of the
quantum connection and $\rho\sigma$ those of the graviton; $\bar\phi GG$, from
$\frac{\kappa}{2}G\star M(\bar\phi)\star G$, has the same form, with the graviton leg now a background
one.

\subsection{Quartic vertices}

Each quartic vertex is given as an elementary block, followed by the symmetrizations that complete
it.

\paragraph{$\phi\phi\bar G\bar G$.} From the gauge fixing, momentum independent; $\mu\nu$ and
$\rho\sigma$ are the pairs of the quantum gravitons and $\gamma\alpha\beta$, $\zeta\eta\theta$ the indices
of the two background connections:
\begin{equation}\label{eq:VffGG1}
\begin{split}
(\mathcal{V}^{\phi\phi\bar G\bar G}_{1})^{\mu\nu\rho\sigma}_{\gamma\alpha\beta\zeta\eta\theta}&=-\frac{\kappa^{2}}{2\xi}\Big[
\frac{4}{(1-d)^{2}}\,\delta^{\mu}_{\beta}\delta^{\rho}_{\theta}\eta^{\nu\sigma}\eta_{\gamma\alpha}\eta_{\zeta\eta}\\
&\quad+\frac{2}{1-d}\left(\delta^{\mu}_{\alpha}\delta^{\nu}_{\beta}\delta^{\rho}_{\theta}\delta^{\sigma}_{\gamma}\eta_{\eta\zeta}
+\delta^{\mu}_{\beta}\delta^{\nu}_{\zeta}\delta^{\rho}_{\eta}\delta^{\sigma}_{\theta}\eta_{\gamma\alpha}\right) +\delta^{\mu}_{\alpha}\delta^{\nu}_{\beta}\delta^{\rho}_{\eta}\delta^{\sigma}_{\theta}\eta_{\gamma\zeta}\Big].
\end{split}
\end{equation}
It is symmetrized first in $\mu\nu$ and $\rho\sigma$, then in $\alpha\beta$ and $\eta\theta$, each time
with a factor $\frac{1}{4}$, and the vertex is the sum of the result over the exchange of the two
graviton pairs and of the two connection triples.

\paragraph{$\bar\phi\bar\phi\bar cc$.} From the ghost action; $\mu\nu$ and $\rho\sigma$ are the pairs of
the background gravitons, of momenta $k$ and $r$, $\alpha\beta$ the ghost indices and $p$ the momentum
of the ghost of index $\beta$:
\begin{equation}\label{eq:Vphiphicc1}
\begin{split}
\tensor*{(\mathcal{V}_{\bar\phi\bar\phi\bar cc})}{_{\mu\nu\rho\sigma}^{\alpha\beta}}(k,r,p)
&=\kappa^{2}k_{\upsilon}\,\tensor{\mathcal{D}}{^{\alpha}_{\lambda\xi}^{\upsilon}_{\mu\nu}}\Big[\tensor{I}{^{\lambda\xi}_{\rho\sigma}}(p+r)^{\beta} -p_{\zeta}\left(\eta^{\lambda\beta}\tensor{I}{^{\zeta\xi}_{\rho\sigma}}+\eta^{\xi\beta}\tensor{I}{^{\lambda\zeta}_{\rho\sigma}}\right)\Big].
\end{split}
\end{equation}
The vertex is the sum over the exchange of the two background gravitons, with their momenta. The
exchange $\alpha\leftrightarrow\beta$ is not a symmetry: the two ghost legs are not equivalent.

\paragraph{$\bar\phi\bar\phi\phi\phi$.} From the gauge fixing; $(k;\alpha\beta)$ and $(r;\gamma\delta)$ are
the background gravitons and $(p;\mu\nu)$, $(q;\rho\sigma)$ the quantum ones:
\begin{equation}\label{eq:Vphiphiff1}
\begin{split}
\tensor*{(\mathcal{V}^{\bar\phi\bar\phi\phi\phi}_{1})}{_{\alpha\beta\gamma\delta}^{\mu\nu\rho\sigma}}(k,r,p,q)
&=\frac{\kappa^{2}}{2\xi}\Big[p^{\mu}q^{\rho}\,\eta_{\alpha\gamma}\delta^{\nu}_{\beta}\delta^{\sigma}_{\delta}
-p^{\mu}k_{\xi}\,\delta^{\nu}_{\gamma}\,\tensor{\mathcal{D}}{_{\delta}^{\rho\sigma\xi}_{\alpha\beta}}\\
&\quad-q^{\rho}r_{\zeta}\,\delta^{\sigma}_{\beta}\,\tensor{\mathcal{D}}{_{\alpha}^{\mu\nu\zeta}_{\gamma\delta}}
+k_{\xi}r_{\zeta}\,\eta_{\lambda\pi}\,\tensor{\mathcal{D}}{^{\lambda\mu\nu\xi}_{\alpha\beta}}\,
\tensor{\mathcal{D}}{^{\pi\rho\sigma\zeta}_{\gamma\delta}}\Big].
\end{split}
\end{equation}
The four terms are the ways of taking two background fields from Eq.~\eqref{eq:gf}: the
second-order term of $\bar h_{\nu\sigma}$, its first-order term times the background part of either
covariant derivative, and the background parts of both derivatives. The block is symmetrized in
the background pairs and then in the quantum pairs, each time with a factor $\frac{1}{4}$, and the
vertex is the sum over the exchange of the two background gravitons and of the two quantum ones,
each with its momentum.

\paragraph{$\bar\phi\bar G\phi\phi$.} From the gauge fixing; $(k;\alpha\beta)$ is the background
graviton, $\zeta\eta\theta$ the indices of the background connection and $(p;\mu\nu)$, $(q;\rho\sigma)$ the
quantum gravitons:
\begin{equation}\label{eq:VphiGff1}
\begin{split}
\tensor*{(\mathcal{V}^{\bar\phi\bar G\phi\phi}_{1})}{_{\alpha\beta\zeta\eta\theta}^{\mu\nu\rho\sigma}}(k,p,q)
&=-\frac{i\kappa^{2}}{2\xi}\Big[\frac{4}{(1-d)^{2}}\,k_{\upsilon}\,\eta^{\nu\sigma}\eta_{\eta\zeta}
\Big(\tensor{D}{^{\upsilon}_{\alpha\beta}^{\chi}_{\chi}^{\rho}}\delta^{\mu}_{\theta}
+\tensor{D}{^{\upsilon}_{\alpha\beta}^{\chi}_{\chi}^{\mu}}\delta^{\rho}_{\theta}\Big)\\
&\quad+\frac{2}{1-d}\,k_{\upsilon}\Big(
\tensor{D}{^{\upsilon}_{\alpha\beta}^{\chi}_{\chi}^{\rho}}\delta^{\mu}_{\eta}\delta^{\nu}_{\theta}\delta^{\sigma}_{\zeta}
+\tensor{D}{^{\upsilon}_{\alpha\beta}^{\nu\rho\sigma}}\delta^{\mu}_{\theta}\eta_{\zeta\eta}
+\tensor{D}{^{\upsilon}_{\alpha\beta}^{\sigma\mu\nu}}\delta^{\rho}_{\theta}\eta_{\zeta\eta}\\
&\quad+\tensor{D}{^{\upsilon}_{\alpha\beta}^{\chi}_{\chi}^{\mu}}\delta^{\rho}_{\eta}\delta^{\sigma}_{\theta}\delta^{\nu}_{\zeta}\Big)
+k_{\upsilon}\Big(\delta^{\mu}_{\eta}\delta^{\nu}_{\theta}\tensor{D}{_{\zeta}^{\rho\sigma\upsilon}_{\alpha\beta}}
+\delta^{\rho}_{\eta}\delta^{\sigma}_{\theta}\tensor{D}{_{\zeta}^{\mu\nu\upsilon}_{\alpha\beta}}\Big)\\
&\quad-\frac{2}{1-d}\,\delta^{\nu}_{\alpha}\delta^{\sigma}_{\beta}\eta_{\eta\zeta}
\left(\delta^{\mu}_{\theta}\,q^{\rho}+\delta^{\rho}_{\theta}\,p^{\mu}\right)
-\eta_{\beta\zeta}\delta^{\nu}_{\alpha}\delta^{\rho}_{\eta}\delta^{\sigma}_{\theta}\,p^{\mu} -\delta^{\sigma}_{\beta}\delta^{\mu}_{\eta}\delta^{\nu}_{\theta}\eta_{\zeta\alpha}\,q^{\rho}\Big].
\end{split}
\end{equation}
The terms proportional to $k$ come from the background parts of both covariant derivatives, one
supplying $\bar G$ and the other $D\star\partial\bar\phi$; those proportional to $p$ or $q$ come from
one derivative supplying $\bar G$ and the first-order term of $\bar h_{\nu\sigma}$ supplying $\bar\phi$.
The block is symmetrized in $\alpha\beta$ and $\eta\theta$, then in the quantum pairs, each time with a
factor $\frac{1}{4}$, and the vertex is the sum over the exchange of the two quantum gravitons.

\paragraph{$\bar\phi\bar G\bar G\phi\phi$.} From the gauge fixing; this quintic vertex enters only
$\delta_{\bar\phi}\delta^{3}_{\bar G}S_{\rm eff}$ (Sec.~\ref{sec:fourpointtest}). The part of
$\bar\nabla_{\mu}\phi^{\mu\nu}$ linear in $\bar G$ is $\kappa\mathcal{B}^{\nu}$, with
\begin{equation}
\mathcal{B}^{\nu}(\bar G,\phi)\equiv\bar G^{\nu}_{\mu\lambda}\,\phi^{\mu\lambda}-\frac{2}{d-1}\,\bar G^{\rho}_{\rho\mu}\,\phi^{\mu\nu},
\label{eq:Bnu}
\end{equation}
and the first-order term of $\bar h_{\nu\sigma}=\eta_{\nu\sigma}-\kappa\bar\phi_{\nu\sigma}+O(\bar\phi^{2})$ in
Eq.~\eqref{eq:gf} gives
\begin{equation}
\mathcal{L}_{\rm gf}\big|_{\bar\phi\bar G\bar G\phi\phi}=\frac{\kappa^{3}}{2\xi}\,\bar\phi_{\nu\sigma}\,
\mathcal{B}^{\nu}(\bar G,\phi)\,\mathcal{B}^{\sigma}(\bar G,\phi),
\label{eq:Lquintic}
\end{equation}
with the indices of $\bar\phi$ lowered by $\eta$. The vertex follows by functional differentiation, with the
symmetrizations of $\phi\phi\bar G\bar G$; the same expansion with $\eta_{\nu\sigma}$ in place of
$-\kappa\bar\phi_{\nu\sigma}$ gives Eq.~\eqref{eq:VffGG1}.

\paragraph{$\bar\phi\bar G\bar cc$.} From the ghost action; $(k;\rho\sigma)$ is the background graviton,
$\lambda\mu\nu$ the indices of the background connection, $\alpha\beta$ the ghost indices and $p$ the
momentum of the ghost of index $\beta$:
\begin{equation}\label{eq:VphiGcc1}
\begin{split}
\tensor*{(\mathcal{V}^{\bar\phi\bar G\bar cc}_{1})}{_{\lambda}^{\mu\nu}_{\rho\sigma}^{\alpha\beta}}(k,p)
&=-i\kappa^{2}\Big\{(k+p)^{\beta}\Big[\frac{2}{1-d}\,\delta^{\nu}_{\lambda}\,\tensor{I}{^{\mu\alpha}_{\rho\sigma}}
+\delta^{\alpha}_{\lambda}\,\tensor{I}{^{\mu\nu}_{\rho\sigma}}\Big]\\
&\quad-p_{\zeta}\,\tensor{I}{^{\mu\zeta}_{\rho\sigma}}\Big(\frac{2}{1-d}\,\delta^{\nu}_{\lambda}\eta^{\alpha\beta}
+\delta^{\alpha}_{\lambda}\eta^{\nu\beta}\Big)\\
&\quad-p_{\zeta}\,\eta^{\mu\beta}\Big[\frac{2}{1-d}\,\delta^{\nu}_{\lambda}\,\tensor{I}{^{\zeta\alpha}_{\rho\sigma}}
+\delta^{\alpha}_{\lambda}\,\tensor{I}{^{\zeta\nu}_{\rho\sigma}}\Big]\Big\}.
\end{split}
\end{equation}
The vertex is its symmetrization in the lower pair $\mu\nu$ of the connection, with a factor
$\frac{1}{2}$; as for $\bar\phi\bar\phi\bar cc$, $\alpha\leftrightarrow\beta$ is not a symmetry.
\clearpage
\twocolumngrid
\section{Diagrams, symmetry factors and channel structure}
\label{app:diagrams}

With Feynman rules already symmetrized in the identical fields of each vertex, the symmetry factor of
a connected diagram $\Gamma$ is $S_{\Gamma}=1/\lvert{\rm Aut}(\Gamma)\rvert$, where ${\rm Aut}(\Gamma)$ is the group
of permutations of the vertices and internal lines of $\Gamma$, including the exchange of the two ends of
a line, that leave the graph identical to itself. Each labeled external leg is kept fixed, each line
keeps its field species and, on ghost lines, the orientation of the ghost number. $S_{\Gamma}$ corrects
the overcounting of the Wick contractions that, once the vertices are symmetrized, produce the same
diagram. The Grassmann sign of a closed ghost loop is not part of $S_{\Gamma}$ and is counted
separately. With the three external legs fixed, no triangle has a nontrivial automorphism; in the
bosonic bubbles the two internal graviton lines can be exchanged, so $S_{\Gamma}=1/2$, while in the ghost
bubbles the orientation forbids the exchange and $S_{\Gamma}=1$.

Tables~\ref{tab:diagGG}--\ref{tab:diagphiphi} list the diagrams of the three-point functions, with the
momentum labels of Sec.~\ref{sec:threepoint}. A pair of vertices such as
$\bar G\bar G\phi\phi$--$\bar\phi\phi\phi$ denotes the bubble formed by these two vertices, and the
channel of a diagram is the momentum that flows through its most singular denominator. The third column gives
the channel of the bubbles and the special features of each diagram; the diagrams marked transverse are
transverse, one by one, in the leg that carries the inhomogeneous variation.

\begin{table}[b]
\caption{Diagrams of $\delta_{\bar\phi}\delta^{2}_{\bar G}S_{\rm eff}$. The variation is on the $\bar\phi$
leg, of momentum $k$. In the triangles (2)--(4) the internal $G$ line occupies each of the three edges.}
\label{tab:diagGG}
\begin{ruledtabular}
\begin{tabular}{cllc}
Diagram & Loop & Channel; status & $S_{\Gamma}$\\
\hline
1 & $\phi\phi\phi$ triangle & $C_{0}$ & $1$\\
2, 4 & $G\phi\phi$ triangle & $p_{2}$, $p_{1}$ & $1$\\
3 & $G\phi\phi$ triangle & $k$; transverse & $1$\\
5, 6 & ghost triangle & $C_{0}$; sign $-1$ & $1$\\
7, 8 & $\bar\phi\bar G\bar cc$--$\bar G\bar cc$ & sign $-1$ & $1$\\
9 & $G\phi G$ triangle & scaleless & $1$\\
10 & $\bar G\bar G\phi\phi$--$\bar\phi\phi\phi$ & $k$; transverse & $1/2$\\
11, 12 & $\bar\phi\bar G\phi\phi$--$\bar G\phi\phi$ & & $1/2$\\
\end{tabular}
\end{ruledtabular}
\end{table}

\begin{table}[b]
\caption{Diagrams of $\delta^{2}_{\bar\phi}\delta_{\bar G}S_{\rm eff}$. The $\bar G$ leg carries $k$, and the
variation is on the $\bar\phi$ leg $p_{2}$.}
\label{tab:diagphiG}
\begin{ruledtabular}
\begin{tabular}{cllc}
Diagram & Loop & Channel; status & $S_{\Gamma}$\\
\hline
1 & $\phi\phi\phi$ triangle & $C_{0}$ & $1$\\
2, 3, 4 & $G\phi\phi$ triangle & $k$, $p_{2}$, $p_{1}$; (3) transverse & $1$\\
5, 6 & $G\phi G$ triangle & scaleless & $1$\\
7, 8 & ghost triangle & $C_{0}$; sign $-1$ & $1$\\
9 & $\bar\phi\bar\phi\bar cc$--$\bar G\bar cc$ & $k$; sign $-1$ & $1$\\
10, 11 & $\bar\phi\bar G\bar cc$--$\bar\phi\bar cc$ & $p_{1}$, $p_{2}$; (11) transverse & $1$\\
12 & $\bar\phi\bar\phi\phi\phi$--$\bar G\phi\phi$ & $k$ & $1/2$\\
13, 14 & $\bar\phi\bar G\phi\phi$--$\bar\phi\phi\phi$ & $p_{1}$, $p_{2}$; (14) transverse & $1/2$\\
\end{tabular}
\end{ruledtabular}
\end{table}

\begin{table}[b]
\caption{Diagrams of $\delta^{3}_{\bar\phi}S_{\rm eff}$. The variation is on the leg $k$. The last four
vanish by rule (iv) of Sec.~\ref{sec:counting}.}
\label{tab:diagphiphi}
\begin{ruledtabular}
\begin{tabular}{cllc}
Diagram & Loop & Channel; status & $S_{\Gamma}$\\
\hline
1 & $\phi\phi\phi$ triangle & $C_{0}$ & $1$\\
2, 3 & ghost triangle & $C_{0}$; sign $-1$ & $1$\\
4--6 & $G\phi\phi$ triangle & $k$, $p_{1}$, $p_{2}$ & $1$\\
7--9 & $\bar\phi\bar\phi\phi\phi$--$\bar\phi\phi\phi$ & $k$, $p_{1}$, $p_{2}$ & $1/2$\\
10--12 & $\bar\phi\bar\phi\bar cc$--$\bar\phi\bar cc$ & $k$, $p_{1}$, $p_{2}$; sign $-1$ & $1$\\
13--15 & $GG\phi$ triangle & scaleless & \\
16 & $GGG$ triangle & no momentum & \\
\end{tabular}
\end{ruledtabular}
\end{table}

\paragraph{Exchange of legs.} In $\delta_{\bar\phi}\delta^{2}_{\bar G}S_{\rm eff}$ the exchange of the two
$\bar G$ legs is the operation $(p_{1},a)\leftrightarrow(-p_{2},b)$, where $a$ and $b$ stand for the index
sets of the legs; it involves momenta and indices, not momenta alone, and the sign accounts for $p_{2}$
being outgoing. It relates the pairs $(7,8)$ and $(11,12)$ of Table~\ref{tab:diagGG}. The two
orientations $(5)$ and $(6)$ of the ghost triangle are related in the same way, with the orientation of
the loop held fixed: reversing the orientation is not an automorphism, but what generates the second
diagram. In $\delta^{2}_{\bar\phi}\delta_{\bar G}S_{\rm eff}$ the exchange of the two $\bar\phi$ legs leaves
$(1)$, $(2)$, $(9)$ and $(12)$ invariant and exchanges the pairs $(3,4)$, $(5,6)$, $(7,8)$, $(10,11)$ and
$(13,14)$. In $\delta^{3}_{\bar\phi}S_{\rm eff}$ the exchange of the legs $p_{1}$ and $p_{2}$ closes the set of
sixteen diagrams, six of which are invariant.

\paragraph{Channel structure.} The two-point terms of the identities are attached to the legs that do
not carry the variation, and the momenta that flow through them are those of these legs. In
Eqs.~\eqref{eq:ward3GG} and~\eqref{eq:ward3phiphi} they reach only the channels $p_{1}$ and $p_{2}$, and in
Eq.~\eqref{eq:ward3phiG} only the channels $k$ and $p_{1}$. The channel of the varied leg therefore closes
on the three-point diagrams alone. In $\delta_{\bar\phi}\delta^{2}_{\bar G}S_{\rm eff}$ and $\delta^{2}_{\bar\phi}\delta_{\bar G}S_{\rm eff}$, the
$\phi GG$ and $\phi\phi G$ cases, the contraction vanishes diagram by diagram, in arbitrary $d$, which is the transversality of the diagrams so marked. Their symmetry factors
enter the identities multiplied by zero and are not tested. In $\delta^{2}_{\bar\phi}\delta_{\bar G}S_{\rm eff}$
the nine nonvanishing diagrams outside the channel $p_{2}$ contribute to the identity in the channels to
which they belong, and changing the symmetry factor of any one of them, with the others held fixed, would
break the closure of its channel; this does not exclude compensating errors. The function as a whole is not transverse; only the marked diagrams are.

In $\delta^{3}_{\bar\phi}S_{\rm eff}$ there is a single graviton triangle against two ghost triangles,
because the ghost line is oriented and its two cyclic orders are distinct diagrams. The
diagram $(16)$, with three internal $G$ lines, is allowed here, since its vertices are $\bar\phi GG$, which
exists, and not $\bar GGG$; it vanishes because $M(\bar\phi)$ and the three auxiliary propagators, all
multiples of the constant tensor $D$, do not depend on the loop momentum, and the diagram is a
scaleless power divergence.

\begin{figure*}[tbp]
\centering
\setlength{\tabcolsep}{1pt}
\begin{tabular}{ccccccc}
\begin{tikzpicture}[scale=0.42]
\node[inner sep=1pt] (e1) at (0,2.5) {\scriptsize $k$};
\coordinate (T) at (0,1.05);
\coordinate (L) at (-1.25,-0.72);
\coordinate (R) at (1.25,-0.72);
\node[inner sep=1pt] (e2) at (-2.2,-1.3) {\scriptsize $p_{1}$};
\node[inner sep=1pt] (e3) at (2.2,-1.3) {\scriptsize $p_{2}$};
\draw[aux,bgtick] (e1) -- (T);
\draw[aux,bgtick] (e2) -- (L);
\draw[aux,bgtick] (e3) -- (R);
\draw[gh,ghar] (T) -- (R);
\draw[gh,ghar] (R) -- (L);
\draw[gh,ghar] (L) -- (T);
\node at (0,-2.05) {\scriptsize (a)};
\end{tikzpicture} &
\begin{tikzpicture}[scale=0.42]
\node[inner sep=1pt] (e1) at (-2.4,1.1) {\scriptsize $k$};
\node[inner sep=1pt] (e2) at (-2.4,-1.1) {\scriptsize $p_{1}$};
\coordinate (L) at (-1,0);
\coordinate (R) at (1,0);
\node[inner sep=1pt] (e3) at (2.5,0) {\scriptsize $p_{2}$};
\draw[aux,bgtick] (e1) -- (L);
\draw[aux,bgtick] (e2) -- (L);
\draw[grav] (L) to[bend left=90,looseness=1.5] (R);
\draw[grav] (R) to[bend left=90,looseness=1.5] (L);
\draw[aux,bgtick] (e3) -- (R);
\node at (0,-1.85) {\scriptsize (b)};
\end{tikzpicture} &
\begin{tikzpicture}[scale=0.42]
\node[inner sep=1pt] (e1) at (0,2.5) {\scriptsize $k$};
\coordinate (T) at (0,1.05);
\coordinate (L) at (-1.25,-0.72);
\coordinate (R) at (1.25,-0.72);
\node[inner sep=1pt] (e2) at (-2.2,-1.3) {\scriptsize $p_{1}$};
\node[inner sep=1pt] (e3) at (2.2,-1.3) {\scriptsize $p_{2}$};
\draw[aux,bgtick] (e1) -- (T);
\draw[aux,bgtick] (e2) -- (L);
\draw[aux,bgtick] (e3) -- (R);
\draw[grav] (T) -- (L);
\draw[aux] (L) -- (R);
\draw[grav] (R) -- (T);
\node at (0,-2.05) {\scriptsize (c)};
\end{tikzpicture} &
\begin{tikzpicture}[scale=0.42]
\node[inner sep=1pt] (e1) at (0,2.5) {\scriptsize $k$};
\coordinate (T) at (0,1.05);
\coordinate (L) at (-1.25,-0.72);
\coordinate (R) at (1.25,-0.72);
\node[inner sep=1pt] (e2) at (-2.2,-1.3) {\scriptsize $p_{1}$};
\node[inner sep=1pt] (e3) at (2.2,-1.3) {\scriptsize $p_{2}$};
\draw[aux,bgtick] (e1) -- (T);
\draw[aux,bgtick] (e2) -- (L);
\draw[aux,bgtick] (e3) -- (R);
\draw[grav] (T) -- (L);
\draw[grav] (L) -- (R);
\draw[grav] (R) -- (T);
\node at (0,-2.05) {\scriptsize (d)};
\end{tikzpicture} &
 &
 &
 \\[2pt]
\begin{tikzpicture}[scale=0.42]
\coordinate (e1) at (-2,2);
\coordinate (e2) at (2,2);
\coordinate (e3) at (2,-2);
\coordinate (e4) at (-2,-2);
\coordinate (A) at (-1,1);
\coordinate (B) at (1,1);
\coordinate (C) at (1,-1);
\coordinate (D) at (-1,-1);
\draw[aux,bgtick] (e1) -- (A);
\draw[aux,bgtick] (e2) -- (B);
\draw[aux,bgtick] (e3) -- (C);
\draw[aux,bgtick] (e4) -- (D);
\draw[gh,ghar] (A) -- (B);
\draw[gh,ghar] (B) -- (C);
\draw[gh,ghar] (C) -- (D);
\draw[gh,ghar] (D) -- (A);
\node at (0,-3.45) {\scriptsize (e)};
\end{tikzpicture} &
\begin{tikzpicture}[scale=0.42]
\coordinate (e1) at (-2,2);
\coordinate (e2) at (2,2);
\coordinate (e3) at (2,-2);
\coordinate (e4) at (-2,-2);
\coordinate (A) at (-1,1);
\coordinate (B) at (1,1);
\coordinate (C) at (1,-1);
\coordinate (D) at (-1,-1);
\draw[aux,bgtick] (e1) -- (A);
\draw[aux,bgtick] (e2) -- (B);
\draw[aux,bgtick] (e3) -- (C);
\draw[aux,bgtick] (e4) -- (D);
\draw[grav] (A) -- (B);
\draw[grav] (B) -- (C);
\draw[grav] (C) -- (D);
\draw[grav] (D) -- (A);
\node at (0,-3.45) {\scriptsize (f)};
\end{tikzpicture} &
\begin{tikzpicture}[scale=0.42]
\coordinate (e1) at (-2,2);
\coordinate (e2) at (2,2);
\coordinate (e3) at (2,-2);
\coordinate (e4) at (-2,-2);
\coordinate (A) at (-1,1);
\coordinate (B) at (1,1);
\coordinate (C) at (1,-1);
\coordinate (D) at (-1,-1);
\draw[aux,bgtick] (e1) -- (A);
\draw[aux,bgtick] (e2) -- (B);
\draw[aux,bgtick] (e3) -- (C);
\draw[aux,bgtick] (e4) -- (D);
\draw[aux] (A) -- (B);
\draw[grav] (B) -- (C);
\draw[grav] (C) -- (D);
\draw[grav] (D) -- (A);
\node at (0,-3.45) {\scriptsize (g)};
\end{tikzpicture} &
\begin{tikzpicture}[scale=0.42]
\coordinate (e1) at (-2,2);
\coordinate (e2) at (2,2);
\coordinate (e3) at (2,-2);
\coordinate (e4) at (-2,-2);
\coordinate (A) at (-1,1);
\coordinate (B) at (1,1);
\coordinate (C) at (1,-1);
\coordinate (D) at (-1,-1);
\draw[aux,bgtick] (e1) -- (A);
\draw[aux,bgtick] (e2) -- (B);
\draw[aux,bgtick] (e3) -- (C);
\draw[aux,bgtick] (e4) -- (D);
\draw[aux] (A) -- (B);
\draw[grav] (B) -- (C);
\draw[aux] (C) -- (D);
\draw[grav] (D) -- (A);
\node at (0,-3.45) {\scriptsize (h)};
\end{tikzpicture} &
\begin{tikzpicture}[scale=0.42]
\coordinate (e1) at (-2.5,0.8);
\coordinate (e2) at (-2.5,-0.8);
\coordinate (e3) at (2.2,2);
\coordinate (e4) at (2.2,-2);
\coordinate (Q) at (-1.1,0);
\coordinate (U) at (1,1);
\coordinate (L) at (1,-1);
\draw[aux,bgtick] (e1) -- (Q);
\draw[aux,bgtick] (e2) -- (Q);
\draw[aux,bgtick] (e3) -- (U);
\draw[aux,bgtick] (e4) -- (L);
\draw[grav] (Q) -- (U);
\draw[grav] (U) -- (L);
\draw[grav] (L) -- (Q);
\node at (0,-3.45) {\scriptsize (i)};
\end{tikzpicture} &
\begin{tikzpicture}[scale=0.42]
\coordinate (e1) at (-2.5,0.8);
\coordinate (e2) at (-2.5,-0.8);
\coordinate (e3) at (2.2,2);
\coordinate (e4) at (2.2,-2);
\coordinate (Q) at (-1.1,0);
\coordinate (U) at (1,1);
\coordinate (L) at (1,-1);
\draw[aux,bgtick] (e1) -- (Q);
\draw[aux,bgtick] (e2) -- (Q);
\draw[aux,bgtick] (e3) -- (U);
\draw[aux,bgtick] (e4) -- (L);
\draw[grav] (Q) -- (U);
\draw[aux] (U) -- (L);
\draw[grav] (L) -- (Q);
\node at (0,-3.45) {\scriptsize (j)};
\end{tikzpicture} &
\begin{tikzpicture}[scale=0.42]
\coordinate (e1) at (-2.4,1.2);
\coordinate (e2) at (-2.4,-1.2);
\coordinate (e3) at (2.4,1.2);
\coordinate (e4) at (2.4,-1.2);
\coordinate (L) at (-1,0);
\coordinate (R) at (1,0);
\draw[aux,bgtick] (e1) -- (L);
\draw[aux,bgtick] (e2) -- (L);
\draw[aux,bgtick] (e3) -- (R);
\draw[aux,bgtick] (e4) -- (R);
\draw[grav] (L) to[bend left=90,looseness=1.5] (R);
\draw[grav] (R) to[bend left=90,looseness=1.5] (L);
\node at (0,-3.45) {\scriptsize (k)};
\end{tikzpicture}
\end{tabular}
\caption{Representatives of the nonvanishing one-loop diagrams of $\delta^{3}_{\bar G}S_{\rm eff}$, (a)--(d), and of the seven families of $\delta^{4}_{\bar G}S_{\rm eff}$, (e)--(k): (a) ghost triangle, (b) bubble with the quartic gauge-fixing vertex, (c) triangle with one $G$ line, (d) graviton triangle; (e) ghost box, (f) graviton box, (g) box with one $G$ line, (h) box with two $G$ lines, (i) graviton triangle, (j) triangle with one $G$ line, (k) bubble. The other diagrams of each family differ by the assignment of the external legs and, for ghost loops, by the orientation.}
\label{fig:diagG}
\end{figure*}
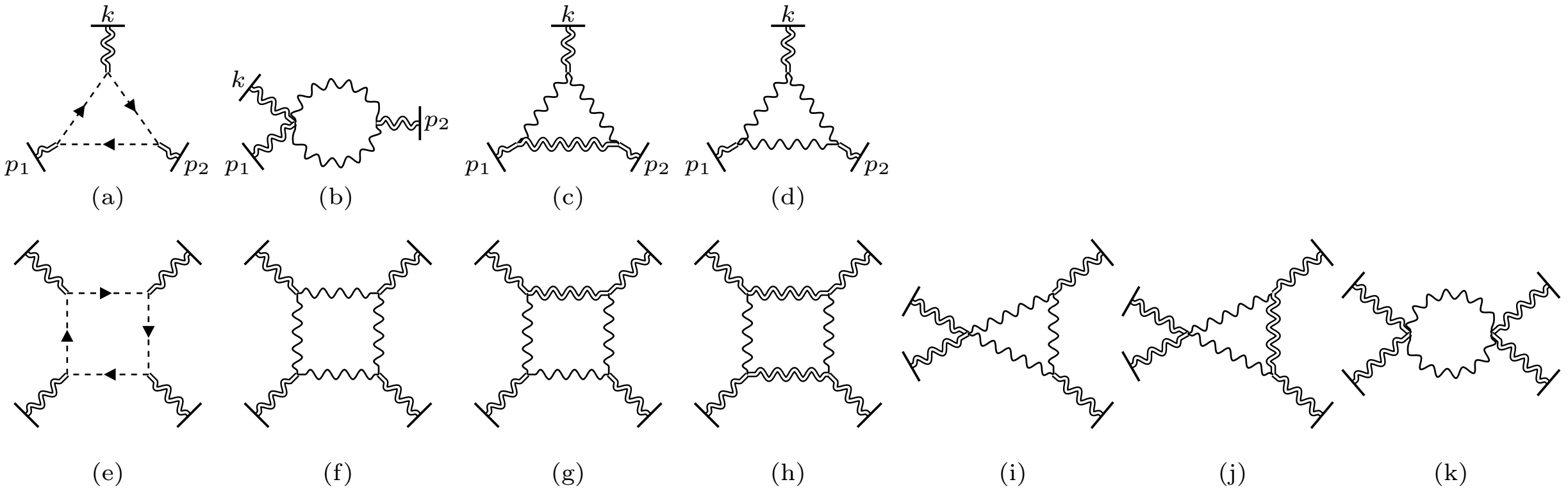

\paragraph{Four-point function of the background connection.} The diagrams of
$\delta^{4}_{\bar G}S_{\rm eff}$ are organized in the seven families of Sec.~\ref{sec:Q4}; the diagrams of
each family differ by the assignment of the four external legs to the vertices and, in the ghost loops,
by the orientation.
Figures~\ref{fig:diagGG}--\ref{fig:diagG} show the diagrams.

\begin{figure*}[tbp]
\centering
\setlength{\tabcolsep}{1pt}
\begin{tabular}{cccccc}
\begin{tikzpicture}[scale=0.42]
\node[inner sep=1pt] (e1) at (0,2.5) {\scriptsize $k$};
\coordinate (T) at (0,1.05);
\coordinate (L) at (-1.25,-0.72);
\coordinate (R) at (1.25,-0.72);
\node[inner sep=1pt] (e2) at (-2.2,-1.3) {\scriptsize $p_{1}$};
\node[inner sep=1pt] (e3) at (2.2,-1.3) {\scriptsize $p_{2}$};
\draw[grav,bgtick] (e1) -- (T);
\draw[aux,bgtick] (e2) -- (L);
\draw[aux,bgtick] (e3) -- (R);
\draw[grav] (T) -- (L);
\draw[grav] (L) -- (R);
\draw[grav] (R) -- (T);
\node at (0,-2.85) {\scriptsize (1)};
\end{tikzpicture} &
\begin{tikzpicture}[scale=0.42]
\node[inner sep=1pt] (e1) at (0,2.5) {\scriptsize $k$};
\coordinate (T) at (0,1.05);
\coordinate (L) at (-1.25,-0.72);
\coordinate (R) at (1.25,-0.72);
\node[inner sep=1pt] (e2) at (-2.2,-1.3) {\scriptsize $p_{1}$};
\node[inner sep=1pt] (e3) at (2.2,-1.3) {\scriptsize $p_{2}$};
\draw[grav,bgtick] (e1) -- (T);
\draw[aux,bgtick] (e2) -- (L);
\draw[aux,bgtick] (e3) -- (R);
\draw[aux] (T) -- (L);
\draw[grav] (L) -- (R);
\draw[grav] (R) -- (T);
\node at (0,-2.85) {\scriptsize (2)};
\end{tikzpicture} &
\begin{tikzpicture}[scale=0.42]
\node[inner sep=1pt] (e1) at (0,2.5) {\scriptsize $k$};
\coordinate (T) at (0,1.05);
\coordinate (L) at (-1.25,-0.72);
\coordinate (R) at (1.25,-0.72);
\node[inner sep=1pt] (e2) at (-2.2,-1.3) {\scriptsize $p_{1}$};
\node[inner sep=1pt] (e3) at (2.2,-1.3) {\scriptsize $p_{2}$};
\draw[grav,bgtick] (e1) -- (T);
\draw[aux,bgtick] (e2) -- (L);
\draw[aux,bgtick] (e3) -- (R);
\draw[grav] (T) -- (L);
\draw[aux] (L) -- (R);
\draw[grav] (R) -- (T);
\node at (0,-2.85) {\scriptsize (3)$^{*}$};
\end{tikzpicture} &
\begin{tikzpicture}[scale=0.42]
\node[inner sep=1pt] (e1) at (0,2.5) {\scriptsize $k$};
\coordinate (T) at (0,1.05);
\coordinate (L) at (-1.25,-0.72);
\coordinate (R) at (1.25,-0.72);
\node[inner sep=1pt] (e2) at (-2.2,-1.3) {\scriptsize $p_{1}$};
\node[inner sep=1pt] (e3) at (2.2,-1.3) {\scriptsize $p_{2}$};
\draw[grav,bgtick] (e1) -- (T);
\draw[aux,bgtick] (e2) -- (L);
\draw[aux,bgtick] (e3) -- (R);
\draw[grav] (T) -- (L);
\draw[grav] (L) -- (R);
\draw[aux] (R) -- (T);
\node at (0,-2.85) {\scriptsize (4)};
\end{tikzpicture} &
\begin{tikzpicture}[scale=0.42]
\node[inner sep=1pt] (e1) at (0,2.5) {\scriptsize $k$};
\coordinate (T) at (0,1.05);
\coordinate (L) at (-1.25,-0.72);
\coordinate (R) at (1.25,-0.72);
\node[inner sep=1pt] (e2) at (-2.2,-1.3) {\scriptsize $p_{1}$};
\node[inner sep=1pt] (e3) at (2.2,-1.3) {\scriptsize $p_{2}$};
\draw[grav,bgtick] (e1) -- (T);
\draw[aux,bgtick] (e2) -- (L);
\draw[aux,bgtick] (e3) -- (R);
\draw[gh,ghar] (T) -- (L);
\draw[gh,ghar] (L) -- (R);
\draw[gh,ghar] (R) -- (T);
\node at (0,-2.85) {\scriptsize (5)};
\end{tikzpicture} &
\begin{tikzpicture}[scale=0.42]
\node[inner sep=1pt] (e1) at (0,2.5) {\scriptsize $k$};
\coordinate (T) at (0,1.05);
\coordinate (L) at (-1.25,-0.72);
\coordinate (R) at (1.25,-0.72);
\node[inner sep=1pt] (e2) at (-2.2,-1.3) {\scriptsize $p_{1}$};
\node[inner sep=1pt] (e3) at (2.2,-1.3) {\scriptsize $p_{2}$};
\draw[grav,bgtick] (e1) -- (T);
\draw[aux,bgtick] (e2) -- (L);
\draw[aux,bgtick] (e3) -- (R);
\draw[gh,ghar] (T) -- (R);
\draw[gh,ghar] (R) -- (L);
\draw[gh,ghar] (L) -- (T);
\node at (0,-2.85) {\scriptsize (6)};
\end{tikzpicture} \\[2pt]
\begin{tikzpicture}[scale=0.42]
\node[inner sep=1pt] (e1) at (-2.4,1.1) {\scriptsize $k$};
\node[inner sep=1pt] (e2) at (-2.4,-1.1) {\scriptsize $p_{1}$};
\coordinate (L) at (-1,0);
\coordinate (R) at (1,0);
\node[inner sep=1pt] (e3) at (2.5,0) {\scriptsize $p_{2}$};
\draw[grav,bgtick] (e1) -- (L);
\draw[aux,bgtick] (e2) -- (L);
\draw[gh,ghar] (L) to[bend left=90,looseness=1.5] (R);
\draw[gh,ghar] (R) to[bend left=90,looseness=1.5] (L);
\draw[aux,bgtick] (e3) -- (R);
\node at (0,-2.55) {\scriptsize (7)};
\end{tikzpicture} &
\begin{tikzpicture}[scale=0.42]
\node[inner sep=1pt] (e1) at (-2.4,1.1) {\scriptsize $k$};
\node[inner sep=1pt] (e2) at (-2.4,-1.1) {\scriptsize $p_{2}$};
\coordinate (L) at (-1,0);
\coordinate (R) at (1,0);
\node[inner sep=1pt] (e3) at (2.5,0) {\scriptsize $p_{1}$};
\draw[grav,bgtick] (e1) -- (L);
\draw[aux,bgtick] (e2) -- (L);
\draw[gh,ghar] (L) to[bend left=90,looseness=1.5] (R);
\draw[gh,ghar] (R) to[bend left=90,looseness=1.5] (L);
\draw[aux,bgtick] (e3) -- (R);
\node at (0,-2.55) {\scriptsize (8)};
\end{tikzpicture} &
\begin{tikzpicture}[scale=0.42]
\node[inner sep=1pt] (e1) at (0,2.5) {\scriptsize $k$};
\coordinate (T) at (0,1.05);
\coordinate (L) at (-1.25,-0.72);
\coordinate (R) at (1.25,-0.72);
\node[inner sep=1pt] (e2) at (-2.2,-1.3) {\scriptsize $p_{1}$};
\node[inner sep=1pt] (e3) at (2.2,-1.3) {\scriptsize $p_{2}$};
\draw[grav,bgtick] (e1) -- (T);
\draw[aux,bgtick] (e2) -- (L);
\draw[aux,bgtick] (e3) -- (R);
\draw[aux] (T) -- (L);
\draw[grav] (L) -- (R);
\draw[aux] (R) -- (T);
\node at (0,-2.85) {\scriptsize (9)};
\end{tikzpicture} &
\begin{tikzpicture}[scale=0.42]
\node[inner sep=1pt] (e1) at (-2.4,1.1) {\scriptsize $p_{1}$};
\node[inner sep=1pt] (e2) at (-2.4,-1.1) {\scriptsize $p_{2}$};
\coordinate (L) at (-1,0);
\coordinate (R) at (1,0);
\node[inner sep=1pt] (e3) at (2.5,0) {\scriptsize $k$};
\draw[aux,bgtick] (e1) -- (L);
\draw[aux,bgtick] (e2) -- (L);
\draw[grav] (L) to[bend left=90,looseness=1.5] (R);
\draw[grav] (R) to[bend left=90,looseness=1.5] (L);
\draw[grav,bgtick] (e3) -- (R);
\node at (0,-2.55) {\scriptsize (10)$^{*}$};
\end{tikzpicture} &
\begin{tikzpicture}[scale=0.42]
\node[inner sep=1pt] (e1) at (-2.4,1.1) {\scriptsize $k$};
\node[inner sep=1pt] (e2) at (-2.4,-1.1) {\scriptsize $p_{1}$};
\coordinate (L) at (-1,0);
\coordinate (R) at (1,0);
\node[inner sep=1pt] (e3) at (2.5,0) {\scriptsize $p_{2}$};
\draw[grav,bgtick] (e1) -- (L);
\draw[aux,bgtick] (e2) -- (L);
\draw[grav] (L) to[bend left=90,looseness=1.5] (R);
\draw[grav] (R) to[bend left=90,looseness=1.5] (L);
\draw[aux,bgtick] (e3) -- (R);
\node at (0,-2.55) {\scriptsize (11)};
\end{tikzpicture} &
\begin{tikzpicture}[scale=0.42]
\node[inner sep=1pt] (e1) at (-2.4,1.1) {\scriptsize $k$};
\node[inner sep=1pt] (e2) at (-2.4,-1.1) {\scriptsize $p_{2}$};
\coordinate (L) at (-1,0);
\coordinate (R) at (1,0);
\node[inner sep=1pt] (e3) at (2.5,0) {\scriptsize $p_{1}$};
\draw[grav,bgtick] (e1) -- (L);
\draw[aux,bgtick] (e2) -- (L);
\draw[grav] (L) to[bend left=90,looseness=1.5] (R);
\draw[grav] (R) to[bend left=90,looseness=1.5] (L);
\draw[aux,bgtick] (e3) -- (R);
\node at (0,-2.55) {\scriptsize (12)};
\end{tikzpicture}
\end{tabular}
\caption{One-loop diagrams of $\delta_{\bar\phi}\delta^{2}_{\bar G}S_{\rm eff}$, numbered as in Table~\ref{tab:diagGG}. Momentum labels as in Sec.~\ref{sec:threepoint}; an asterisk marks the diagrams that are transverse in the leg carrying the inhomogeneous variation.}
\label{fig:diagGG}
\end{figure*}
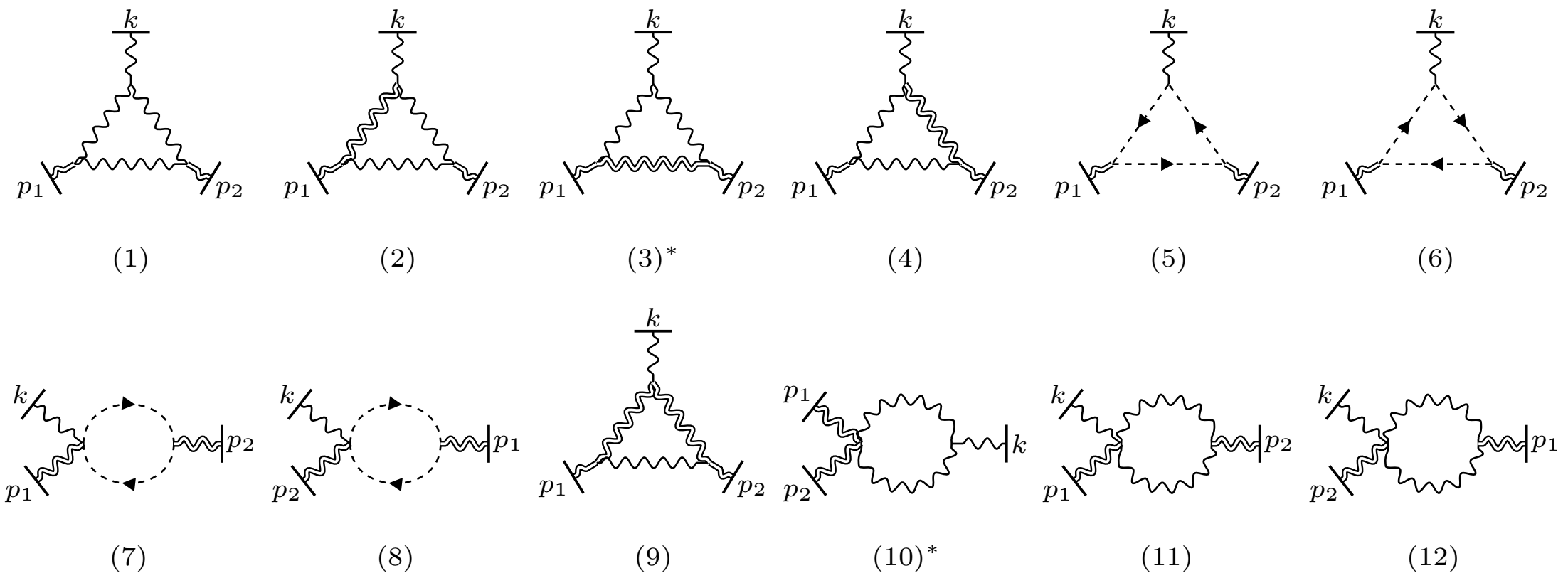

\begin{figure*}[tbp]
\centering
\setlength{\tabcolsep}{1pt}
\begin{tabular}{ccccccc}
\begin{tikzpicture}[scale=0.42]
\node[inner sep=1pt] (e1) at (0,2.5) {\scriptsize $k$};
\coordinate (T) at (0,1.05);
\coordinate (L) at (1.25,-0.72);
\coordinate (R) at (-1.25,-0.72);
\node[inner sep=1pt] (e2) at (2.2,-1.3) {\scriptsize $p_{2}$};
\node[inner sep=1pt] (e3) at (-2.2,-1.3) {\scriptsize $p_{1}$};
\draw[aux,bgtick] (e1) -- (T);
\draw[grav,bgtick] (e2) -- (L);
\draw[grav,bgtick] (e3) -- (R);
\draw[grav] (T) -- (L);
\draw[grav] (L) -- (R);
\draw[grav] (R) -- (T);
\node at (0,-2.85) {\scriptsize (1)};
\end{tikzpicture} &
\begin{tikzpicture}[scale=0.42]
\node[inner sep=1pt] (e1) at (0,2.5) {\scriptsize $k$};
\coordinate (T) at (0,1.05);
\coordinate (L) at (1.25,-0.72);
\coordinate (R) at (-1.25,-0.72);
\node[inner sep=1pt] (e2) at (2.2,-1.3) {\scriptsize $p_{2}$};
\node[inner sep=1pt] (e3) at (-2.2,-1.3) {\scriptsize $p_{1}$};
\draw[aux,bgtick] (e1) -- (T);
\draw[grav,bgtick] (e2) -- (L);
\draw[grav,bgtick] (e3) -- (R);
\draw[grav] (T) -- (L);
\draw[aux] (L) -- (R);
\draw[grav] (R) -- (T);
\node at (0,-2.85) {\scriptsize (2)};
\end{tikzpicture} &
\begin{tikzpicture}[scale=0.42]
\node[inner sep=1pt] (e1) at (0,2.5) {\scriptsize $k$};
\coordinate (T) at (0,1.05);
\coordinate (L) at (1.25,-0.72);
\coordinate (R) at (-1.25,-0.72);
\node[inner sep=1pt] (e2) at (2.2,-1.3) {\scriptsize $p_{2}$};
\node[inner sep=1pt] (e3) at (-2.2,-1.3) {\scriptsize $p_{1}$};
\draw[aux,bgtick] (e1) -- (T);
\draw[grav,bgtick] (e2) -- (L);
\draw[grav,bgtick] (e3) -- (R);
\draw[grav] (T) -- (L);
\draw[grav] (L) -- (R);
\draw[aux] (R) -- (T);
\node at (0,-2.85) {\scriptsize (3)$^{*}$};
\end{tikzpicture} &
\begin{tikzpicture}[scale=0.42]
\node[inner sep=1pt] (e1) at (0,2.5) {\scriptsize $k$};
\coordinate (T) at (0,1.05);
\coordinate (L) at (1.25,-0.72);
\coordinate (R) at (-1.25,-0.72);
\node[inner sep=1pt] (e2) at (2.2,-1.3) {\scriptsize $p_{2}$};
\node[inner sep=1pt] (e3) at (-2.2,-1.3) {\scriptsize $p_{1}$};
\draw[aux,bgtick] (e1) -- (T);
\draw[grav,bgtick] (e2) -- (L);
\draw[grav,bgtick] (e3) -- (R);
\draw[aux] (T) -- (L);
\draw[grav] (L) -- (R);
\draw[grav] (R) -- (T);
\node at (0,-2.85) {\scriptsize (4)};
\end{tikzpicture} &
\begin{tikzpicture}[scale=0.42]
\node[inner sep=1pt] (e1) at (0,2.5) {\scriptsize $k$};
\coordinate (T) at (0,1.05);
\coordinate (L) at (1.25,-0.72);
\coordinate (R) at (-1.25,-0.72);
\node[inner sep=1pt] (e2) at (2.2,-1.3) {\scriptsize $p_{2}$};
\node[inner sep=1pt] (e3) at (-2.2,-1.3) {\scriptsize $p_{1}$};
\draw[aux,bgtick] (e1) -- (T);
\draw[grav,bgtick] (e2) -- (L);
\draw[grav,bgtick] (e3) -- (R);
\draw[aux] (T) -- (L);
\draw[aux] (L) -- (R);
\draw[grav] (R) -- (T);
\node at (0,-2.85) {\scriptsize (5)};
\end{tikzpicture} &
\begin{tikzpicture}[scale=0.42]
\node[inner sep=1pt] (e1) at (0,2.5) {\scriptsize $k$};
\coordinate (T) at (0,1.05);
\coordinate (L) at (1.25,-0.72);
\coordinate (R) at (-1.25,-0.72);
\node[inner sep=1pt] (e2) at (2.2,-1.3) {\scriptsize $p_{2}$};
\node[inner sep=1pt] (e3) at (-2.2,-1.3) {\scriptsize $p_{1}$};
\draw[aux,bgtick] (e1) -- (T);
\draw[grav,bgtick] (e2) -- (L);
\draw[grav,bgtick] (e3) -- (R);
\draw[grav] (T) -- (L);
\draw[aux] (L) -- (R);
\draw[aux] (R) -- (T);
\node at (0,-2.85) {\scriptsize (6)};
\end{tikzpicture} &
\begin{tikzpicture}[scale=0.42]
\node[inner sep=1pt] (e1) at (0,2.5) {\scriptsize $k$};
\coordinate (T) at (0,1.05);
\coordinate (L) at (1.25,-0.72);
\coordinate (R) at (-1.25,-0.72);
\node[inner sep=1pt] (e2) at (2.2,-1.3) {\scriptsize $p_{2}$};
\node[inner sep=1pt] (e3) at (-2.2,-1.3) {\scriptsize $p_{1}$};
\draw[aux,bgtick] (e1) -- (T);
\draw[grav,bgtick] (e2) -- (L);
\draw[grav,bgtick] (e3) -- (R);
\draw[gh,ghar] (T) -- (L);
\draw[gh,ghar] (L) -- (R);
\draw[gh,ghar] (R) -- (T);
\node at (0,-2.85) {\scriptsize (7)};
\end{tikzpicture} \\[2pt]
\begin{tikzpicture}[scale=0.42]
\node[inner sep=1pt] (e1) at (0,2.5) {\scriptsize $k$};
\coordinate (T) at (0,1.05);
\coordinate (L) at (1.25,-0.72);
\coordinate (R) at (-1.25,-0.72);
\node[inner sep=1pt] (e2) at (2.2,-1.3) {\scriptsize $p_{2}$};
\node[inner sep=1pt] (e3) at (-2.2,-1.3) {\scriptsize $p_{1}$};
\draw[aux,bgtick] (e1) -- (T);
\draw[grav,bgtick] (e2) -- (L);
\draw[grav,bgtick] (e3) -- (R);
\draw[gh,ghar] (T) -- (R);
\draw[gh,ghar] (R) -- (L);
\draw[gh,ghar] (L) -- (T);
\node at (0,-2.85) {\scriptsize (8)};
\end{tikzpicture} &
\begin{tikzpicture}[scale=0.42]
\node[inner sep=1pt] (e1) at (-2.4,1.1) {\scriptsize $p_{2}$};
\node[inner sep=1pt] (e2) at (-2.4,-1.1) {\scriptsize $p_{1}$};
\coordinate (L) at (-1,0);
\coordinate (R) at (1,0);
\node[inner sep=1pt] (e3) at (2.5,0) {\scriptsize $k$};
\draw[grav,bgtick] (e1) -- (L);
\draw[grav,bgtick] (e2) -- (L);
\draw[gh,ghar] (L) to[bend left=90,looseness=1.5] (R);
\draw[gh,ghar] (R) to[bend left=90,looseness=1.5] (L);
\draw[aux,bgtick] (e3) -- (R);
\node at (0,-2.55) {\scriptsize (9)};
\end{tikzpicture} &
\begin{tikzpicture}[scale=0.42]
\node[inner sep=1pt] (e1) at (-2.4,1.1) {\scriptsize $k$};
\node[inner sep=1pt] (e2) at (-2.4,-1.1) {\scriptsize $p_{2}$};
\coordinate (L) at (-1,0);
\coordinate (R) at (1,0);
\node[inner sep=1pt] (e3) at (2.5,0) {\scriptsize $p_{1}$};
\draw[aux,bgtick] (e1) -- (L);
\draw[grav,bgtick] (e2) -- (L);
\draw[gh,ghar] (L) to[bend left=90,looseness=1.5] (R);
\draw[gh,ghar] (R) to[bend left=90,looseness=1.5] (L);
\draw[grav,bgtick] (e3) -- (R);
\node at (0,-2.55) {\scriptsize (10)};
\end{tikzpicture} &
\begin{tikzpicture}[scale=0.42]
\node[inner sep=1pt] (e1) at (-2.4,1.1) {\scriptsize $k$};
\node[inner sep=1pt] (e2) at (-2.4,-1.1) {\scriptsize $p_{1}$};
\coordinate (L) at (-1,0);
\coordinate (R) at (1,0);
\node[inner sep=1pt] (e3) at (2.5,0) {\scriptsize $p_{2}$};
\draw[aux,bgtick] (e1) -- (L);
\draw[grav,bgtick] (e2) -- (L);
\draw[gh,ghar] (L) to[bend left=90,looseness=1.5] (R);
\draw[gh,ghar] (R) to[bend left=90,looseness=1.5] (L);
\draw[grav,bgtick] (e3) -- (R);
\node at (0,-2.55) {\scriptsize (11)$^{*}$};
\end{tikzpicture} &
\begin{tikzpicture}[scale=0.42]
\node[inner sep=1pt] (e1) at (-2.4,1.1) {\scriptsize $p_{2}$};
\node[inner sep=1pt] (e2) at (-2.4,-1.1) {\scriptsize $p_{1}$};
\coordinate (L) at (-1,0);
\coordinate (R) at (1,0);
\node[inner sep=1pt] (e3) at (2.5,0) {\scriptsize $k$};
\draw[grav,bgtick] (e1) -- (L);
\draw[grav,bgtick] (e2) -- (L);
\draw[grav] (L) to[bend left=90,looseness=1.5] (R);
\draw[grav] (R) to[bend left=90,looseness=1.5] (L);
\draw[aux,bgtick] (e3) -- (R);
\node at (0,-2.55) {\scriptsize (12)};
\end{tikzpicture} &
\begin{tikzpicture}[scale=0.42]
\node[inner sep=1pt] (e1) at (-2.4,1.1) {\scriptsize $k$};
\node[inner sep=1pt] (e2) at (-2.4,-1.1) {\scriptsize $p_{2}$};
\coordinate (L) at (-1,0);
\coordinate (R) at (1,0);
\node[inner sep=1pt] (e3) at (2.5,0) {\scriptsize $p_{1}$};
\draw[aux,bgtick] (e1) -- (L);
\draw[grav,bgtick] (e2) -- (L);
\draw[grav] (L) to[bend left=90,looseness=1.5] (R);
\draw[grav] (R) to[bend left=90,looseness=1.5] (L);
\draw[grav,bgtick] (e3) -- (R);
\node at (0,-2.55) {\scriptsize (13)};
\end{tikzpicture} &
\begin{tikzpicture}[scale=0.42]
\node[inner sep=1pt] (e1) at (-2.4,1.1) {\scriptsize $k$};
\node[inner sep=1pt] (e2) at (-2.4,-1.1) {\scriptsize $p_{1}$};
\coordinate (L) at (-1,0);
\coordinate (R) at (1,0);
\node[inner sep=1pt] (e3) at (2.5,0) {\scriptsize $p_{2}$};
\draw[aux,bgtick] (e1) -- (L);
\draw[grav,bgtick] (e2) -- (L);
\draw[grav] (L) to[bend left=90,looseness=1.5] (R);
\draw[grav] (R) to[bend left=90,looseness=1.5] (L);
\draw[grav,bgtick] (e3) -- (R);
\node at (0,-2.55) {\scriptsize (14)$^{*}$};
\end{tikzpicture}
\end{tabular}
\caption{One-loop diagrams of $\delta^{2}_{\bar\phi}\delta_{\bar G}S_{\rm eff}$, numbered as in Table~\ref{tab:diagphiG}. Momentum labels as in Sec.~\ref{sec:threepoint}; an asterisk marks the diagrams that are transverse in the leg carrying the inhomogeneous variation.}
\label{fig:diagphiG}
\end{figure*}
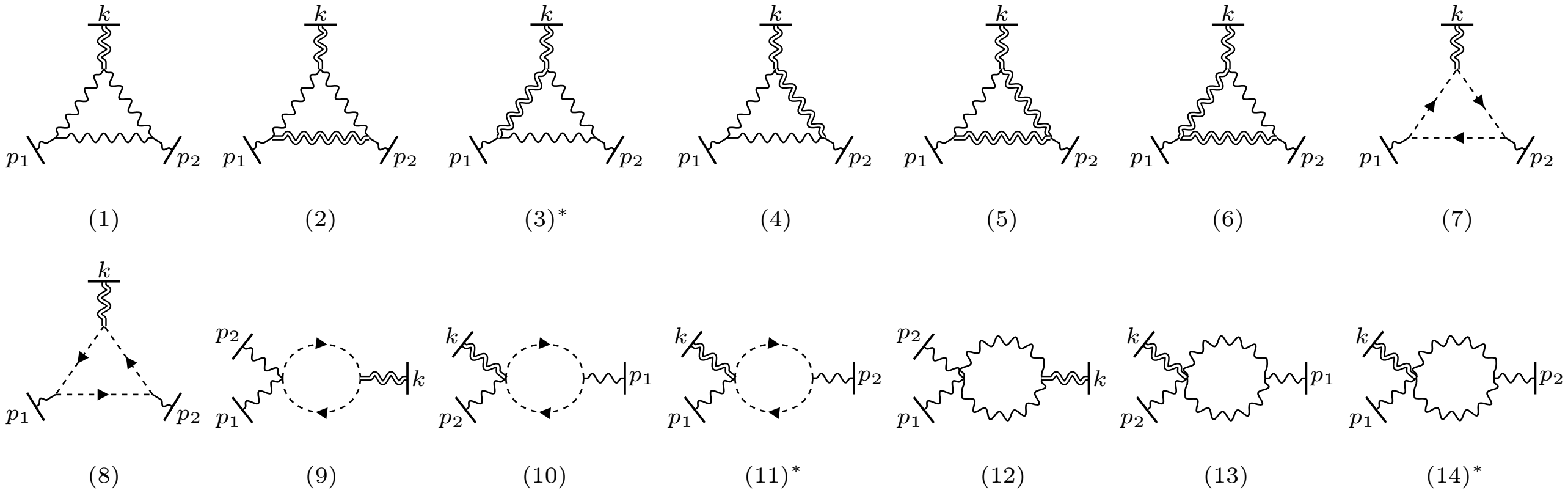

\begin{figure*}[tbp]
\centering
\setlength{\tabcolsep}{1pt}
\begin{tabular}{cccccccc}
\begin{tikzpicture}[scale=0.37]
\node[inner sep=1pt] (e1) at (0,2.5) {\scriptsize $k$};
\coordinate (T) at (0,1.05);
\coordinate (L) at (-1.25,-0.72);
\coordinate (R) at (1.25,-0.72);
\node[inner sep=1pt] (e2) at (-2.2,-1.3) {\scriptsize $p_{1}$};
\node[inner sep=1pt] (e3) at (2.2,-1.3) {\scriptsize $p_{2}$};
\draw[grav,bgtick] (e1) -- (T);
\draw[grav,bgtick] (e2) -- (L);
\draw[grav,bgtick] (e3) -- (R);
\draw[grav] (T) -- (L);
\draw[grav] (L) -- (R);
\draw[grav] (R) -- (T);
\node at (0,-2.05) {\scriptsize (1)};
\end{tikzpicture} &
\begin{tikzpicture}[scale=0.37]
\node[inner sep=1pt] (e1) at (0,2.5) {\scriptsize $k$};
\coordinate (T) at (0,1.05);
\coordinate (L) at (-1.25,-0.72);
\coordinate (R) at (1.25,-0.72);
\node[inner sep=1pt] (e2) at (-2.2,-1.3) {\scriptsize $p_{1}$};
\node[inner sep=1pt] (e3) at (2.2,-1.3) {\scriptsize $p_{2}$};
\draw[grav,bgtick] (e1) -- (T);
\draw[grav,bgtick] (e2) -- (L);
\draw[grav,bgtick] (e3) -- (R);
\draw[gh,ghar] (T) -- (L);
\draw[gh,ghar] (L) -- (R);
\draw[gh,ghar] (R) -- (T);
\node at (0,-2.05) {\scriptsize (2)};
\end{tikzpicture} &
\begin{tikzpicture}[scale=0.37]
\node[inner sep=1pt] (e1) at (0,2.5) {\scriptsize $k$};
\coordinate (T) at (0,1.05);
\coordinate (L) at (-1.25,-0.72);
\coordinate (R) at (1.25,-0.72);
\node[inner sep=1pt] (e2) at (-2.2,-1.3) {\scriptsize $p_{1}$};
\node[inner sep=1pt] (e3) at (2.2,-1.3) {\scriptsize $p_{2}$};
\draw[grav,bgtick] (e1) -- (T);
\draw[grav,bgtick] (e2) -- (L);
\draw[grav,bgtick] (e3) -- (R);
\draw[gh,ghar] (T) -- (R);
\draw[gh,ghar] (R) -- (L);
\draw[gh,ghar] (L) -- (T);
\node at (0,-2.05) {\scriptsize (3)};
\end{tikzpicture} &
\begin{tikzpicture}[scale=0.37]
\node[inner sep=1pt] (e1) at (0,2.5) {\scriptsize $k$};
\coordinate (T) at (0,1.05);
\coordinate (L) at (-1.25,-0.72);
\coordinate (R) at (1.25,-0.72);
\node[inner sep=1pt] (e2) at (-2.2,-1.3) {\scriptsize $p_{1}$};
\node[inner sep=1pt] (e3) at (2.2,-1.3) {\scriptsize $p_{2}$};
\draw[grav,bgtick] (e1) -- (T);
\draw[grav,bgtick] (e2) -- (L);
\draw[grav,bgtick] (e3) -- (R);
\draw[grav] (T) -- (L);
\draw[aux] (L) -- (R);
\draw[grav] (R) -- (T);
\node at (0,-2.05) {\scriptsize (4)};
\end{tikzpicture} &
\begin{tikzpicture}[scale=0.37]
\node[inner sep=1pt] (e1) at (0,2.5) {\scriptsize $k$};
\coordinate (T) at (0,1.05);
\coordinate (L) at (-1.25,-0.72);
\coordinate (R) at (1.25,-0.72);
\node[inner sep=1pt] (e2) at (-2.2,-1.3) {\scriptsize $p_{1}$};
\node[inner sep=1pt] (e3) at (2.2,-1.3) {\scriptsize $p_{2}$};
\draw[grav,bgtick] (e1) -- (T);
\draw[grav,bgtick] (e2) -- (L);
\draw[grav,bgtick] (e3) -- (R);
\draw[grav] (T) -- (L);
\draw[grav] (L) -- (R);
\draw[aux] (R) -- (T);
\node at (0,-2.05) {\scriptsize (5)};
\end{tikzpicture} &
\begin{tikzpicture}[scale=0.37]
\node[inner sep=1pt] (e1) at (0,2.5) {\scriptsize $k$};
\coordinate (T) at (0,1.05);
\coordinate (L) at (-1.25,-0.72);
\coordinate (R) at (1.25,-0.72);
\node[inner sep=1pt] (e2) at (-2.2,-1.3) {\scriptsize $p_{1}$};
\node[inner sep=1pt] (e3) at (2.2,-1.3) {\scriptsize $p_{2}$};
\draw[grav,bgtick] (e1) -- (T);
\draw[grav,bgtick] (e2) -- (L);
\draw[grav,bgtick] (e3) -- (R);
\draw[aux] (T) -- (L);
\draw[grav] (L) -- (R);
\draw[grav] (R) -- (T);
\node at (0,-2.05) {\scriptsize (6)};
\end{tikzpicture} &
\begin{tikzpicture}[scale=0.37]
\node[inner sep=1pt] (e1) at (-2.4,1.1) {\scriptsize $p_{1}$};
\node[inner sep=1pt] (e2) at (-2.4,-1.1) {\scriptsize $p_{2}$};
\coordinate (L) at (-1,0);
\coordinate (R) at (1,0);
\node[inner sep=1pt] (e3) at (2.5,0) {\scriptsize $k$};
\draw[grav,bgtick] (e1) -- (L);
\draw[grav,bgtick] (e2) -- (L);
\draw[grav] (L) to[bend left=90,looseness=1.5] (R);
\draw[grav] (R) to[bend left=90,looseness=1.5] (L);
\draw[grav,bgtick] (e3) -- (R);
\node at (0,-1.85) {\scriptsize (7)};
\end{tikzpicture} &
\begin{tikzpicture}[scale=0.37]
\node[inner sep=1pt] (e1) at (-2.4,1.1) {\scriptsize $k$};
\node[inner sep=1pt] (e2) at (-2.4,-1.1) {\scriptsize $p_{2}$};
\coordinate (L) at (-1,0);
\coordinate (R) at (1,0);
\node[inner sep=1pt] (e3) at (2.5,0) {\scriptsize $p_{1}$};
\draw[grav,bgtick] (e1) -- (L);
\draw[grav,bgtick] (e2) -- (L);
\draw[grav] (L) to[bend left=90,looseness=1.5] (R);
\draw[grav] (R) to[bend left=90,looseness=1.5] (L);
\draw[grav,bgtick] (e3) -- (R);
\node at (0,-1.85) {\scriptsize (8)};
\end{tikzpicture} \\[2pt]
\begin{tikzpicture}[scale=0.37]
\node[inner sep=1pt] (e1) at (-2.4,1.1) {\scriptsize $k$};
\node[inner sep=1pt] (e2) at (-2.4,-1.1) {\scriptsize $p_{1}$};
\coordinate (L) at (-1,0);
\coordinate (R) at (1,0);
\node[inner sep=1pt] (e3) at (2.5,0) {\scriptsize $p_{2}$};
\draw[grav,bgtick] (e1) -- (L);
\draw[grav,bgtick] (e2) -- (L);
\draw[grav] (L) to[bend left=90,looseness=1.5] (R);
\draw[grav] (R) to[bend left=90,looseness=1.5] (L);
\draw[grav,bgtick] (e3) -- (R);
\node at (0,-1.85) {\scriptsize (9)};
\end{tikzpicture} &
\begin{tikzpicture}[scale=0.37]
\node[inner sep=1pt] (e1) at (-2.4,1.1) {\scriptsize $p_{1}$};
\node[inner sep=1pt] (e2) at (-2.4,-1.1) {\scriptsize $p_{2}$};
\coordinate (L) at (-1,0);
\coordinate (R) at (1,0);
\node[inner sep=1pt] (e3) at (2.5,0) {\scriptsize $k$};
\draw[grav,bgtick] (e1) -- (L);
\draw[grav,bgtick] (e2) -- (L);
\draw[gh,ghar] (L) to[bend left=90,looseness=1.5] (R);
\draw[gh,ghar] (R) to[bend left=90,looseness=1.5] (L);
\draw[grav,bgtick] (e3) -- (R);
\node at (0,-1.85) {\scriptsize (10)};
\end{tikzpicture} &
\begin{tikzpicture}[scale=0.37]
\node[inner sep=1pt] (e1) at (-2.4,1.1) {\scriptsize $k$};
\node[inner sep=1pt] (e2) at (-2.4,-1.1) {\scriptsize $p_{2}$};
\coordinate (L) at (-1,0);
\coordinate (R) at (1,0);
\node[inner sep=1pt] (e3) at (2.5,0) {\scriptsize $p_{1}$};
\draw[grav,bgtick] (e1) -- (L);
\draw[grav,bgtick] (e2) -- (L);
\draw[gh,ghar] (L) to[bend left=90,looseness=1.5] (R);
\draw[gh,ghar] (R) to[bend left=90,looseness=1.5] (L);
\draw[grav,bgtick] (e3) -- (R);
\node at (0,-1.85) {\scriptsize (11)};
\end{tikzpicture} &
\begin{tikzpicture}[scale=0.37]
\node[inner sep=1pt] (e1) at (-2.4,1.1) {\scriptsize $k$};
\node[inner sep=1pt] (e2) at (-2.4,-1.1) {\scriptsize $p_{1}$};
\coordinate (L) at (-1,0);
\coordinate (R) at (1,0);
\node[inner sep=1pt] (e3) at (2.5,0) {\scriptsize $p_{2}$};
\draw[grav,bgtick] (e1) -- (L);
\draw[grav,bgtick] (e2) -- (L);
\draw[gh,ghar] (L) to[bend left=90,looseness=1.5] (R);
\draw[gh,ghar] (R) to[bend left=90,looseness=1.5] (L);
\draw[grav,bgtick] (e3) -- (R);
\node at (0,-1.85) {\scriptsize (12)};
\end{tikzpicture} &
\begin{tikzpicture}[scale=0.37]
\node[inner sep=1pt] (e1) at (0,2.5) {\scriptsize $k$};
\coordinate (T) at (0,1.05);
\coordinate (L) at (-1.25,-0.72);
\coordinate (R) at (1.25,-0.72);
\node[inner sep=1pt] (e2) at (-2.2,-1.3) {\scriptsize $p_{1}$};
\node[inner sep=1pt] (e3) at (2.2,-1.3) {\scriptsize $p_{2}$};
\draw[grav,bgtick] (e1) -- (T);
\draw[grav,bgtick] (e2) -- (L);
\draw[grav,bgtick] (e3) -- (R);
\draw[aux] (T) -- (L);
\draw[grav] (L) -- (R);
\draw[aux] (R) -- (T);
\node at (0,-2.05) {\scriptsize (13)};
\end{tikzpicture} &
\begin{tikzpicture}[scale=0.37]
\node[inner sep=1pt] (e1) at (0,2.5) {\scriptsize $k$};
\coordinate (T) at (0,1.05);
\coordinate (L) at (-1.25,-0.72);
\coordinate (R) at (1.25,-0.72);
\node[inner sep=1pt] (e2) at (-2.2,-1.3) {\scriptsize $p_{1}$};
\node[inner sep=1pt] (e3) at (2.2,-1.3) {\scriptsize $p_{2}$};
\draw[grav,bgtick] (e1) -- (T);
\draw[grav,bgtick] (e2) -- (L);
\draw[grav,bgtick] (e3) -- (R);
\draw[grav] (T) -- (L);
\draw[aux] (L) -- (R);
\draw[aux] (R) -- (T);
\node at (0,-2.05) {\scriptsize (14)};
\end{tikzpicture} &
\begin{tikzpicture}[scale=0.37]
\node[inner sep=1pt] (e1) at (0,2.5) {\scriptsize $k$};
\coordinate (T) at (0,1.05);
\coordinate (L) at (-1.25,-0.72);
\coordinate (R) at (1.25,-0.72);
\node[inner sep=1pt] (e2) at (-2.2,-1.3) {\scriptsize $p_{1}$};
\node[inner sep=1pt] (e3) at (2.2,-1.3) {\scriptsize $p_{2}$};
\draw[grav,bgtick] (e1) -- (T);
\draw[grav,bgtick] (e2) -- (L);
\draw[grav,bgtick] (e3) -- (R);
\draw[aux] (T) -- (L);
\draw[aux] (L) -- (R);
\draw[grav] (R) -- (T);
\node at (0,-2.05) {\scriptsize (15)};
\end{tikzpicture} &
\begin{tikzpicture}[scale=0.37]
\node[inner sep=1pt] (e1) at (0,2.5) {\scriptsize $k$};
\coordinate (T) at (0,1.05);
\coordinate (L) at (-1.25,-0.72);
\coordinate (R) at (1.25,-0.72);
\node[inner sep=1pt] (e2) at (-2.2,-1.3) {\scriptsize $p_{1}$};
\node[inner sep=1pt] (e3) at (2.2,-1.3) {\scriptsize $p_{2}$};
\draw[grav,bgtick] (e1) -- (T);
\draw[grav,bgtick] (e2) -- (L);
\draw[grav,bgtick] (e3) -- (R);
\draw[aux] (T) -- (L);
\draw[aux] (L) -- (R);
\draw[aux] (R) -- (T);
\node at (0,-2.05) {\scriptsize (16)};
\end{tikzpicture}
\end{tabular}
\caption{One-loop diagrams of $\delta^{3}_{\bar\phi}S_{\rm eff}$, numbered as in Table~\ref{tab:diagphiphi}; the variation is on the leg $k$. Diagrams (13)--(16) vanish.}
\label{fig:diagphiphi}
\end{figure*}
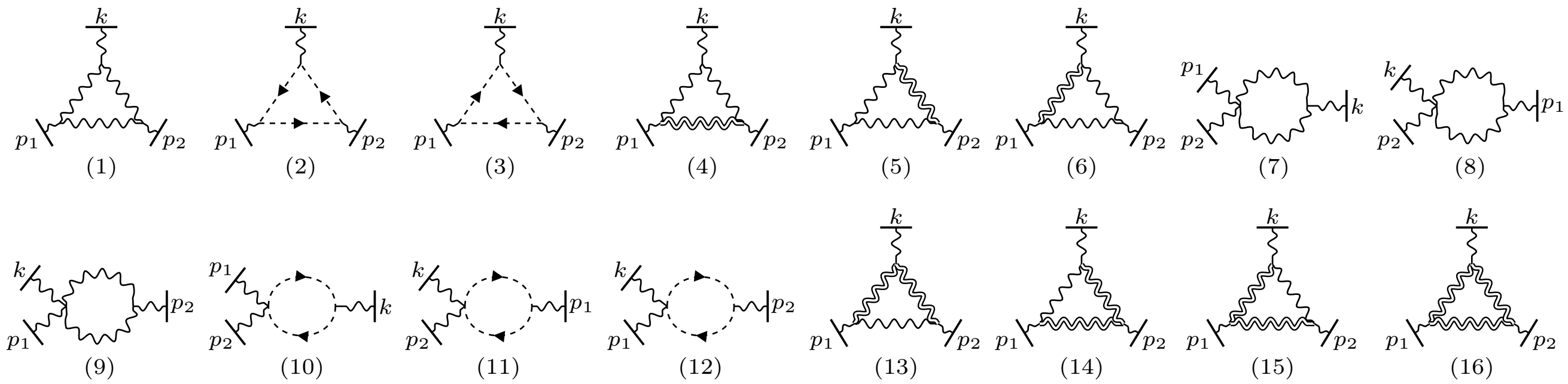

\section{Extraction of the ultraviolet poles}
\label{app:poles}

\paragraph{Degree of the pole.} The one-loop pole of a function with $n_{G}$ external $\bar G$
legs is a homogeneous polynomial of degree $4-n_{G}$ in the external momenta
(Sec.~\ref{sec:poles}): four for $\delta^{2}_{\bar\phi}S_{\rm eff}$ and $\delta^{3}_{\bar\phi}S_{\rm eff}$,
three for $\delta_{\bar\phi}\delta_{\bar G}S_{\rm eff}$ and $\delta^{2}_{\bar\phi}\delta_{\bar G}S_{\rm eff}$, two
for $\delta^{2}_{\bar G}S_{\rm eff}$ and $\delta_{\bar\phi}\delta^{2}_{\bar G}S_{\rm eff}$, one for
$\delta^{3}_{\bar G}S_{\rm eff}$ and zero for $\delta^{4}_{\bar G}S_{\rm eff}$.

\paragraph{Expansion in the external momenta.} Expanding the integrand of a diagram in the
external momenta, which amounts to the region of large loop momentum, each term becomes an
integral $\int d^{d}l\,l^{\mu_{1}}\cdots l^{\mu_{2j}}/(l^{2})^{m}$ without scale. In dimensional
regularization these integrals vanish, but only because their ultraviolet and infrared poles
cancel. The ultraviolet pole of the original diagram is well defined and does not depend on how
the infrared region is regulated; it can be isolated, for instance, by giving all propagators a
common auxiliary mass, which removes the infrared divergence without changing the ultraviolet
one. The counterterm of degree $4-n_{G}$ in the external momenta comes from the logarithmically
divergent terms of the expansion, of total degree $-d$ in $l$, and only these are kept. Terms of
lower degree converge in the ultraviolet. Terms of higher degree diverge as powers; with the
auxiliary mass they may produce poles proportional to powers of that mass, which are artifacts of
the infrared regulator and are distinguished from, and discarded with respect to, the result we
seek. It therefore suffices to keep the order $4-n_{G}$ in the external momenta.

\paragraph{Logarithmic tadpole rule.} The retained terms are reduced by symmetric integration,
and the ultraviolet pole of the resulting logarithmic integral is that of $B_{0}$:
\begin{multline}
\int\frac{d^{d}l}{(2\pi)^{d}}\,\frac{l^{\mu_{1}}\cdots l^{\mu_{2j}}}{(l^{2})^{j+2}}\bigg|_{\rm UV\ pole}\\
=\frac{i}{8\pi^{2}(4-d)}\,
\frac{\eta^{\{\mu_{1}\mu_{2}}\cdots\eta^{\mu_{2j-1}\mu_{2j}\}}}{d(d+2)\cdots(d+2j-2)},
\label{eq:logtadpole}
\end{multline}
where the braces denote the sum over the $(2j-1)!!$ pairings of the indices. Odd powers of $l$
vanish, and $d=4$ may be taken in the denominator of the residue. For the three two-point functions
the procedure reproduces the pole of the full amplitudes, for all $\xi$.

\paragraph{Exact angular average.} For the recomputation of $\delta^{4}_{\bar G}S_{\rm eff}$, whose
numerators are polynomials of degree at most four in the direction of the loop momentum, the
symmetric integration of Eq.~\eqref{eq:logtadpole}, which holds for any $j$, can be done exactly
on a finite set of directions. After Wick rotation it is the average of the numerator over the unit
sphere $S^{3}$. The $24$ vertices of the $24$-cell
form a spherical $5$-design, so the average over them reproduces the sphere average of every
polynomial of degree up to five exactly. This gives an independent way of computing the pole of
$\delta^{4}_{\bar G}S_{\rm eff}$ at vanishing external momenta in $d=4$: the four legs are contracted
with random integer tensors, each diagram becomes the trace of a product of matrices, and the
average over the $24$ directions is computed in exact arithmetic. The result of this check is
given in Sec.~\ref{sec:vertices}.
\section{Invariant basis, conventions and completeness}
\label{app:basis}

\paragraph{Symbolic vertices.} The rank comparisons of Secs.~\ref{sec:basis} and~\ref{sec:vertices} are
symbolic and exact, and were made with xAct~\cite{xAct,MartinGarcia2008,Nutma2014}; the numerical checks
of completeness are the modular counts and the cross-check in $\phi GGG$ described at the end of this appendix. Each invariant is
written with the covariant objects of Sec.~\ref{sec:geometry}, in the exact
parametrization~\eqref{eq:ctparam} and with the conventions of Sec.~\ref{sec:blocks}. The fields are
replaced by plane waves with symbolic polarizations and momenta, momentum conservation is imposed, and
the vertex is the term multilinear in the polarizations, brought to canonical form. Imposing momentum
conservation identifies automatically the contractions related by integration by parts. Equalities and
ranks are then equalities and ranks between polynomials in the canonical monomials, with rational
coefficients.

\paragraph{Contractions.} For each class of Eq.~\eqref{eq:classes} we generated all the complete
contractions of its indices and discarded those related by the symmetries of the factors. The
$945$ complete contractions of the ten indices of $\Riem\,QQ$ and of $\nabla Q\,QQ$ reduce in this way to
$75$ and $95$ distinct ones, and the $10395$ contractions of the twelve indices of $Q^{4}$ to $69$; the
classes $\mathcal{R}^{2}$, $\mathcal{R}\nabla Q$ and $(\nabla Q)^{2}$ have $16$, $27$ and $25$ distinct
contractions. Of these, $3$ contractions of $\mathcal{R}\nabla Q$ and $14$ of $\Riem\,QQ$ vanish identically,
since they contract the antisymmetric pair of the curvature with a symmetric pair of $Q$; they do not
affect the ranks. The bases of Tables~\ref{tab:dQ2}, \ref{tab:RQQ}, \ref{tab:QQdQ} and~\ref{tab:Q4} were
chosen from the simplest to the most complex, with the contractions with more traces first, discarding
the dependent ones. In the $GGG$ parts of the invariants $\mathcal{R}QQ$, where the curvature of $\Gamma$
does not have the symmetries of the Levi-Civita curvature, $\Ric_{\mu\nu}=\Riem^{\rho}{}_{\mu\rho\nu}$, and
in the invariants of Table~\ref{tab:RQQ} that contain the Ricci tensor, $j=6,\ldots,16$, its indices are
contracted in the order written there: the first index of the Ricci tensor goes to the first index of $Q$
not contracted with $W$ or $V$, or to the first factor $Q$ when there are two. The same definitions were used in the three fits, of the $r_{j}$, $s_{k}$ and $u_{k}$. This convention is material. The $\phi GG$ sector sees only
$\mathcal{R}^{(1)}[\phi]$, the linearized curvature of the Levi-Civita connection, whose Ricci tensor is
symmetric, so the $r_{j}$ do not distinguish the two orders; the $GGG$ parts do, and through them the
other order would change the $s_{k}$ and, in turn, the $u_{k}$. Tables~\ref{tab:QQdQ} and~\ref{tab:Q4} hold
for the convention stated here. In these sectors the invariants may be contracted with $\eta$: with
only $G$ legs the metric is flat, and in $\phi GG$ only $\mathcal{R}^{(1)}[\phi]\,Q^{(1)}Q^{(1)}$ contributes.

\paragraph{Four-dimensional identities.} From ten indices on, that is, for three legs with one or two
derivatives or for four $G$ legs, the monomials satisfy identities that hold only in $d=4$. The
antisymmetrization of five indices vanishes in four dimensions,
\begin{equation}
\delta^{\mu_{1}\mu_{2}\mu_{3}\mu_{4}\mu_{5}}_{\nu_{1}\nu_{2}\nu_{3}\nu_{4}\nu_{5}}=0,
\label{eq:delta5}
\end{equation}
and this generalized delta, contracted with the fields and momenta of a sector, gives linear relations
among its monomials~\cite{Edgar2002}. A calculation in generic $d$ does not see them, and the pole,
whose residue is computed in $d=4$, fits the basis only after they are included. We generated all of
them with xTras~\cite{Nutma2014} and included them in each comparison. There are one, two, four and four
in the sectors $\phi\phi\phi$, $\phi\phi G$, $\phi GG$ and $GGG$, and none in the two-point sectors, which
have only eight indices. In $GGGG$ ten of the twelve indices of $Q^{4}$ enter Eq.~\eqref{eq:delta5} and the
remaining pair is contracted directly; xTras generates five relations of this kind, four of which are
independent. This is why the $69$ contractions $Q^{4}$, independent in generic $d$, have rank $65$ in
$d=4$. The single identity of the $\phi\phi\phi$ sector is the cubic Gauss--Bonnet identity: for general
symmetric polarizations it takes the Lovelock form
$\delta^{\mu_{1}\cdots\mu_{5}}_{\nu_{1}\cdots\nu_{5}}\,\phi_{1}\phi_{2}\phi_{3}\,pppp$, each polarization carrying one
upper and one lower index, and for rank-one polarizations $\phi_{i}=e_{i}e_{i}$ it reduces to the Gram
determinant of $(e_{1},e_{2},e_{3},p_{1},p_{2})$. In the control with the Levi-Civita connection, the cubic
vertices of $\mathcal{E}$ and $\mathcal{E}_{\varepsilon\varepsilon}$ vanish modulo this identity, not identically.

\paragraph{Sensitivity.} The three comparisons that are tests, the parameter-free equalities for
$\delta^{3}_{\bar\phi}S_{\rm eff}$ and $\delta^{2}_{\bar\phi}\delta_{\bar G}S_{\rm eff}$ and the membership of the
remainder of $\delta_{\bar\phi}\delta^{2}_{\bar G}S_{\rm eff}$ in the $21$-dimensional space of
$\mathcal{R}^{(1)}[\phi]\,QQ$, were repeated with the sign of the lower-order expansion reversed. The
equalities then fail and the remainder increases the rank, so these three tests discriminate. For
$\delta^{3}_{\bar G}S_{\rm eff}$ and $\delta^{4}_{\bar G}S_{\rm eff}$ the reversal is not informative, since any
local remainder fits.

\paragraph{Completeness in the computed sectors.} Since each class first appears in a sector of its own, the spans of the basis
and of the candidate contractions were compared in the calculated sectors. In each sector we computed the vertices of all the contractions of
the classes that reach it, expanded with all their metric and density factors, and compared the rank of
their span with that of the basis. Up to cubic order the comparison was also made jointly, with the
vertices of the seven sectors of two and three legs in a single matrix, together with the
four-dimensional identities of each sector; the joint rank, $101$, is that of the basis. The results are given in Sec.~\ref{sec:completeness}. These
checks establish the equality of the spans in the sectors $\phi\phi$, $\phi G$, $GG$, the four cubic sectors and
$GGGG$.

\paragraph{Global completeness.} The basis is complete among all local functionals with the properties of
the counterterms, whatever their form. We work in $d=4$, where the identities~\eqref{eq:delta5} hold
identically, with local functionals of the background fields that are formal power series about
$\bar h=\eta$, $\bar G=0$, of even parity and with dimensionless coefficients, homogeneous of degree four
when each derivative and each factor $\kappa\bar G$ has degree one and $\kappa\bar\phi$ has degree zero. This
is the counting of the invariants~\eqref{eq:classes}: $\bar\Gamma$, Eq.~\eqref{eq:ctparam}, is linear in
$\kappa\bar G$ and $\kappa\,\partial\bar\phi$, and $\bar h=\eta+\kappa\bar\phi$. A term with $p$ fields $\bar\phi$ and
$q$ fields $\bar G$ carries $4-q$ derivatives. We show that every such functional that is invariant under
background diffeomorphisms is, up to a boundary term, a combination of the $166$ invariants of the basis.

\paragraph{Lowest-order invariance.} Expand an invariant functional $I=\sum_{N}I_{N}$ in the number $N$ of
background fields, and let $I_{N}$ be its first term that is not a boundary term. The
transformations~\eqref{eq:deltaphibar} and~\eqref{eq:deltaGbar} split as $\delta=\delta_{\rm inh}+\delta_{\rm hom}$:
$\delta_{\rm inh}$ shifts $\bar\phi$ by the field-independent term $\delta\bar\phi|_{\rm inh}$ and leaves $\bar G$
unchanged, while $\delta_{\rm hom}$, which contains the Lie derivatives and $\Xi$, preserves the number of
fields. At order $N-1$ in the fields, $\delta I=0$ gives $\delta_{\rm inh}I_{N}+\delta_{\rm hom}I_{N-1}=0$ modulo
divergences; since $I_{N-1}$ is a boundary term, so is its variation, and
\begin{equation}
\delta_{\rm inh}I_{N}=0
\label{eq:dinhF}
\end{equation}
up to a boundary term. Since $\delta_{\rm inh}$ removes one field $\bar\phi$ and does not change the number of
fields $\bar G$, the part $I_{p,q}$ of $I_{N}$ with $p$ fields $\bar\phi$ and $q=N-p$ fields $\bar G$ satisfies
Eq.~\eqref{eq:dinhF} by itself. We denote by $V_{p,q}$ the space of local vertices of the sector $(p,q)$ that
are invariant under $\delta_{\rm inh}$, and $V_{N}\equiv\bigoplus_{p+q=N}V_{p,q}$.

\paragraph{Induction.} By the same argument, the lowest term of a combination of the basis whose terms of
lower order are boundary terms lies in $V_{N}$. If these lowest terms span $V_{N}$ for every $N$, the lowest
term of $I$ is cancelled by such a combination, and the procedure, repeated order by order, leaves a
boundary term. The rank comparisons give the dimension of the span of these lowest terms exactly: $22$ at
$N=2$; $101-22=79$ at $N=3$, since the combinations whose quadratic vertices vanish have cubic vertices of
joint rank $79$; and $65$ at $N=4$, from $Q^{4}$ in $GGGG$. With a single field the vertices vanish by momentum
conservation. Since the basis already supplies that many independent elements of $V_{N}$, it suffices to
show that
\begin{gather}
\dim V_{2}\leq22,\qquad \dim V_{3}\leq79,\qquad \dim V_{4}\leq65,\nonumber\\
V_{N}=0\quad(N\geq5).
\label{eq:dimV}
\end{gather}

\paragraph{Sectors without $\bar\phi$.} For $p=0$, $\delta_{\rm inh}$ acts trivially, and $V_{0,q}$ is the whole
space of local vertices with $q$ fields $\bar G$ and $4-q$ derivatives. Since $Q^{(1)}$ is an invertible linear
function of $\bar G$, these are the vertices of the contractions $(\nabla Q)^{2}$, $QQ\nabla Q$ and $Q^{4}$, and
$\dim V_{0,2}=16$, $\dim V_{0,3}=58$ and $\dim V_{0,4}=65$ (Secs.~\ref{sec:dQ2}, \ref{sec:QQdQ}
and~\ref{sec:Q4}). There are none for $q\geq5$.

\paragraph{Sectors with $\bar\phi$.} For $p\geq1$ we use the Euler--Lagrange derivative
$E_{\mu\nu}\equiv\delta I_{p,q}/\delta\bar\phi^{\mu\nu}$, a local function~\cite{Henneaux2000}. Writing
Eq.~\eqref{eq:dinhF} as $\int d^{4}x\,E_{\mu\nu}\,\delta\bar\phi^{\mu\nu}|_{\rm inh}=0$ for arbitrary $\omega$ and
integrating by parts gives the Noether identity
\begin{equation}
\partial_{\mu}T^{\mu\nu}=0,\qquad
T^{\mu\nu}\equiv E^{\mu\nu}-\tfrac{1}{2}\eta^{\mu\nu}\eta^{\rho\sigma}E_{\rho\sigma},
\label{eq:noetherE}
\end{equation}
identically in the fields, with indices raised by $\eta$. The tensor $T$ is symmetric, and in $d=4$ the same
trace reversal recovers $E$ from $T$. Differentiating the same relation with respect to $\bar\phi$, and using
that $\delta\bar\phi|_{\rm inh}$ does not depend on the fields and that second functional derivatives are
symmetric, gives $\delta_{\rm inh}E_{\mu\nu}=0$: the Euler--Lagrange derivative is invariant as a local
function, not only up to a total derivative.\footnote{The integrand of $I_{N}$ need not be invariant itself,
and Eq.~\eqref{eq:dinhF} alone does not allow it to be written in terms of the linearized curvatures: without
antifields, the cohomology of $d$ in the space of invariant polynomials is not
trivial~\cite{Boulanger2000}. The Euler--Lagrange derivative avoids this.} The functions of the fields that
are invariant under linearized diffeomorphisms are the functions of the linearized Riemann tensor and its
derivatives~\cite{Boulanger2000}, and $\bar G$ is inert under $\delta_{\rm inh}$, so $E_{\mu\nu}$ is a polynomial
in $\mathcal{R}^{(1)}[\bar\phi]$, the linearized Riemann tensor of the metric defined by $\bar h$, in $\bar G$ and in
their derivatives. Each factor $\mathcal{R}^{(1)}$ is linear in $\bar\phi$ and carries two derivatives, and the
Euler--Lagrange derivative removes one $\bar\phi$ without changing the number of derivatives. Therefore
$E_{\mu\nu}$ is built from $p-1$ factors $\mathcal{R}^{(1)}$, $q$ factors $\bar G$ and
\begin{equation}
e=4-q-2(p-1)=6-2p-q
\label{eq:ecount}
\end{equation}
further derivatives, and it vanishes when $e<0$. The Bianchi identities of $\mathcal{R}^{(1)}$ relate monomials
with the same number of derivatives and do not affect this count. When $E_{\mu\nu}=0$, the part $I_{p,q}$ is a
boundary term: its integrand $f_{p,q}$ is homogeneous of degree $p$ in $\bar\phi$ and its derivatives, so
\begin{equation}
p\,f_{p,q}=\bar\phi^{\mu\nu}E_{\mu\nu}+\partial_{\mu}j^{\mu}
\label{eq:homog}
\end{equation}
for some local $j^{\mu}$.

Equation~\eqref{eq:ecount} leaves, for $p\geq1$, the sectors $\phi\phi$ ($e=2$) and $\phi G$ ($e=3$) at $N=2$;
$\phi\phi\phi$ ($e=0$), $\phi\phi G$ ($e=1$) and $\phi GG$ ($e=2$) at $N=3$; $\phi\phi GG$ ($e=0$) and $\phi GGG$
($e=1$) at $N=4$; $\phi GGGG$ ($e=0$) at $N=5$; and none for $N\geq6$. We treat them in turn.
\begin{itemize}
\item $\phi GGGG$ and $\phi\phi GG$ ($e=0$). No derivative of $\bar G$ occurs in $T$, so in $\partial_{\mu}T^{\mu\nu}$ the
first derivatives $\partial_{\mu}\bar G^{A}$, where $A$ labels the components of $\bar G$, occur only in
$(\partial T^{\mu\nu}/\partial\bar G^{A})\,\partial_{\mu}\bar G^{A}$. They are independent jet coordinates, so
Eq.~\eqref{eq:noetherE} requires $\partial T^{\mu\nu}/\partial\bar G^{A}=0$. Being homogeneous of positive degree in
$\bar G$, $T$ vanishes, and so does $E$.
\item $\phi GGG$ ($e=1$). Here $T^{\mu\nu}=A^{\mu\nu\rho}_{A}(\bar G)\,\partial_{\rho}\bar G^{A}$, with $A^{\mu\nu\rho}_{A}$
quadratic in $\bar G$ and symmetric in $\mu\nu$. The coefficients of the independent second derivatives
$\partial_{\mu}\partial_{\rho}\bar G^{A}$ in Eq.~\eqref{eq:noetherE} require $A^{\mu\nu\rho}_{A}=-A^{\rho\nu\mu}_{A}$. A tensor
symmetric in its first two indices and antisymmetric in the first and third vanishes, so $T=0$.
\item $\phi\phi G$ ($e=1$). Schematically $T=a\,\mathcal{R}^{(1)}\partial\bar G+b\,(\partial\mathcal{R}^{(1)})\,\bar G$, with
constant coefficients. The second derivatives of $\bar G$ come only from the first term, and the argument
for $\phi GGG$ gives $a=0$; the argument for $e=0$ then gives $T=0$.
\item $\phi GG$ and $\phi G$ ($e=2$ and $3$). Here $p=1$ and $T$ depends on $\bar G$ alone. The vertices
$\mathcal{R}^{(1)}[\phi]\,QQ$ and $\mathcal{R}^{(1)}[\phi]\,\nabla Q$ are invariant, and they span spaces of
dimensions $21$ (Sec.~\ref{sec:cubicunknowns}) and $4$ (Sec.~\ref{sec:X}). The exact counts below give
$\dim V_{1,2}=21$ and $\dim V_{1,1}=4$, so these are all the invariant vertices of the two sectors. For
smooth tensor fields, an identically conserved symmetric tensor is of the form
$T^{\mu\nu}=\partial_{\rho}\partial_{\sigma}U^{\mu\rho\nu\sigma}$, with $U$ having the symmetries of the Riemann
tensor~\cite{DuboisViolette2002}, which would give the same result if $U$ could be taken local in the
fields; we do not rely on that local version.
\item $\phi\phi$ ($e=2$). The invariant quadratic functionals of $\bar\phi$ with four derivatives are the
curvature-squared ones, of which two are independent in $d=4$ (Sec.~\ref{sec:phiphi}).
\item $\phi\phi\phi$ ($e=0$). Here $T$ is quadratic in $\mathcal{R}^{(1)}$, without derivatives. The exact count
below gives $\dim V_{3,0}=0$ in $d=4$. In $d=5$ there is one invariant vertex, the cubic vertex of the
Gauss--Bonnet density, which vanishes identically in four dimensions.
\end{itemize}
Collecting, $\dim V_{2}=2+4+16=22$, $\dim V_{3}=0+0+21+58=79$ and $\dim V_{4}=65$, all the sectors of $V_{4}$
with $\bar\phi$ being empty, while $V_{N}=0$ for $N\geq5$. Equation~\eqref{eq:dimV} holds, and the basis is
complete among the background-invariant local functionals of dimension four and even parity.

\paragraph{Exact counts.} The most general vertex of a sector is a combination of all the Lorentz monomials,
products of scalar products, multilinear in the polarizations and homogeneous of degree $4-q$ in the
independent momenta. By the first fundamental theorem of invariant theory for the orthogonal group, these
monomials span all the parity-even invariant polynomials, and momentum conservation identifies the monomials
related by integrations by parts. The polarizations are taken of rank one, $\phi^{\mu\nu}\to u^{\mu}u^{\nu}$ and
$G^{\lambda}{}_{\mu\nu}\to w^{\lambda}v_{\mu}v_{\nu}$, which loses nothing by multilinearity, and the vertices are
symmetrized over identical legs, so that it suffices to vary one leg $\bar\phi$. The variation $\delta_{\rm inh}$
replaces $u^{\mu}u^{\nu}$ by $\eta^{\mu\nu}(p\cdot\omega)-p^{\mu}\omega^{\nu}-p^{\nu}\omega^{\mu}$, with $p$ the momentum
of the leg. For the sectors $\phi G$, $\phi\phi\phi$, $\phi\phi G$ and $\phi GG$ we expanded each symmetrized
monomial and its variation as explicit polynomials in the components of the vectors in $d=4$, so that the
four-dimensional identities hold automatically. If $B$ and $A$ are the integer matrices of the coefficients
of the monomials and of their variations, the vertex space has dimension $\operatorname{rank}B$, and, since
every combination that vanishes identically also has a vanishing variation,
$\dim V_{p,q}=\operatorname{rank}B-\operatorname{rank}A$. The four sectors have $11$, $219$, $318$ and $395$
Lorentz monomials before symmetrization, and the ranks, exact and without sampling, are $\operatorname{rank}B=11$, $44$, $158$ and $209$, and $\operatorname{rank}A=7$, $44$, $158$ and $188$, which gives
$\dim V_{1,1}=4$, $\dim V_{3,0}=0$, $\dim V_{2,1}=0$ and $\dim V_{1,2}=21$. The same computation in $d=5$, with the same monomials, gives
$\operatorname{rank}B=45$ and $\operatorname{rank}A=44$ in $\phi\phi\phi$, one invariant vertex, that of the
Gauss--Bonnet density, and $\operatorname{rank}B=212$ and $\operatorname{rank}A=190$ in $\phi GG$, $22$ invariant
vertices: one of the invariant vertices of each sector vanishes only by a four-dimensional identity.

\paragraph{Modular counts.} The dimensions of all the spaces $V_{p,q}$ were also computed with the same
vertices, evaluated at randomly chosen points using exact arithmetic modulo the prime $1000003$. These
counts are numerical checks of the analytic statements above, and no error probability is assigned to them
from the size of the prime alone. In $d=4$ they give $2$, $4$ and $16$ in
$\phi\phi$, $\phi G$ and $GG$; $0$, $0$, $21$ and $58$ in $\phi\phi\phi$, $\phi\phi G$, $\phi GG$ and $GGG$; and $0$ in the
four quartic sectors with $\bar\phi$ and $65$ in $GGGG$, in agreement with each entry of Eq.~\eqref{eq:dimV}
and with the exact counts. As a further control, in the sector
$\phi\phi GG$ with four momenta, restricted to the vertices in which the polarizations of $G$ are contracted
among themselves, the count gives the $15=3\times5$ invariant vertices $\mathcal{R}^{(1)}\mathcal{R}^{(1)}GG$.

\paragraph{Cross-check in $\phi GGG$.} Independently, the joint comparison was extended to $\phi GGG$ by
sampling. The vertices of the $238$ contractions of the first five classes in that sector ($217$ of them
with a nonvanishing vertex there: besides the $17$ that vanish identically, four, two of
$\mathcal{R}^{2}$, one of $\mathcal{R}\nabla Q$ and one of $(\nabla Q)^{2}$, have a vanishing $\phi GGG$
vertex), evaluated in $d=4$ at $320$ random integer points with independent momenta on the three $G$
legs, leave the joint rank at $101$. Equivalently, each of the $137$ contractions outside a maximal
independent set, once its expansion in the lower sectors is subtracted in terms of that set, vanishes at all
these points in $\phi GGG$.
\onecolumngrid
\section{Tensor bases and coefficients of the two-point functions}
\label{app:tensor}

All coefficients below multiply $B_{0}(k^{2})$ as in Sec.~\ref{sec:twopoint}; superscripts $(\phi)$ and
$(c)$ denote the bosonic and ghost loops.

\subsection{Graviton self-energy}

The structures of Eq.~\eqref{eq:Piphi} are
\begin{gather}
T^{\phi\phi}_{1}\equiv\eta^{\alpha\mu}\eta^{\beta\nu}+\eta^{\alpha\nu}\eta^{\beta\mu},\qquad
T^{\phi\phi}_{2}\equiv\eta^{\alpha\beta}\eta^{\mu\nu},\qquad
T^{\phi\phi}_{3}\equiv\eta^{\mu\nu}k^{\alpha}k^{\beta}+\eta^{\alpha\beta}k^{\mu}k^{\nu},\nonumber\\
T^{\phi\phi}_{4}\equiv\eta^{\alpha\mu}k^{\beta}k^{\nu}+\eta^{\alpha\nu}k^{\beta}k^{\mu}
+\eta^{\beta\mu}k^{\alpha}k^{\nu}+\eta^{\beta\nu}k^{\alpha}k^{\mu},\qquad
T^{\phi\phi}_{5}\equiv k^{\alpha}k^{\beta}k^{\mu}k^{\nu},
\label{eq:Tphiphi}
\end{gather}
and the free coefficients are
\begingroup\small\allowdisplaybreaks
\begin{align*}
a_{1}^{(\phi)} &= \frac{1}{64 (d-1)}\Big[4 d^3+4 d^2-23 d-8 -8 (d-1) \left(d^2-3 d-2\right)\xi +4 (d-1) \left(d^2-7 d+16\right)\xi^{2}\Big],\\
a_{2}^{(\phi)} &= -\frac{1}{64 (d-2)^2 (d-1)}\Big[(d-2) \left(8 d^3-9 d^2-62 d-16\right) -16 \left(d^4-8 d^3+17 d^2-4 d-4\right)\xi\nonumber\\&\qquad+8 (d-4) (d-1) \left(d^2-7 d+14\right)\xi^{2}\Big],\\
a_{3}^{(\phi)} &= \frac{1}{64 (d-2) (d-1)}\Big[(d-2) d (17 d+16) -32 d \left(d^2-4 d+2\right)\xi +16 (d-1) \left(d^2-6 d+12\right)\xi^{2}\Big],\\
a_{1}^{(c)} &= -\frac{2 d^2-3 d-6}{16 (d-1) (d+1)},\\
a_{2}^{(c)} &= \frac{3 d^2-22 d-28}{16 (d-2) (d-1) (d+1)},\\
a_{3}^{(c)} &= -\frac{d^2+16 d+16}{16 (d-1) (d+1)}.
\end{align*}
\endgroup
Summing the two loops, the coefficients~\eqref{eq:cphiphi} are
\begingroup\small\allowdisplaybreaks
\begin{align*}
c_{R}^{(\phi\phi)} &= \frac{1}{128 (d-1) (d+1)}\Big[d^4-37 d^3+2 d^2+108 d+64 -32 (d-2) (d+1)\xi +16 (d-4) (d-1) (d+1)\xi^{2}\Big],\\
c_{\mathrm{Ric}}^{(\phi\phi)} &= \frac{1}{16 (d-1) (d+1)}\Big[4 d^4+8 d^3-27 d^2-19 d+16 -8 (d-1) (d+1) \left(d^2-3 d-2\right)\xi\nonumber\\&\qquad+4 (d-1) (d+1) \left(d^2-7 d+16\right)\xi^{2}\Big].
\end{align*}
\endgroup

\subsection{Mixed function}

The structures of Sec.~\ref{sec:phiG} are
\begin{gather}
T^{\phi G}_{1}\equiv k^{2}k^{\lambda}(\eta^{\mu\alpha}\eta^{\nu\beta}+\eta^{\nu\alpha}\eta^{\mu\beta}),\qquad
T^{\phi G}_{2}\equiv k^{2}k^{\lambda}\eta^{\alpha\beta}\eta^{\mu\nu},\qquad
T^{\phi G}_{3}\equiv k^{2}\eta^{\alpha\beta}(\eta^{\mu\lambda}k^{\nu}+\eta^{\nu\lambda}k^{\mu}),\nonumber\\
T^{\phi G}_{4}\equiv k^{\alpha}k^{\beta}(\eta^{\mu\lambda}k^{\nu}+\eta^{\nu\lambda}k^{\mu}),\qquad
T^{\phi G}_{5}\equiv\eta^{\mu\nu}k^{\alpha}k^{\beta}k^{\lambda},\qquad
T^{\phi G}_{6}\equiv k^{2}(\eta^{\mu\lambda}\eta^{\nu\beta}k^{\alpha}+\eta^{\mu\lambda}\eta^{\nu\alpha}k^{\beta}+\mu\leftrightarrow\nu),\nonumber\\
T^{\phi G}_{7}\equiv k^{2}(\eta^{\alpha\lambda}\eta^{\beta\nu}k^{\mu}+\eta^{\alpha\lambda}\eta^{\beta\mu}k^{\nu}+\alpha\leftrightarrow\beta),\qquad
T^{\phi G}_{8}\equiv k^{\lambda}(\eta^{\beta\nu}k^{\alpha}k^{\mu}+\eta^{\beta\mu}k^{\alpha}k^{\nu}+\alpha\leftrightarrow\beta),\nonumber\\
T^{\phi G}_{9}\equiv(\eta^{\alpha\lambda}k^{\beta}+\eta^{\beta\lambda}k^{\alpha})k^{\mu}k^{\nu},\qquad
T^{\phi G}_{10}\equiv k^{2}\eta^{\mu\nu}(\eta^{\alpha\lambda}k^{\beta}+\eta^{\beta\lambda}k^{\alpha}),\nonumber\\
T^{\phi G}_{11}\equiv\eta^{\alpha\beta}k^{\lambda}k^{\mu}k^{\nu},\qquad
T^{\phi G}_{12}\equiv k^{\alpha}k^{\beta}k^{\lambda}k^{\mu}k^{\nu}/k^{2}.
\label{eq:TphiG}
\end{gather}
The five free coefficients of Eqs.~\eqref{eq:wardS} are
\begingroup\small\allowdisplaybreaks
\begin{align*}
k_{1}^{(\phi)} &= -\frac{1}{8 (d-1)}\Big[(d-2) \left(d^2+1\right) -(2 d^3-11 d^2+14 d-2)\xi +(d^3-8 d^2+22 d-14)\xi^{2}\Big],\\
k_{2}^{(\phi)} &= \frac{1}{16 (d-2) (d-1)}\Big[(d-2) \left(5 d^2+3 d-6\right) -2 \left(5 d^3-23 d^2+18 d+4\right)\xi +(5 d^3-37 d^2+96 d-60)\xi^{2}\Big],\\
k_{3}^{(\phi)} &= \frac{d^2 \xi ^2+6 d^2 \xi -8 d^2-7 d \xi ^2-32 d \xi +16 d+6 \xi ^2+26 \xi +2}{16 (d-1)^2},\\
k_{7}^{(\phi)} &= -\frac{4 d^2 \xi -5 d^2-2 d \xi ^2-18 d \xi +11 d+4 \xi ^2+12 \xi -2}{16 (d-1)},\\
k_{11}^{(\phi)} &= -\frac{(d-4) (\xi -1) \left(d^2 \xi -d^2-4 d \xi +d+2\right)}{8 (d-2) (d-1)},\\
k_{1}^{(c)} &= -\frac{1}{4 (d-1)},\\
k_{2}^{(c)} &= -\frac{3}{8 (d-1)},\\
k_{3}^{(c)} &= \frac{2 d-7}{8 (d-1)^2},\\
k_{7}^{(c)} &= -\frac{d-2}{8 (d-1)},
\end{align*}
\endgroup
with $k_{11}^{(c)}=0$. Summing the two loops, the coefficients of the decomposition of
Sec.~\ref{sec:basis} are
\begingroup\small\allowdisplaybreaks
\begin{align*}
C_{R} &= \frac{1}{32 (d-1)}\Big[d^3+d^2-22 d+24 -2 \left(d^3-d^2-10 d+8\right)\xi +(d-2)^2 (d-1)\xi^{2}\Big],\\
C_{\mathrm{Ric}} &= \frac{1}{4 (d-1)}\Big[(d-3) \left(d^2-4 d+2\right) -(2 d-7) \left(d^2-4 d+2\right)\xi +(d^3-8 d^2+20 d-10)\xi^{2}\Big],\\
C_{\mathrm{Riem}} &= -\frac{4 d^2 \xi -5 d^2-2 d \xi ^2-18 d \xi +13 d+4 \xi ^2+12 \xi -6}{8 (d-1)},\\
C_{X} &= -\frac{d^2 \xi ^2-2 d^2 \xi +2 d^2-3 d \xi ^2+4 d \xi -6 d-2 \xi ^2+2 \xi }{16 (d-1)^2}.
\end{align*}
\endgroup
and the coefficients of the counterterms of Sec.~\ref{sec:basis} that contribute at two points,
besides $a_{R}=c^{(\phi\phi)}_{R}$ and $a_{\Ric}=c^{(\phi\phi)}_{\Ric}$, are
\begingroup\small\allowdisplaybreaks
\begin{align*}
b_{V} &= \frac{1}{64 (d-1) (d+1)}\Big[3 d^4+45 d^3-86 d^2-100 d+32 -8 (d+1) \left(d^3-d^2-14 d+16\right)\xi +4 (d-1) (d+1) \left(d^2-8 d+20\right)\xi^{2}\Big],\\
b_{W} &= \frac{1}{64 (d-2) (d-1) (d+1)}\Big[(d-2) \left(11 d^3-13 d^2-12 d+16\right) -8 (d-3) (d+1) \left(2 d^2-5 d-2\right)\xi\nonumber\\&\qquad+8 (d+1) \left(d^3-7 d^2+14 d-4\right)\xi^{2}\Big],\\
b_{Y} &= -\frac{1}{8 (d-1) (d+1)}\Big[12 d^3-23 d^2-23 d+16 -12 (d+1) \left(d^2-4 d+2\right)\xi +4 (d-2) (d+1)\xi^{2}\Big],\\
b_{Z} &= -\frac{1}{16 (d-1) (d+1)}\Big[8 d^3-9 d^2-5 d+8 -4 (d+1) \left(d^2-6 d+6\right)\xi +4 (d-2) (d+1)\xi^{2}\Big].
\end{align*}
\endgroup

\subsection{Connection self-energy}

The local and nonlocal structures of Eq.~\eqref{eq:GammaGG} are
\begin{gather}
T^{GG}_{1}\equiv k^{2}\eta^{\lambda\gamma}\eta^{\mu\nu}\eta^{\alpha\beta},\quad
T^{GG}_{2}\equiv k^{2}\eta^{\lambda\gamma}\eta^{\mu\alpha}\eta^{\nu\beta},\quad
T^{GG}_{3}\equiv k^{2}\eta^{\lambda\alpha}\eta^{\mu\gamma}\eta^{\nu\beta},\quad
T^{GG}_{4}\equiv k^{2}\eta^{\lambda\mu}\eta^{\gamma\alpha}\eta^{\nu\beta},\nonumber\\
T^{GG}_{5}\equiv k^{2}\eta^{\lambda\alpha}\eta^{\gamma\beta}\eta^{\mu\nu},\quad
T^{GG}_{6}\equiv k^{\lambda}k^{\gamma}\eta^{\mu\alpha}\eta^{\nu\beta},\quad
T^{GG}_{7}\equiv k^{\mu}k^{\nu}\eta^{\lambda\gamma}\eta^{\alpha\beta},\quad
T^{GG}_{8}\equiv k^{\lambda}k^{\alpha}\eta^{\mu\nu}\eta^{\gamma\beta},\nonumber\\
T^{GG}_{9}\equiv k^{\mu}k^{\nu}\eta^{\lambda\alpha}\eta^{\gamma\beta},\quad
T^{GG}_{10}\equiv k^{\lambda}k^{\alpha}\eta^{\mu\gamma}\eta^{\nu\beta},\quad
T^{GG}_{11}\equiv k^{\mu}k^{\alpha}\eta^{\lambda\nu}\eta^{\gamma\beta},\quad
T^{GG}_{12}\equiv k^{\mu}k^{\alpha}\eta^{\lambda\gamma}\eta^{\nu\beta},\nonumber\\
T^{GG}_{13}\equiv k^{\mu}k^{\alpha}\eta^{\lambda\beta}\eta^{\gamma\nu},\quad
T^{GG}_{14}\equiv k^{\lambda}k^{\mu}\eta^{\gamma\alpha}\eta^{\nu\beta},\quad
T^{GG}_{15}\equiv k^{\lambda}k^{\mu}\eta^{\gamma\nu}\eta^{\alpha\beta},\quad
T^{GG}_{16}\equiv k^{\lambda}k^{\gamma}\eta^{\mu\nu}\eta^{\alpha\beta},
\label{eq:TGG}
\end{gather}
\begin{gather}
N_{1}\equiv\frac{k^{\lambda}k^{\gamma}k^{\alpha}k^{\beta}\eta^{\mu\nu}}{k^{2}},\quad
N_{2}\equiv\frac{k^{\mu}k^{\gamma}k^{\alpha}k^{\beta}\eta^{\lambda\nu}}{k^{2}},\quad
N_{3}\equiv\frac{k^{\lambda}k^{\mu}k^{\alpha}k^{\beta}\eta^{\gamma\nu}}{k^{2}},\nonumber\\
N_{4}\equiv\frac{k^{\lambda}k^{\mu}k^{\gamma}k^{\alpha}\eta^{\nu\beta}}{k^{2}},\quad
N_{5}\equiv\frac{k^{\mu}k^{\nu}k^{\alpha}k^{\beta}\eta^{\lambda\gamma}}{k^{2}},\quad
N_{6}\equiv\frac{k^{\lambda}k^{\mu}k^{\nu}k^{\gamma}k^{\alpha}k^{\beta}}{k^{4}}.
\label{eq:NGG}
\end{gather}
The bosonic loop gives
\begingroup\small\allowdisplaybreaks
\begin{align*}
g_{1}^{(\phi)} &= -\frac{1}{16 (d-2) (d-1) (d+1)}\Big[3 d^3-11 d^2+5 d+16 -(3 d^3-5 d-8)\xi -(d-2) d\xi^{2}\Big],\\
g_{2}^{(\phi)} &= \frac{1}{16 (d-2) (d-1) (d+1)}\Big[4 d^4-17 d^3+24 d^2+7 d-32 -(4 d^4-13 d^3+18 d^2-9 d-32)\xi -2 (d-2) (d+2)\xi^{2}\Big],\\
g_{3}^{(\phi)} &= \frac{1}{16 (d-2) (d-1) (d+1)}\Big[4 d^4-47 d^3+102 d^2+13 d-128 -(4 d^4-19 d^3+2 d^2+17 d+16)\xi -4 (d-2)\xi^{2}\Big],\\
g_{4}^{(\phi)} &= -\frac{1}{16 (d-2) (d-1)^3 (d+1)}\Big[3 d^5-28 d^4+114 d^3-172 d^2-45 d+224 -(d-1) d \left(3 d^3-d^2+5 d-39\right)\xi\nonumber\\&\qquad+4 (d-2) (d-1)^2\xi^{2}\Big],\\
g_{5}^{(\phi)} &= \frac{1}{8 (d-2) (d-1)^2 (d+1)}\Big[3 d^4+13 d^3-73 d^2+25 d+96 -(3 d^4-9 d^3+17 d^2+13 d-40)\xi +2 (d-2) (d-1) d\xi^{2}\Big],\\
g_{6}^{(\phi)} &= \frac{1}{16 (d-1) (d+1)}\Big[8 d^4-37 d^3+44 d^2+27 d-56 -(16 d^4-101 d^3+118 d^2+103 d-120)\xi\nonumber\\&\qquad+2 \left(4 d^4-28 d^3+55 d^2+32 d-52\right)\xi^{2}\Big],\\
g_{7}^{(\phi)} &= \frac{d^3 \xi -d^3-d^2 \xi ^2-6 d^2 \xi +3 d^2-d \xi -3 d+4 \xi ^2-4}{8 (d-1) (d+1)},\\
g_{8}^{(\phi)} &= \frac{1}{4 (d-1)^2 (d+1)}\Big[(d-2) \left(11 d^2-5 d-14\right) -2 (d-1) \left(7 d^2-18 d-22\right)\xi -(d-2) (d-1) (d+2)\xi^{2}\Big],\\
g_{9}^{(\phi)} &= -\frac{1}{8 (d-2) (d-1)^2 (d+1)}\Big[(d-2) \left(3 d^4-7 d^3+27 d^2-31 d-56\right) -(d-1) \left(3 d^4-12 d^3+7 d^2+18 d+32\right)\xi\nonumber\\&\qquad+2 (d-2)^2 (d-1) (d+2)\xi^{2}\Big],\\
g_{10}^{(\phi)} &= -\frac{1}{8 (d-1) (d+1)}\Big[35 d^3-82 d^2-17 d+88 -(27 d^3-118 d^2-17 d+104)\xi +4 (d-2) (2 d+3)\xi^{2}\Big],\\
g_{11}^{(\phi)} &= -\frac{1}{4 (d-2) (d-1)^3 (d+1)}\Big[(d-2)^2 \left(5 d^2-14 d-15\right) -(d-1) \left(d^4-6 d^3+d^2+20 d-24\right)\xi\nonumber\\&\qquad-(d-2)^2 (d-1)^2 d\xi^{2}\Big],\\
g_{12}^{(\phi)} &= \frac{1}{16 (d-2) (d-1) (d+1)}\Big[(d-2) \left(3 d^3+6 d^2-25 d-16\right) +(d^4-8 d^3+45 d^2-66 d-48)\xi +4 (d-2)^2 (2 d+1)\xi^{2}\Big],\\
g_{13}^{(\phi)} &= \frac{1}{4 (d-2) (d-1) (d+1)}\Big[(d-2) \left(2 d^3+5 d^2-14 d-14\right) -(d^4-5 d^3+2 d^2+18 d-8)\xi +(d-2)^2 d\xi^{2}\Big],\\
g_{14}^{(\phi)} &= \frac{1}{8 (d-2) (d-1)^2 (d+1)}\Big[3 (d-2) (d-1) \left(d^3-6 d^2+5 d+16\right) -(3 d^5-31 d^4+71 d^3+23 d^2-178 d+48)\xi\nonumber\\&\qquad-4 (d-2)^2 (d-1)\xi^{2}\Big],\\
g_{15}^{(\phi)} &= \frac{1}{8 (d-1) (d+1)}\Big[13 d^3-32 d^2-11 d+40 -(13 d^3-42 d^2-27 d+40)\xi +2 (d-2) d\xi^{2}\Big],\\
g_{16}^{(\phi)} &= \frac{1}{16 (d-1) (d+1)}\Big[(d-2) \left(8 d^2-11 d-20\right) -2 \left(8 d^3-27 d^2-2 d+36\right)\xi +(8 d^3-23 d^2-10 d+24)\xi^{2}\Big],\\
n_{1}^{(\phi)} &= \frac{1}{8 (d-1) (d+1)}\Big[(d-4) (d-2) \left(4 d^2-7 d-12\right) -2 (d-4) \left(4 d^3-16 d^2-d+22\right)\xi +(d-4) \left(4 d^3-17 d^2-4 d+20\right)\xi^{2}\Big],\\
n_{2}^{(\phi)} &= \frac{1}{4 (d-1)^2 (d+1)}\Big[(d-4) \left(d^3-5 d^2+6 d+6\right) -2 (d-4) \left(d^3-4 d^2+4 d+3\right)\xi +(d-4) (d-2) (d-1) d\xi^{2}\Big],\\
n_{3}^{(\phi)} &= \frac{(d-4) (\xi -1) \left(d^3-2 d^2 \xi -8 d^2+9 d+8 \xi +12\right)}{8 (d-1) (d+1)},\\
n_{4}^{(\phi)} &= -\frac{1}{16 (d-1) (d+1)}\Big[3 (d-4) \left(5 d^3-14 d^2+d+16\right) -(d-4) \left(31 d^3-106 d^2-9 d+104\right)\xi\nonumber\\&\qquad+4 (d-4) \left(4 d^3-16 d^2-3 d+14\right)\xi^{2}\Big],\\
n_{5}^{(\phi)} &= -\frac{(d-4) d (\xi -1) \left(d^2-d \xi -d+2 \xi -5\right)}{16 (d-1) (d+1)},\\
n_{6}^{(\phi)} &= \frac{(d-6) (d-4) (d-2) d (\xi -1)^2}{16 (d-1) (d+1)},
\end{align*}
\endgroup
and the ghost loop, independent of $\xi$ and without the structures $N_{j}$,
\begingroup\small\allowdisplaybreaks
\begin{align*}
g_{1}^{(c)} &= -\frac{1}{4 (d-1)},&\qquad g_{3}^{(c)} &= -\frac{1}{d-1},&\qquad g_{4}^{(c)} &= \frac{3 (d-2)}{(d-1)^3},\\
g_{5}^{(c)} &= -\frac{d-3}{(d-1)^2},&\qquad g_{9}^{(c)} &= \frac{2 (d-2)}{(d-1)^2},&\qquad g_{11}^{(c)} &= \frac{(d-4) (d-2)}{(d-1)^3},\\
g_{13}^{(c)} &= -\frac{d-2}{d-1},&\qquad g_{14}^{(c)} &= -\frac{2 (d-2)}{(d-1)^2},&\qquad g_{15}^{(c)} &= \frac{d-2}{d-1},\\
g_{16}^{(c)} &= -\frac{d-2}{4 (d-1)},
\end{align*}
\endgroup
with the remaining $g_{i}^{(c)}$ equal to zero.
\twocolumngrid
\vspace*{2\baselineskip}
\section{Coefficient tables}
\label{app:tables}

Tables~\ref{tab:QQdQ} and~\ref{tab:Q4} give the coefficients of the invariants $QQ\nabla Q$ and $Q^{4}$ fixed by
$\delta^{3}_{\bar G}S_{\rm eff}$ and $\delta^{4}_{\bar G}S_{\rm eff}$ (Secs.~\ref{sec:QQdQ} and~\ref{sec:Q4}), with the
normalization of Eq.~\eqref{eq:Sdiv}. Like the $r_{j}$ of Table~\ref{tab:RQQ}, they depend on the choice of basis
at each order and on the conventions of Sec.~\ref{sec:blocks}; another choice redistributes the values without
changing the action.

\begin{table*}[p]
\caption{Coefficients $s_{k}$ of the $58$ invariants $QQ\nabla Q$, in the sense $\sum_{I}c_{I}I\supset\sum_{k}s_{k}\,(\sqrt{-g})^{1-w_{k}}\,\mathcal{O}^{QQ\nabla Q}_{k}$, in $d=4$ and $\xi=1$. Shorthand, not the Einstein convention: all indices are written down and each repeated pair is contracted with the inverse metric, as in Table~\ref{tab:dQ2}. $\nabla\cdot W=\nabla^{\lambda}W_{\lambda}$, $\nabla\cdot V=\nabla_{\mu}V^{\mu}$, $A$ and $B$ are the divergences~\eqref{eq:AB}, and the first index of $Q$ is its derivative index. $\nabla$ acts on $Q_{\lambda}{}^{\mu\nu}$, $V^{\nu}$ and $W_{\lambda}$ before any index is moved or contracted with the metric, which matters because $\nabla g\neq0$ (Sec.~\ref{sec:blocks}); and the weight $w_{k}$ counts one for each $Q$, $V$, $A$ or $B$ and zero for each $W$.}
\label{tab:QQdQ}
\renewcommand{\arraystretch}{1.7}
\begin{ruledtabular}
\begin{tabular}{rlc@{\qquad}rlc}
$k$ & invariant & $s_{k}$ & $k$ & invariant & $s_{k}$\\
\hline
1 & $(\nabla\!\cdot\! W)\,V_{\delta}\,W_{\delta}$ & $\frac{16}{15}$ & 30 & $\nabla_{\alpha}\,V_{\gamma}\,Q_{\alpha\gamma\delta}\,W_{\delta}$ & $\frac{5}{12}$\\
2 & $(\nabla\!\cdot\! W)\,V_{\delta}\,V_{\delta}$ & $-\frac{1}{16}$ & 31 & $\nabla_{\alpha}\,V_{\gamma}\,Q_{\gamma\alpha\delta}\,W_{\delta}$ & $\frac{71}{120}$\\
3 & $(\nabla\!\cdot\! W)\,W_{\gamma}\,W_{\gamma}$ & $-\frac{147}{320}$ & 32 & $\nabla_{\alpha}\,V_{\gamma}\,Q_{\gamma\alpha\delta}\,V_{\delta}$ & $-\frac{1}{12}$\\
4 & $(\nabla\!\cdot\! V)\,V_{\delta}\,W_{\delta}$ & $-\frac{13}{120}$ & 33 & $\nabla_{\alpha}\,V_{\gamma}\,Q_{\delta\alpha\gamma}\,W_{\delta}$ & $-\frac{1}{2}$\\
5 & $(\nabla\!\cdot\! V)\,V_{\delta}\,V_{\delta}$ & $-\frac{1}{8}$ & 34 & $\nabla_{\alpha}\,W_{\beta}\,Q_{\beta\alpha\delta}\,W_{\delta}$ & $-\frac{97}{80}$\\
6 & $(\nabla\!\cdot\! V)\,W_{\gamma}\,W_{\gamma}$ & $\frac{23}{96}$ & 35 & $A_{\beta\gamma}\,Q_{\beta\delta\epsilon}\,Q_{\gamma\delta\epsilon}$ & $\frac{11}{60}$\\
7 & $A_{\beta\gamma}\,W_{\beta}\,W_{\gamma}$ & $\frac{89}{240}$ & 36 & $A_{\beta\gamma}\,Q_{\beta\delta\epsilon}\,Q_{\delta\gamma\epsilon}$ & $-\frac{7}{30}$\\
8 & $A_{\beta\gamma}\,W_{\beta}\,V_{\gamma}$ & $\frac{1}{4}$ & 37 & $A_{\beta\gamma}\,Q_{\delta\beta\epsilon}\,Q_{\delta\gamma\epsilon}$ & $-\frac{7}{40}$\\
9 & $A_{\beta\gamma}\,V_{\beta}\,V_{\gamma}$ & $\frac{1}{24}$ & 38 & $A_{\beta\gamma}\,Q_{\delta\beta\epsilon}\,Q_{\epsilon\gamma\delta}$ & $\frac{19}{120}$\\
10 & $B_{\beta\gamma}\,W_{\beta}\,W_{\gamma}$ & $-\frac{19}{30}$ & 39 & $B_{\beta\gamma}\,Q_{\beta\delta\epsilon}\,Q_{\gamma\delta\epsilon}$ & $\frac{41}{120}$\\
11 & $B_{\beta\gamma}\,W_{\beta}\,V_{\gamma}$ & $-\frac{7}{24}$ & 40 & $B_{\beta\gamma}\,Q_{\beta\delta\epsilon}\,Q_{\delta\gamma\epsilon}$ & $-\frac{1}{60}$\\
12 & $B_{\beta\gamma}\,V_{\beta}\,W_{\gamma}$ & $\frac{7}{40}$ & 41 & $B_{\beta\gamma}\,Q_{\delta\beta\epsilon}\,Q_{\gamma\delta\epsilon}$ & $-\frac{1}{60}$\\
13 & $B_{\beta\gamma}\,V_{\beta}\,V_{\gamma}$ & $-\frac{1}{4}$ & 42 & $B_{\beta\gamma}\,Q_{\delta\beta\epsilon}\,Q_{\delta\gamma\epsilon}$ & $-\frac{7}{30}$\\
14 & $\nabla_{\alpha}\,V_{\gamma}\,W_{\alpha}\,W_{\gamma}$ & $\frac{1}{60}$ & 43 & $B_{\beta\gamma}\,Q_{\delta\beta\epsilon}\,Q_{\epsilon\gamma\delta}$ & $\frac{1}{10}$\\
15 & $\nabla_{\alpha}\,V_{\gamma}\,V_{\alpha}\,W_{\gamma}$ & $-\frac{1}{24}$ & 44 & $\nabla_{\alpha}\,V_{\gamma}\,Q_{\alpha\delta\epsilon}\,Q_{\gamma\delta\epsilon}$ & $\frac{1}{24}$\\
16 & $(\nabla\!\cdot\! W)\,Q_{\gamma\delta\epsilon}\,Q_{\gamma\delta\epsilon}$ & $-\frac{13}{160}$ & 45 & $\nabla_{\alpha}\,V_{\gamma}\,Q_{\alpha\delta\epsilon}\,Q_{\delta\gamma\epsilon}$ & $-\frac{1}{3}$\\
17 & $(\nabla\!\cdot\! W)\,Q_{\gamma\delta\epsilon}\,Q_{\delta\gamma\epsilon}$ & $\frac{1}{10}$ & 46 & $\nabla_{\alpha}\,Q_{\beta\gamma\delta}\,Q_{\alpha\beta\gamma}\,W_{\delta}$ & $-\frac{1}{12}$\\
18 & $A_{\beta\gamma}\,Q_{\beta\gamma\delta}\,W_{\delta}$ & $-\frac{61}{120}$ & 47 & $\nabla_{\alpha}\,W_{\beta}\,Q_{\alpha\delta\epsilon}\,Q_{\beta\delta\epsilon}$ & $\frac{3}{20}$\\
19 & $A_{\beta\gamma}\,Q_{\beta\gamma\delta}\,V_{\delta}$ & $0$ & 48 & $\nabla_{\alpha}\,V_{\gamma}\,Q_{\delta\alpha\epsilon}\,Q_{\gamma\delta\epsilon}$ & $-\frac{1}{12}$\\
20 & $A_{\beta\gamma}\,Q_{\delta\beta\gamma}\,W_{\delta}$ & $\frac{23}{120}$ & 49 & $\nabla_{\alpha}\,V_{\gamma}\,Q_{\delta\alpha\epsilon}\,Q_{\delta\gamma\epsilon}$ & $\frac{2}{3}$\\
21 & $A_{\beta\gamma}\,Q_{\delta\beta\gamma}\,V_{\delta}$ & $\frac{23}{60}$ & 50 & $\nabla_{\alpha}\,V_{\gamma}\,Q_{\delta\alpha\epsilon}\,Q_{\epsilon\gamma\delta}$ & $-\frac{1}{3}$\\
22 & $(\nabla\!\cdot\! V)\,Q_{\gamma\delta\epsilon}\,Q_{\gamma\delta\epsilon}$ & $-\frac{59}{240}$ & 51 & $\nabla_{\alpha}\,Q_{\beta\gamma\delta}\,Q_{\beta\alpha\gamma}\,W_{\delta}$ & $-\frac{2}{15}$\\
23 & $(\nabla\!\cdot\! V)\,Q_{\gamma\delta\epsilon}\,Q_{\delta\gamma\epsilon}$ & $\frac{29}{120}$ & 52 & $\nabla_{\alpha}\,Q_{\beta\gamma\delta}\,Q_{\gamma\alpha\beta}\,W_{\delta}$ & $-\frac{137}{240}$\\
24 & $B_{\beta\gamma}\,Q_{\beta\gamma\delta}\,W_{\delta}$ & $\frac{5}{24}$ & 53 & $\nabla_{\alpha}\,W_{\beta}\,Q_{\delta\alpha\epsilon}\,Q_{\beta\delta\epsilon}$ & $-\frac{19}{40}$\\
25 & $B_{\beta\gamma}\,Q_{\beta\gamma\delta}\,V_{\delta}$ & $\frac{1}{4}$ & 54 & $\nabla_{\alpha}\,Q_{\beta\gamma\delta}\,Q_{\alpha\beta\epsilon}\,Q_{\gamma\delta\epsilon}$ & $\frac{7}{12}$\\
26 & $B_{\beta\gamma}\,Q_{\gamma\beta\delta}\,W_{\delta}$ & $-\frac{29}{80}$ & 55 & $\nabla_{\alpha}\,Q_{\beta\gamma\delta}\,Q_{\beta\alpha\epsilon}\,Q_{\gamma\delta\epsilon}$ & $-\frac{11}{12}$\\
27 & $B_{\beta\gamma}\,Q_{\gamma\beta\delta}\,V_{\delta}$ & $\frac{5}{12}$ & 56 & $\nabla_{\alpha}\,Q_{\beta\gamma\delta}\,Q_{\beta\alpha\epsilon}\,Q_{\epsilon\gamma\delta}$ & $\frac{13}{24}$\\
28 & $B_{\beta\gamma}\,Q_{\delta\beta\gamma}\,W_{\delta}$ & $-\frac{7}{15}$ & 57 & $\nabla_{\alpha}\,Q_{\beta\gamma\delta}\,Q_{\epsilon\alpha\beta}\,Q_{\gamma\delta\epsilon}$ & $\frac{1}{2}$\\
29 & $B_{\beta\gamma}\,Q_{\delta\beta\gamma}\,V_{\delta}$ & $\frac{29}{60}$ & 58 & $\nabla_{\alpha}\,Q_{\beta\gamma\delta}\,Q_{\gamma\alpha\epsilon}\,Q_{\beta\delta\epsilon}$ & $\frac{7}{12}$\\
\end{tabular}
\end{ruledtabular}
\end{table*}

\begin{table*}[p]
\caption{Coefficients $u_{k}$ of the $65$ invariants $Q^{4}$, in the sense $\sum_{I}c_{I}I\supset\sum_{k}u_{k}\,(\sqrt{-g})^{1-w_{k}}\,\mathcal{O}^{Q^{4}}_{k}$, in $d=4$ and $\xi=1$, with the notation and weights of Table~\ref{tab:QQdQ}; the indices of each invariant are named in the order in which they appear. Unlike the $s_{k}$, the $u_{k}$ depend on the order of $\nabla$ and $g$ in the invariants $QQ\nabla Q$ that contain $W$, and hold for the convention of Sec.~\ref{sec:blocks}.}
\label{tab:Q4}
\renewcommand{\arraystretch}{1.7}
\begin{ruledtabular}
\begin{tabular}{rlc@{\qquad}rlc}
$k$ & invariant & $u_{k}$ & $k$ & invariant & $u_{k}$\\
\hline
1 & $V_{\alpha}\,W_{\alpha}\,V_{\beta}\,W_{\beta}$ & $\frac{437}{1440}$ & 34 & $V_{\alpha}\,Q_{\alpha\beta\gamma}\,Q_{\delta\beta\epsilon}\,Q_{\delta\gamma\epsilon}$ & $-\frac{113}{720}$\\
2 & $V_{\alpha}\,W_{\alpha}\,V_{\beta}\,V_{\beta}$ & $\frac{149}{720}$ & 35 & $V_{\alpha}\,Q_{\alpha\beta\gamma}\,Q_{\delta\beta\epsilon}\,Q_{\epsilon\gamma\delta}$ & $-\frac{11}{72}$\\
3 & $V_{\alpha}\,W_{\alpha}\,W_{\beta}\,W_{\beta}$ & $-\frac{829}{2880}$ & 36 & $V_{\alpha}\,Q_{\beta\alpha\gamma}\,Q_{\beta\delta\epsilon}\,Q_{\gamma\delta\epsilon}$ & $-\frac{59}{720}$\\
4 & $V_{\alpha}\,V_{\alpha}\,V_{\beta}\,V_{\beta}$ & $-\frac{209}{2880}$ & 37 & $V_{\alpha}\,Q_{\beta\alpha\gamma}\,Q_{\beta\delta\epsilon}\,Q_{\delta\gamma\epsilon}$ & $-\frac{31}{720}$\\
5 & $V_{\alpha}\,V_{\alpha}\,W_{\beta}\,W_{\beta}$ & $-\frac{119}{1440}$ & 38 & $V_{\alpha}\,Q_{\beta\alpha\gamma}\,Q_{\delta\beta\epsilon}\,Q_{\gamma\delta\epsilon}$ & $-\frac{83}{360}$\\
6 & $W_{\alpha}\,W_{\alpha}\,W_{\beta}\,W_{\beta}$ & $\frac{257}{5760}$ & 39 & $V_{\alpha}\,Q_{\beta\alpha\gamma}\,Q_{\delta\beta\epsilon}\,Q_{\delta\gamma\epsilon}$ & $-\frac{133}{720}$\\
7 & $V_{\alpha}\,Q_{\alpha\beta\gamma}\,W_{\beta}\,W_{\gamma}$ & $\frac{3}{20}$ & 40 & $V_{\alpha}\,Q_{\beta\alpha\gamma}\,Q_{\delta\beta\epsilon}\,Q_{\epsilon\gamma\delta}$ & $\frac{37}{360}$\\
8 & $V_{\alpha}\,Q_{\alpha\beta\gamma}\,W_{\beta}\,V_{\gamma}$ & $\frac{5}{48}$ & 41 & $W_{\alpha}\,Q_{\alpha\beta\gamma}\,Q_{\beta\delta\epsilon}\,Q_{\gamma\delta\epsilon}$ & $-\frac{181}{1440}$\\
9 & $V_{\alpha}\,Q_{\alpha\beta\gamma}\,V_{\beta}\,V_{\gamma}$ & $\frac{1}{12}$ & 42 & $W_{\alpha}\,Q_{\alpha\beta\gamma}\,Q_{\beta\delta\epsilon}\,Q_{\delta\gamma\epsilon}$ & $\frac{137}{360}$\\
10 & $V_{\alpha}\,Q_{\beta\alpha\gamma}\,W_{\beta}\,W_{\gamma}$ & $0$ & 43 & $W_{\alpha}\,Q_{\alpha\beta\gamma}\,Q_{\delta\beta\epsilon}\,Q_{\delta\gamma\epsilon}$ & $-\frac{79}{180}$\\
11 & $V_{\alpha}\,Q_{\beta\alpha\gamma}\,W_{\beta}\,V_{\gamma}$ & $\frac{5}{96}$ & 44 & $W_{\alpha}\,Q_{\alpha\beta\gamma}\,Q_{\delta\beta\epsilon}\,Q_{\epsilon\gamma\delta}$ & $-\frac{263}{1440}$\\
12 & $W_{\alpha}\,Q_{\alpha\beta\gamma}\,W_{\beta}\,W_{\gamma}$ & $\frac{37}{240}$ & 45 & $Q_{\alpha\beta\gamma}\,Q_{\alpha\beta\delta}\,Q_{\gamma\delta\epsilon}\,W_{\epsilon}$ & $\frac{77}{360}$\\
13 & $V_{\alpha}\,W_{\alpha}\,Q_{\beta\gamma\delta}\,Q_{\beta\gamma\delta}$ & $-\frac{293}{1440}$ & 46 & $Q_{\alpha\beta\gamma}\,Q_{\alpha\delta\epsilon}\,Q_{\beta\gamma\delta}\,W_{\epsilon}$ & $\frac{31}{720}$\\
14 & $V_{\alpha}\,W_{\alpha}\,Q_{\beta\gamma\delta}\,Q_{\gamma\beta\delta}$ & $\frac{35}{144}$ & 47 & $Q_{\alpha\beta\gamma}\,Q_{\alpha\delta\epsilon}\,Q_{\beta\delta\epsilon}\,W_{\gamma}$ & $\frac{77}{720}$\\
15 & $V_{\alpha}\,Q_{\alpha\beta\gamma}\,Q_{\beta\gamma\delta}\,W_{\delta}$ & $-\frac{97}{720}$ & 48 & $W_{\alpha}\,Q_{\beta\alpha\gamma}\,Q_{\delta\beta\epsilon}\,Q_{\gamma\delta\epsilon}$ & $-\frac{49}{144}$\\
16 & $V_{\alpha}\,Q_{\alpha\beta\gamma}\,Q_{\beta\gamma\delta}\,V_{\delta}$ & $-\frac{14}{45}$ & 49 & $W_{\alpha}\,Q_{\beta\alpha\gamma}\,Q_{\delta\beta\epsilon}\,Q_{\epsilon\gamma\delta}$ & $\frac{83}{360}$\\
17 & $V_{\alpha}\,Q_{\alpha\beta\gamma}\,Q_{\delta\beta\gamma}\,W_{\delta}$ & $-\frac{101}{720}$ & 50 & $Q_{\alpha\beta\gamma}\,Q_{\alpha\beta\gamma}\,Q_{\delta\epsilon\zeta}\,Q_{\delta\epsilon\zeta}$ & $-\frac{69}{640}$\\
18 & $V_{\alpha}\,Q_{\alpha\beta\gamma}\,Q_{\delta\beta\gamma}\,V_{\delta}$ & $\frac{71}{144}$ & 51 & $Q_{\alpha\beta\gamma}\,Q_{\alpha\beta\gamma}\,Q_{\delta\epsilon\zeta}\,Q_{\epsilon\delta\zeta}$ & $\frac{47}{320}$\\
19 & $V_{\alpha}\,V_{\alpha}\,Q_{\beta\gamma\delta}\,Q_{\beta\gamma\delta}$ & $\frac{371}{2880}$ & 52 & $Q_{\alpha\beta\gamma}\,Q_{\alpha\beta\delta}\,Q_{\gamma\epsilon\zeta}\,Q_{\delta\epsilon\zeta}$ & $\frac{7}{160}$\\
20 & $V_{\alpha}\,V_{\alpha}\,Q_{\beta\gamma\delta}\,Q_{\gamma\beta\delta}$ & $-\frac{103}{1440}$ & 53 & $Q_{\alpha\beta\gamma}\,Q_{\alpha\beta\delta}\,Q_{\gamma\epsilon\zeta}\,Q_{\epsilon\delta\zeta}$ & $-\frac{13}{240}$\\
21 & $V_{\alpha}\,Q_{\beta\alpha\gamma}\,Q_{\beta\gamma\delta}\,W_{\delta}$ & $-\frac{83}{360}$ & 54 & $Q_{\alpha\beta\gamma}\,Q_{\alpha\beta\delta}\,Q_{\epsilon\gamma\zeta}\,Q_{\epsilon\delta\zeta}$ & $\frac{73}{240}$\\
22 & $V_{\alpha}\,Q_{\beta\alpha\gamma}\,Q_{\beta\gamma\delta}\,V_{\delta}$ & $-\frac{449}{1440}$ & 55 & $Q_{\alpha\beta\gamma}\,Q_{\alpha\beta\delta}\,Q_{\epsilon\gamma\zeta}\,Q_{\zeta\delta\epsilon}$ & $-\frac{7}{32}$\\
23 & $V_{\alpha}\,Q_{\beta\alpha\gamma}\,Q_{\gamma\beta\delta}\,W_{\delta}$ & $-\frac{269}{360}$ & 56 & $Q_{\alpha\beta\gamma}\,Q_{\alpha\delta\epsilon}\,Q_{\beta\gamma\zeta}\,Q_{\delta\epsilon\zeta}$ & $-\frac{41}{480}$\\
24 & $V_{\alpha}\,Q_{\beta\alpha\gamma}\,Q_{\gamma\beta\delta}\,V_{\delta}$ & $\frac{71}{180}$ & 57 & $Q_{\alpha\beta\gamma}\,Q_{\alpha\delta\epsilon}\,Q_{\beta\gamma\zeta}\,Q_{\zeta\delta\epsilon}$ & $-\frac{47}{240}$\\
25 & $V_{\alpha}\,Q_{\beta\alpha\gamma}\,Q_{\delta\beta\gamma}\,W_{\delta}$ & $\frac{389}{720}$ & 58 & $Q_{\alpha\beta\gamma}\,Q_{\alpha\delta\epsilon}\,Q_{\beta\delta\zeta}\,Q_{\gamma\epsilon\zeta}$ & $-\frac{25}{48}$\\
26 & $W_{\alpha}\,W_{\alpha}\,Q_{\beta\gamma\delta}\,Q_{\beta\gamma\delta}$ & $\frac{73}{576}$ & 59 & $Q_{\alpha\beta\gamma}\,Q_{\alpha\delta\epsilon}\,Q_{\beta\delta\zeta}\,Q_{\epsilon\gamma\zeta}$ & $\frac{29}{96}$\\
27 & $W_{\alpha}\,W_{\alpha}\,Q_{\beta\gamma\delta}\,Q_{\gamma\beta\delta}$ & $-\frac{269}{2880}$ & 60 & $Q_{\alpha\beta\gamma}\,Q_{\alpha\delta\epsilon}\,Q_{\beta\delta\zeta}\,Q_{\zeta\gamma\epsilon}$ & $\frac{1}{48}$\\
28 & $W_{\alpha}\,Q_{\alpha\beta\gamma}\,Q_{\beta\gamma\delta}\,W_{\delta}$ & $\frac{47}{144}$ & 61 & $Q_{\alpha\beta\gamma}\,Q_{\alpha\delta\epsilon}\,Q_{\zeta\beta\gamma}\,Q_{\zeta\delta\epsilon}$ & $-\frac{19}{960}$\\
29 & $W_{\alpha}\,Q_{\alpha\beta\gamma}\,Q_{\delta\beta\gamma}\,W_{\delta}$ & $\frac{49}{2880}$ & 62 & $Q_{\alpha\beta\gamma}\,Q_{\beta\alpha\gamma}\,Q_{\delta\epsilon\zeta}\,Q_{\epsilon\delta\zeta}$ & $\frac{23}{960}$\\
30 & $Q_{\alpha\beta\gamma}\,Q_{\alpha\beta\delta}\,W_{\gamma}\,W_{\delta}$ & $-\frac{257}{1440}$ & 63 & $Q_{\alpha\beta\gamma}\,Q_{\beta\alpha\delta}\,Q_{\gamma\epsilon\zeta}\,Q_{\delta\epsilon\zeta}$ & $\frac{13}{80}$\\
31 & $W_{\alpha}\,Q_{\beta\alpha\gamma}\,Q_{\gamma\beta\delta}\,W_{\delta}$ & $-\frac{379}{720}$ & 64 & $Q_{\alpha\beta\gamma}\,Q_{\beta\alpha\delta}\,Q_{\epsilon\gamma\zeta}\,Q_{\zeta\delta\epsilon}$ & $-\frac{103}{480}$\\
32 & $V_{\alpha}\,Q_{\alpha\beta\gamma}\,Q_{\beta\delta\epsilon}\,Q_{\gamma\delta\epsilon}$ & $\frac{101}{720}$ & 65 & $Q_{\alpha\beta\gamma}\,Q_{\delta\alpha\beta}\,Q_{\gamma\epsilon\zeta}\,Q_{\epsilon\delta\zeta}$ & $\frac{37}{240}$\\
33 & $V_{\alpha}\,Q_{\alpha\beta\gamma}\,Q_{\beta\delta\epsilon}\,Q_{\delta\gamma\epsilon}$ & $-\frac{17}{90}$ &  & & \\
\end{tabular}
\end{ruledtabular}
\end{table*}
 \clearpage %

\bibliographystyle{apsrev4-2}

\end{document}